\documentclass{iopart}
\usepackage{graphicx}
 \usepackage{mathptmx} 
\usepackage{latexsym}
\usepackage[normalem]{ulem}
\usepackage{iopams}

\expandafter\let\csname equation*\endcsname\relax
\expandafter\let\csname endequation*\endcsname\relax

\usepackage{amsmath}

\usepackage{amsfonts,float,subfigure,verbatim,rotating}
\newcommand{\abs}[1]{\left| #1 \right|} 
 
\newcommand{\pd}[2]{\frac{\partial #1}{\partial #2}}

\newcommand{\uvec}[1]{\mathbf{#1}}
\newcommand{\dvec}[1]{\text{\textsf{#1}}}

\newcommand{\fvar}[1]{\frac{\mathrm{d}}{\mathrm{d}\,\alpha} \left[\,#1\,\right]_{\,\alpha=0}}

\begin{document}
\title{Computation of Unconstrained Elastic Equilibria of Complete M\"{o}bius Bands and their Stability}
 \author{Alexander Moore}
   \address{Field of Theoretical and Applied Mechanics, Cornell University, Ithaca, NY}
 \ead{amm456@cornell.edu}
 \author{Timothy J. Healey}
\address{Department of Mathematics, Cornell University, Ithaca, NY}
\submitto{\NL}
\vspace{1pc}

\begin{abstract}
Determining the equilibrium configuration of an elastic M\"{o}bius band is a challenging problem.  In recent years numerical results have been obtained by other investigators, employing first the Kirchhoff theory of rods and later the developable, ruled-surface model of Wunderlich.  In particular, the strategy employed previously for the latter does not deliver an unconstrained equilibrium configuration for the complete strip. Here we present our own systematic approach to the same problem for each of these models, with the ultimate goal of assessing the stability of flip-symmetric configurations.  The presence of pointwise constraints considerably complicates the latter step.  We obtain the first stability results for the problem, numerically demonstrating that such equilibria render the total potential energy a local minimum.  Along the way we introduce a novel regularization for the  for the singular Wunderlich model that delivers unconstrained equilibria for the complete strip, which can then be tested for stability.
\end{abstract}

\section{Introduction}
Sadowsky \cite{Sadowsky1930,Hinz2014} and later Wunderlich \cite{Wunderlich1962} were the first to propose models for determining the equilibrium configurations of elastic M\"{o}bius bands, idealizing them as a developable surfaces. In particular, linear isotropic plate theory is employed in \cite{Wunderlich1962} where an integration across the width yields an energy density per unit length, reminiscent of rod theory.  Obtaining numerical solutions for the latter is challenging and was only taken up recently, cf. \cite{Starostin2007}, \cite{Starostin2014}.  The interpretation of the model derived in \cite{Wunderlich1962} in light of classical Kirchhoff rod theory was made precise in the recent work \cite{Dias2014}.  In an earlier work \cite{Mahadevan1993}, Kirchhoff rod theory was employed to obtain certain smooth shapes of M\"{o}bius strips, in which one cross-sectional moment of inertia of the rod is much larger than the other.

     In both \cite{Mahadevan1993} and \cite{Starostin2007}, \cite{Starostin2014} the equilibrium equations are solved numerically on the half-domain with appropriate boundary conditions; assuming flip symmetry, the full solution is generated by rotation through $\pi$ radians about the symmetry axis.  That is, the entire closed-loop configuration has a single rotational symmetry (by $\pi$ radians) about some fixed axis.  The same approach was employed to obtain configurations of twisted isotropic rods in \cite{Domokos2001}.  There the two ends of a straight isotropic rod are twisted through any relative angle and then seamlessly joined, as compared to the M\"{o}bius band, where the relative angle is necessarily $\pi$ radians.  In \cite{Domokos2001} it is shown that all equilibrium configurations possess flip symmetry, i.e., the above solution procedure can be used without loss of generality.  That argument relies crucially on cross-sectional isotropy.  In particular, we know of no such result here for strips, i.e., other non-symmetrical solutions could exist.

     Given that, we address a more modest but nonetheless important question here, viz., we assess the local stability of the flip-symmetric solution.  Stable or not, this does not rule out the possibility of non-symmetrical solutions.  We consider the two distinct models employed in \cite{Mahadevan1993} and \cite{Starostin2007}, viz., the Kirchhoff rod model and the Wunderlich model, respectively. The former serves as a ``warm-up" for the latter.  Also the Kirchhoff model is a reasonable one for bands made of compliant materials like rubber. In order to test stability, we must first obtain reliable symmetric equilibria for the full closed loop, which is a challenging task. This is particularly true for the developable-surface model, due to the inherent singularity associated with the ruled-surface parametrization employed in  \cite{Wunderlich1962}.  For various reasons, discussed below, we do not employ the formulations of \cite{Mahadevan1993} and \cite{Starostin2007}, \cite{Starostin2014}.  Accordingly, the paper is taken up presenting our systematic formulations for both the numerical computation of symmetric equilibria and the assessment of their stability.  While the latter is certainly new, the former constitutes the first systematic approach to computing \emph{unconstrained} equilibria of complete strips for the developable-surface model of Wunderlich.
     
The outline of this work is as follows. In Section 2 we summarize the well-known field equations for hyperelastic, inextensible, unshearable Cosserat rods, ultimately adopting the constitutive assumption normally attributed to Kirchhoff.  In Section 3 we summarize our formulation, as first presented in \cite{Healey2005}, and compute solutions for the half rod with appropriate boundary conditions engendering flip symmetry.  We avoid the use of Euler angles and their associated singularities as in \cite{Mahadevan1993} and \cite{Starostin2014}; the kinematical description of the finite rotation field here is singularity-free via quaternions. As shown in \cite{Healey2005}, the exploitation of a ``conserved" quantity delivers a complete formulation within the context of a linear space.  As in \cite{Mahadevan1993} we use the ratio of the cross-sectional area moments of inertia as a continuation parameter \textendash\, starting from the well-known flat, circular equilibrium configuration.  In anticipation of our stability results, we then extend all solution fields \textendash \,kinematic and kinetic \textendash \,to the entire closed-loop configuration.  In Section 4 we briefly present our results for flip-symmetric equilibria.

     In a conservative problem such as the one at hand, it's enough to check the positivity of the (reduced) stiffness matrix at an equilibrium to deduce that the total potential energy is a local minimum there, i.e., the configuration is locally stable.  Unfortunately in the case of two-point boundary value problems, such information is not a direct by-product of the code AUTO.  However, the real difficulty here stems from pointwise constraints like inextensibility and unshearability, present in the problem at hand.  In Section 5 we employ the methodology of \cite{Kumar2010} to overcome this.  We first identify the discrete, numerical solution for the closed loop with a finite-element mesh, and then consider its linearization about the equilibrium configuration.  This yields a stiffness matrix in the presence of constraints. A QR-factorization of the constraint matrix enables the determination of the symmetric projected stiffness matrix, defined on the orthogonal complement of the subspace spanned by the constraints.  We then compute the smallest eigenvalues of the projected stiffness matrix.  In this way we numerically verify the stability of all closed-loop solutions found. 

\begin{figure}
\centering
\subfigure[Half-Strip Solution]{\includegraphics[width=.4\textwidth] {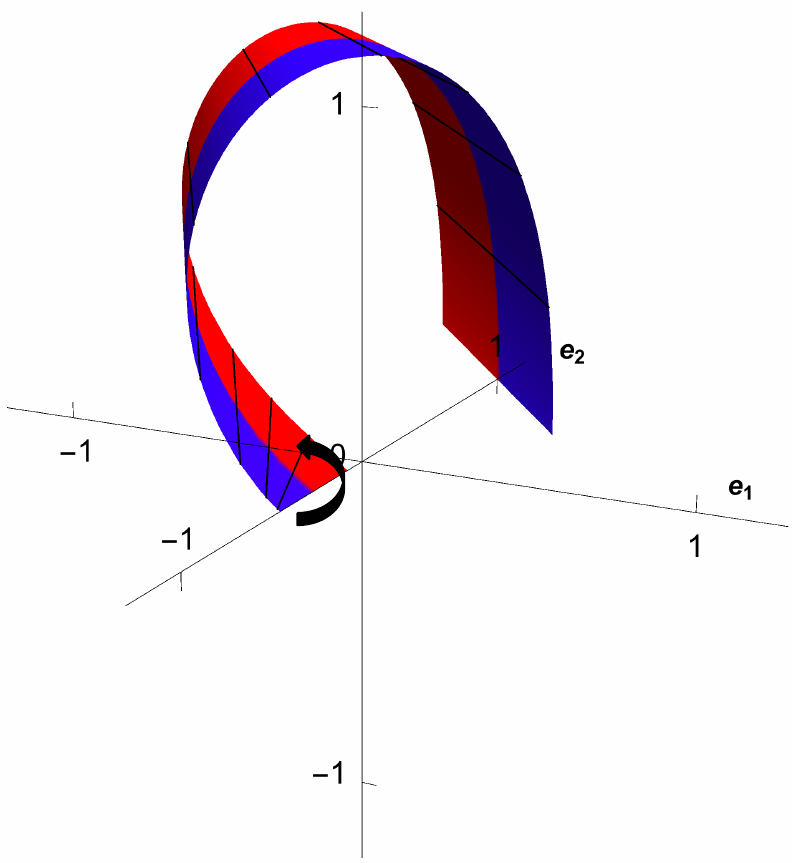}}\hspace{.1\textwidth}
\subfigure[Generated Full Strip]{\includegraphics[width=.4\textwidth]{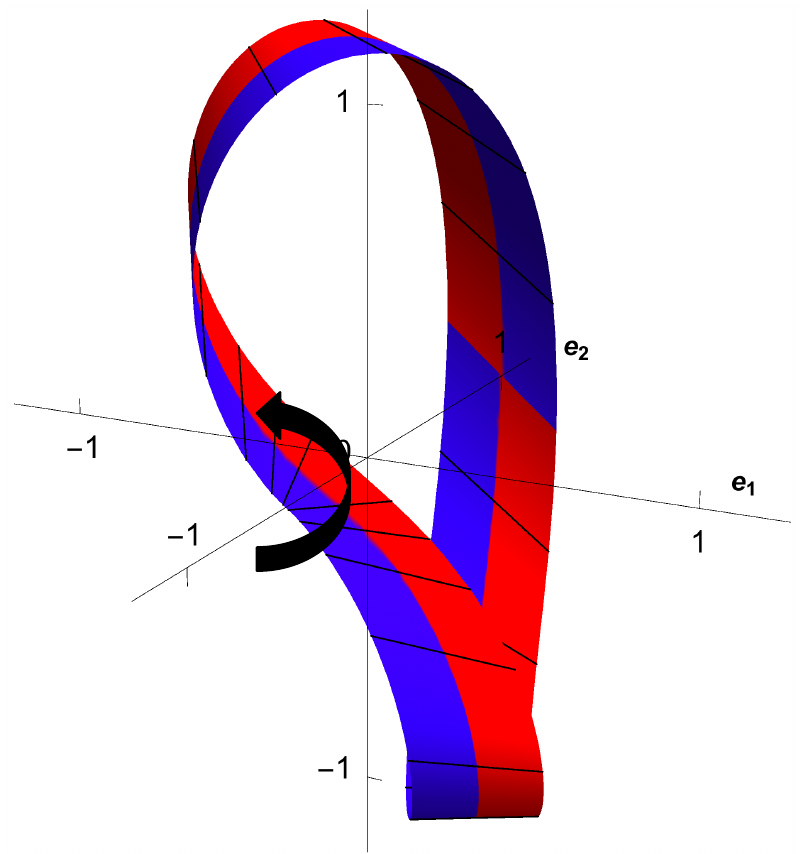}}
\caption{\footnotesize The half-strip solution and the constructed full strip for the developable rod. Indicated are the moments $m_{applied}$ and $2m_{applied}$ in the half strip and full strip respectively caused by specifying a non-zero curvature on the half strip's end. These are external applied moments concentrated at the end point of the half strip.}
\label{fig:dipole}
\end{figure}

     In Section 6 we take up the Wunderlich model, with the same goals in mind as above.  As first noted in \cite{Starostin2007}, but more clearly illuminated in \cite{Dias2014}, the resulting field equations are those of a Cosserat rod in the presence of an additional ``state variable" of a purely geometric nature.  In particular, the governing equation associated with the latter possesses a singularity wherever the curvature of the centerline curve vanishes \cite{Freddi}.  As observed in \cite{Starostin2014}, such a condition necessarily occurs at one end of the half-rod on the symmetry axis.  This renders the numerical determination of complete flip-symmetric configurations and their stability assessment much more difficult.  In \cite{Starostin2014}, a small external curvature is imposed at one end of the half-band in order to overcome the singularity that is otherwise present at that location.   This is equivalent to the presence of a small externally applied moment at that end.  Consequently, \emph{this method does not yield an unconstrained equilibrium configuration}:  When the half configuration is rotated about the symmetry axis through an angle of  $\pi$, the small applied moment is doubled in magnitude and acts externally on the complete band, as shown in Figure  \ref{fig:dipole}.  In the structural mechanics literature this is often referred to as anti-symmetry, cf. \cite{Vanderbilt}. In any case, if the boundary supports for this half-rod are removed, the full band does not satisfy global moment balance.  Here we take a different approach to obtain unconstrained equilibria for the full strip.  We use the same rod formulation described above but now add an internal ``elliptic regularization" associated with the state variable, characterized by a small parameter, to the potential energy derived in \cite{Wunderlich1962}. The resulting Euler-Lagrange equation for the state variable is now singularly perturbed but not singular, and the extension of the solution to a complete configuration does not give rise to an unbalanced external moment.  With this in hand, we carry out the same strategy described above for the Kirchhoff rod, but accompanied by taking the regularizing parameter as small as possible in the continuation scheme.  
     
     Another point of departure from the strategy used in the Kirchhoff model is that a known configuration to initiate continuation here is not at all obvious.  In Section 7 we start from a circularly bent, untwisted half strip of fixed width; this requires the application of an end moment at the hinged end.  We then execute a two-parameter continuation \textendash \, first relaxing the end moment and then aligning the hinge at the appropriate $\pi/2$ orientation.  Hereafter, the width and small regularizing parameter are employed as continuation parameters to obtain half-band configurations.  In Section 8 we present our results for flip-symmetric closed-loop configurations.  In Section 9 we take up the assessment of their stability.  We first extend the development in \cite{Kumar2010} to the more complicated problem at hand.  Then after a careful extension of all computed fields on the half strip to the full closed loop, the computational procedure for obtaining the projected stiffness matrix is the same as above.  Again, we find that all computed flip-symmetric configurations correspond to local energy minima. 

\section{Elastic Rod Formulation}\label{sec:kform}
Let $\left\{\uvec{e}_1,\uvec{e}_2,\uvec{e}_3\right\}$ denote a fixed, right-handed, orthonormal basis for $\text{\bf E}^3$, the translate space for 3-dimensional Euclidean point space. We start by defining the special Cosserat rod with centerline coordinate $s\in\left[0,L\right]$ in a straight, stress-free reference configuration. The position of the rod is defined by the vector-valued function $\mathbf{r}:\left[0,L\right]\rightarrow\mathbb{R}^3$ with the reference configuration's centerline given by $\mathbf{r}_{0}\left(s\right) = s\,\uvec{e}_3$. The cross-sections of the rod in the reference configuration are parallel to the plane $\text{span}\left\{\uvec{e}_1,\uvec{e}_2\right\}$. Let $\mathbf{R}\left(s\right)$ denote the rotation of the cross-sectional plane parallel to $\text{span}\left\{\uvec{e}_1,\uvec{e}_2\right\}$ at $s$ in the undeformed rod.

We define an orthonormal basis field $\left\{\uvec{d}_1,\uvec{d}_2,\uvec{d}_3\right\}$ via
\begin{align}\label{eq:director}
\uvec{d}_i \left(s\right)&= \mathbf{R}\left(s\right)\,\uvec{e}_i \,.
\end{align}
The configuration of the rod is uniquely determined by the functions $\mathbf{r}\left(s\right)$ and $\mathbf{R}\left(s\right)$. The director basis field, $\left\{\uvec{d}_1,\uvec{d}_2,\uvec{d}_3\right\}$ is attached to the centroid of the rod's cross section and defines the orientation of the rod's cross sections. Differentiation of (\ref{eq:director}) yields
\begin{align}\label{eq:dprimes}
\uvec{d}_i^\prime  =\boldsymbol{\kappa}\,\times\,\uvec{d}_i\,,
\end{align}
where $\mathbf{a}\times\mathbf{b}$ denotes the usual right-handed cross product, $\prime$ denotes a derivative with respect to the centerline coordinate, $s$, and $\boldsymbol{\kappa}$ is the axial vector field of the skew symmetric tensor field $\mathbf{R}^\prime\,\mathbf{R}^T$, denoted
\begin{equation}\label{eq:kappaaxial}
\boldsymbol{\kappa}=\text{axial}\left(\mathbf{R}^\prime\,\mathbf{R}^T\right)\,.
\end{equation}
We consider here, \emph{unshearable} and \emph{inextensible} rods, viz., we impose the constraint
\begin{align}\label{eq:unshear_inext}
\mathbf{r}^\prime \equiv\mathbf{d}_3\,.
\end{align}
We also write
\begin{align}\label{eq:rprime}
\uvec{\boldsymbol{\kappa}} & = \kappa_i\,\uvec{d}_i\,,
\end{align}
where here and throughout repeated Latin subscripts imply summation from $1$ to $3$, while repeated Greek subscripts sum from $1$ to $2$. The scalar fields $\kappa_i$, $i=1,2,3$, are the strains of the theory: $\kappa_1, \kappa_2$ are components of the curvature or bending strains, while $\kappa_3$ is the twist or torsional strain.

The vector fields $\uvec{n}\left(s\right)$ and $\uvec{m}\left(s\right)$, denote the internal contact force and contact couple, respectively, acting on the deformed cross section at ``s". 
We write
\begin{align}
\uvec{n}&=n_i\,\uvec{d}_i\,,\\
\uvec{m}&=m_i\,\uvec{d}_i\,,\label{eq:mveckinematics}
\end{align}
where the component fields, $n_i$ and $m_i$, $i=1,2,3$, are the internal forces and moments respectively: $n_1,n_2$ correspond to shear forces; $n_3$ axial force; $m_1,m_2$ correspond to bending moments, and $m_3$ to torque or twisting moment. In the absence of body forces and body couples, the local form of force and moment balance are given by
\begin{align}\label{eq:govfixed1}
\uvec{n}^\prime & = 0\,,\\\label{eq:govfixed2}
\uvec{m}^\prime+\uvec{d}_3\times\uvec{n}&=0\,,
\end{align}
respectively where we have used (\ref{eq:unshear_inext}) \cite{Antman2005}.

Since the tangent vector of the centerline, $\mathbf{r}^\prime$ is constrained to be $\uvec{d}_3$, the rod contact force, $\uvec{n}$, is not constitutively determined. In other words, the contact force serves as a Lagrange multiplier enforcing the unshearable-inextensible constraint (\ref{eq:unshear_inext}).

We define an \emph{objective}, \emph{hyperelastic}, inextensible and unshearable rod as one characterized by the existence of a non-negative $C^2$ function $W:\mathbb{R}^3\rightarrow\left[0,\infty\right)$, called the \emph{stored energy density}, such that 
\begin{equation}\label{eq:hyperm}
m_i=\pd{W}{\kappa_i}\quad, \quad i=1,2,3\,.
\end{equation}
For notational convenience, we denote the following triples of real number via
\begin{align}\label{eq:dvecdef}
\dvec{k}:=\left(\kappa_1,\kappa_2,\kappa_3\right)\quad \dvec{n}:=\left(n_1,n_2,n_3\right)\quad \dvec{m}:=\left(m_1,m_2,m_3\right)
\end{align}
Writing $W\left(\dvec{k}\right)\equiv W\left(\kappa_1,\kappa_2,\kappa_3\right)$, then (\ref{eq:hyperm}) takes the compact form
\begin{equation}\label{eq:comphyperm}
\dvec{m} = \frac{\mathrm{d}W}{\mathrm{d}\dvec{k}}\,.
\end{equation}
With the aid of (\ref{eq:director})-(\ref{eq:dprimes}), (\ref{eq:unshear_inext})-(\ref{eq:rprime}) and (\ref{eq:comphyperm}), the balance equations (\ref{eq:govfixed1}) and (\ref{eq:govfixed2}) take the form
\begin{align}
\dvec{n}^\prime +\dvec{k}\times\dvec{n} &=0\label{eq:neq}\,,\\
\dvec{m}^\prime+\dvec{k}\times\dvec{m}+\hat{\dvec{d}}\times\dvec{n}\label{eq:meq}&=0\,,
\end{align}
respectively with $\hat{\dvec{d}}:=\left(0,0,1\right)$. Equations (\ref{eq:director})-(\ref{eq:dprimes}) and (\ref{eq:rprime}) yield the following compatibility equations:
\begin{align}
\tilde{r}^\prime & = \tilde{R}\,\hat{\dvec{d}}\label{eq:rrprime}\,,\\
\tilde{R}^\prime & = \tilde{R}\,\dvec{K}\label{eq:Rprime}\,,
\end{align}
where $\dvec{K}$ is the unique skew symmetric matrix satisfying $\dvec{k}=\text{axial}\left(\dvec{K}\right)$ and $\tilde{R} $ is the matrix of $\uvec{R}$ relative to the fixed basis $\left\{\uvec{e}_1,\uvec{e}_2,\uvec{e}_3\right\}$. Henceforth, all components written with respect to the fixed $\left\{\uvec{e}_i\right\}$ basis are denoted by an over-tilde, e.g. $\uvec{r}^\prime=\tilde{r}^\prime_i\uvec{e}_i$, $\tilde{r}=\left(\tilde{r}_1,\tilde{r}_2,\tilde{r}_3\right)$, while components expressed with respect to the convected $\left\{\uvec{d}_i\right\}$ basis are written in san-serif font as in (\ref{eq:dvecdef}).

We employ the stored energy density according to the \emph{Kirchhoff model}, viz.,
\begin{align}\label{eq:rodenergy}
W\left(\dvec{k}\right)=\frac{1}{2}\left(\,E\,I_{1}\,\kappa_1^2+E\,I_{2}\,\kappa_2^2+G\,J\,\kappa_3^2\,\right)\,,
\end{align}
where $E$ denotes the Young's modulus of elasticity, $G$ denotes the shear modulus, $I_{\ell}$ is the area moment of inertia, $\ell=1,2$, and $J$ denotes the (weighted) polar area moment of inertia \cite{DenHartog1952}. 

We now normalize the system's variables to non-dimensional form for a strip of length $2\pi$ as follows:
\begin{align}
\bar{s}=2\pi\,\frac{s}{L}\,,\hspace{.04\textwidth} &\hspace{.04\textwidth}\label{eq:norm1k}
\frac{\mathrm{d}}{\mathrm{d}s}=\frac{2\pi}{L}\,\frac{\mathrm{d}}{\mathrm{d}\bar{s}}\,,\hspace{.1\textwidth} 
\bar{r}=\frac{\tilde{r}}{L}\,,
\hspace{.08\textwidth} \bar{\kappa}_i=\frac{L}{2\pi}\,\kappa_i\,,\\&\hspace{.04\textwidth}
\bar{\dvec{n}}_i=\frac{\dvec{n}_i\,L^2}{4\pi^2\,EI_1}\,, \hspace{.1\textwidth} 
\bar{\dvec{m}}_i=\frac{\dvec{m}_i\,L}{2\pi\,EI_1}\,,\label{eq:norm2k}
\end{align}
where $L$ is the original length of the strip. With the exceptions of placing an over-bar above each quantity, the differential equations in (\ref{eq:neq})-(\ref{eq:Rprime}) are unchanged by this normalization. The normalized constitutive relations based on the stored energy density in (\ref{eq:rodenergy}) are
\begin{align}\label{eq:genconst1}
\bar{\dvec{m}}_1&=\bar{\kappa}_1\,,\\
\bar{\dvec{m}}_2&=\lambda\,\bar{\kappa}_2\,,\\
\bar{\dvec{m}}_3&=\gamma\,\bar{\kappa}_3\,,\label{eq:genconst3}
\end{align}
where
 \begin{align}
 \lambda = \frac{I_2}{I_1}\,, \hspace{.05\textwidth}\text{and}\hspace{.05\textwidth}
 \gamma  =  \frac{G\,J}{E\,I_1}\,.
 \end{align}
Assume that the rod's cross section is rectangular with width of length $w$ in the $\uvec{d}_1$ direction and and height of length $h$ in the $\uvec{d}_2$ direction. The parameter $\lambda$ is
\begin{align}
\lambda =  \frac{I_2}{I_1} =\left( \frac{w}{h}\right)^2=\left( \frac{\bar{w}}{\bar{h}}\right)^2\,.
\end{align}
for rectangular cross sections where $\bar{w}$ and $\bar{h}$ are the width and height of the cross section respectively normalized by the rod length $L$. Note that $\lambda=1$ corresponds to a rod with equal bending stiffnesses, while $\lambda<<1$ or $\lambda>>1$ corresponds to a rod with one very compliant bending direction and one very stiff bending direction, e.g. a thin strip. 

 Formulas from strength of materials (cf. \cite{DenHartog1952}) give
\begin{align}
\gamma= \frac{2}{1+\nu}\,.
\end{align}
where $\nu$ is Poisson's ratio. For numerical calculations, $\nu = 1/3$ is used which corresponds to $\gamma=3/2$.
For clarity we now remove the overbars, with the understanding that all quantities are henceforth normalized according to (\ref{eq:norm1k})-(\ref{eq:norm2k}).

\section{Elastic Rod Solution Method}\label{sec:solutionmethod}
Following the approach in \cite{Domokos2001}, \cite{Mahadevan1993} and \cite{Starostin2007}, we search for closed loop solutions of (\ref{eq:neq})-(\ref{eq:Rprime}) that posses a \emph{flip symmetry} about, say, the $\uvec{e}_2$, axis. That is, we suppose that a rotation by $180$ degrees of the closed rod about the $\uvec{e}_2$ axis leaves the configuration unchanged. Hence we solve (\ref{eq:neq})-(\ref{eq:Rprime})  for half of the rod with appropriate boundary conditions (detailed in section \ref{sec:bc}), and generate a full loop solution by symmetry (detailed in section \ref{sec:fullrodrecon}). 

The resulting two-point boundary value problem for the half rod is solved on $\left[0,\pi\right]$ using numerical continuation via the software package AUTO  \cite{Doedel2009}. From this half solution, we generate a solution for the full M\"{o}bius strip on $\left[0,2\pi\right]$. As in \cite{Mahadevan1993}, the continuation is started from the equilibrium configuration of a twisted rod with equal bending stiffnesses, and then the path of equilibria is followed as the constitutive parameter $\lambda$ increases with $\gamma$ fixed.
\subsection{Parameterization}
Following the treatment in \cite{Healey2005}, $\tilde{R}$ in (\ref{eq:rrprime}) is parameterized via quaternions, thus avoiding the usual singularities associated with Euler angles. Accordingly (\ref{eq:rrprime})-(\ref{eq:Rprime}) are replaced by
 \begin{align}
\tilde{r}^\prime &= \tilde{R}\left(\uvec{q}\right)\,\hat{\dvec{d}}\label{eq:req}\,,\\
\uvec{q}^\prime & = \tilde{A}\left(\uvec{q}\right)\,\dvec{k}\label{eq:qeq}\,,
\end{align}
respectively, with $\uvec{q} := \left(q_0,q_1,q_2,q_3\right)$ subject to the normalization
\begin{align}
 q_0^2+q_1^2+q_2^2+q_3^2=1\label{eq:qconstr}\,.
 \end{align}
The quaternion parameterization of the rotation matrix, $\tilde{R}\left(\uvec{q}\right)$, and the quaternion differential equation matrix $\tilde{A}\left(\uvec{q}\right)$ are  (cf. \cite{Healey2005}).
 \begin{align}
 \tilde{R}\left(\uvec{q}\right) &=\label{eq:rqmatrix} 2\begin{pmatrix}q_0^2+q_1^2-1/2&q_1q_2-q_0q_3&q_1q_3+q_0q_2\\q_1q_2+q_0q_3&q_0^2+q_2^2-1/2&q_2q_3-q_0q_1\\q_1q_3-q_0q_2&q_2q_3+q_0q_1&q_0^2+q_3^2-1/2\end{pmatrix}\,,\\\nonumber \\
\tilde{A}\left(\uvec{q}\right) &= \frac{1}{2}\begin{pmatrix}-q_1&-q_2&-q_3\\ q_0&-q_3&q_2\\q_3&q_0&-q_1\\-q_2&q_1&q_0\end{pmatrix}\,.
 \end{align}
In general, an accurate numerical solution of (\ref{eq:req})-(\ref{eq:qeq}) (satisfying reasonable boundary conditions) need not satisfy (\ref{eq:qconstr}) with accuracy. We follow the approach in \cite{Healey2005} and replace (\ref{eq:qeq}) with the augmented equation containing a multiplier $\mu\in\mathbb{R}$:
\begin{align}\label{eq:qeqnew}
\uvec{q}^\prime & = \tilde{A}\left(\uvec{q}\right)\,\dvec{k}+\mu\,\uvec{q}\,.
\end{align}
Use of (\ref{eq:qeqnew}) ensures that (\ref{eq:qconstr}) will be satisfied identically along the entire length of the rod whenever (\ref{eq:qconstr}) is merely enforced on the boundary points. In practice, it turns out that the multiplier $\mu$ takes on numerical values close to zero (typically $\mu=O\left(10^{-8}\right)$), cf. \cite{Healey2005}. 

Combining (\ref{eq:neq})-(\ref{eq:meq}), (\ref{eq:genconst1})-(\ref{eq:genconst3}), (\ref{eq:req}) and (\ref{eq:qeqnew}) we arrive at the full governing system:
\begin{align}\label{eq:sysstart}
\dvec{n}^\prime +\tilde{\dvec{k}}\left(\dvec{m}\right)\times\dvec{n} &=0\,,\\
\dvec{m}^\prime+\tilde{\dvec{k}}\left(\dvec{m}\right)\times\dvec{m}+\hat{\dvec{d}}\times\dvec{n}&=0\,,\\
 \tilde{r}^\prime - \tilde{R}\left(\uvec{q}\right)\,\hat{\dvec{d}} & = 0\,,\\
\uvec{q}^\prime-\tilde{A}\left(\uvec{q}\right)\,\tilde{\dvec{k}}\left(\dvec{m}\right)-\mu\,\uvec{q} & = 0\,.\label{eq:sysend}
 \end{align}
where
\begin{align}\label{eq:sysend2}
\tilde{\dvec{k}}\left(\dvec{m}\right):=\left(m_1,\frac{m_2}{\lambda},\frac{m_3}{\gamma}\right)\,.
\end{align}
\subsection{Boundary Conditions}\label{sec:bc}
\begin{figure}
\centering
\includegraphics[width=.6\textwidth]{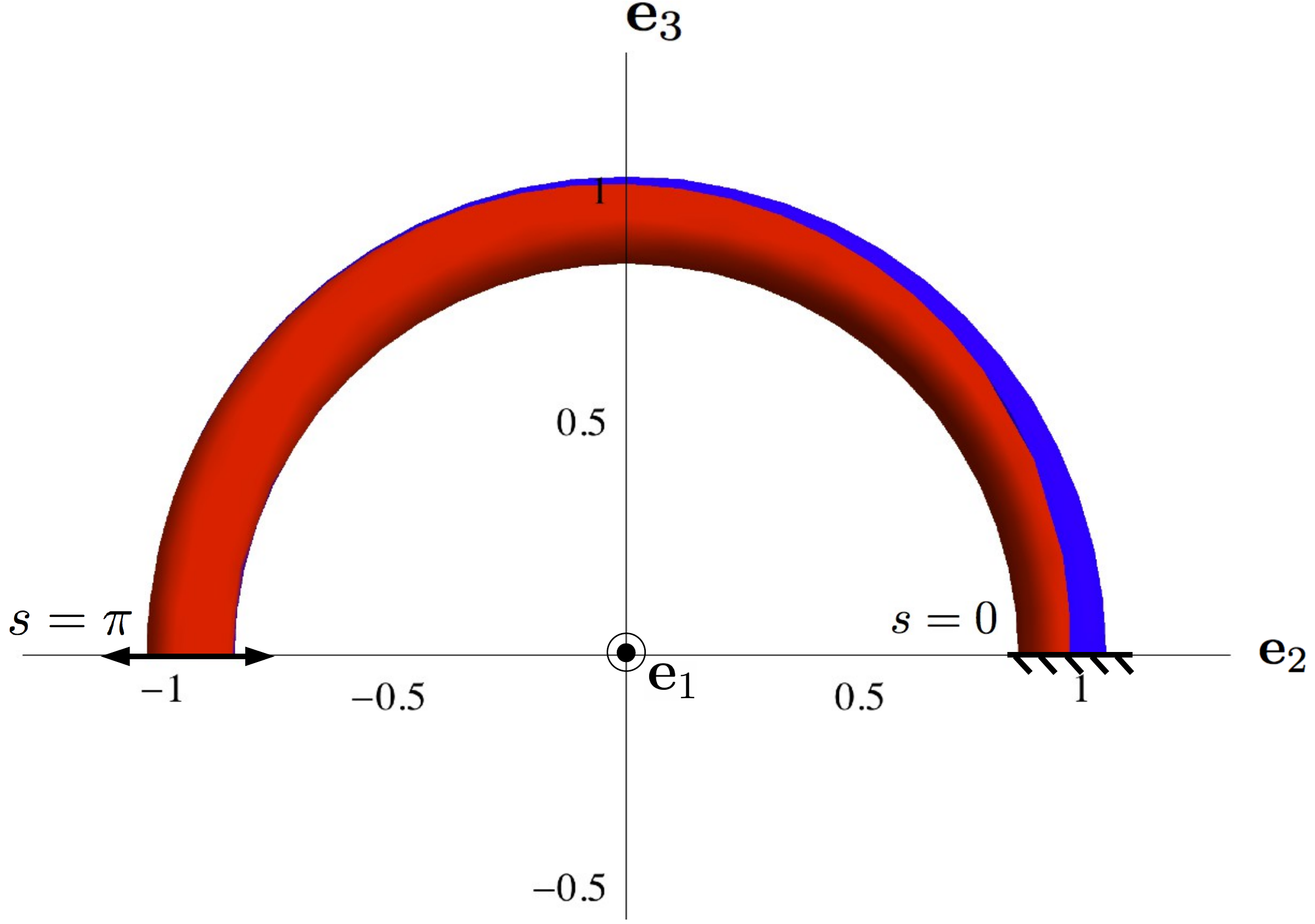}
\caption{\footnotesize The rod in the initial equilibrium configuration used to start the continuation calculations. It has a fixed end at $s=0$ and a hinged end free to slide along the $\uvec{e}_2$ axis at $s=\pi$. The coloring is presented to show orientation of the material points of the rod as the cross-sections rotate clockwise about $\uvec{d}_3$ from $s=0$ to $s=\pi$. The complete closed rod is formed via a reflection about the $\uvec{e}_2$ axis.}
\label{fig:bcsetup}
\end{figure}
Equations (\ref{eq:sysstart})-(\ref{eq:sysend}) constitute a system of first order ODE's in 14 unknowns $\left(\bar{r},\dvec{n},\dvec{m},\uvec{q},\mu\right)$ and the material parameters $\lambda$, $\gamma$. Thus the problem requires 14 boundary conditions. We consider a half-rod of length $\pi$ (so that the M\"{o}bius strip has total length $2\pi$).
The position and orientation of the rod at $s=0$ are fixed at a particular point in $\mathbb{R}^3$, yielding
\begin{align}\label{eq:bc0start}
&\mathbf{r}\left(0\right)\cdot\uvec{e}_1 =\tilde{r}_1\left(0\right) = 0\,,\\
&\mathbf{r}\left(0\right)\cdot\uvec{e}_2 =\tilde{r}_2\left(0\right) = 1\,,\\
&\mathbf{r}\left(0\right)\cdot\uvec{e}_3 =\tilde{r}_3\left(0\right) = 0\,,\\
&\uvec{q}\left(0\right)  = \left(q_0,q_1,q_2,q_3\right) = \left(1,0,0,0\right)\label{eq:bc0end}\,.
\end{align}
This gives seven boundary conditions. We note that (\ref{eq:bc0end}) fulfills (\ref{eq:qconstr}) at the ``left" boundary point.

For the boundary conditions at $s=\pi$, we place a perfect hinge parallel to $\uvec{e}_2$, along which the rod may freely slide and about which freely rotate. From this, the boundary conditions at $s=\pi$ become
\begin{align}\label{eq:bc1start}
&\mathbf{r}\left(\pi\right)\cdot\uvec{e}_1 =\tilde{r}_1\left(\pi\right) = 0\,,\\\label{eq:bc1mid2}
&\mathbf{r}\left(\pi\right)\cdot\uvec{e}_3 =\tilde{r}_3\left(\pi\right) = 0\,, \\
&\uvec{n}\left(\pi\right)\cdot\uvec{e}_2  = \dvec{n}_1\left(\pi\right)  =0\,, \label{eq:bc1mid}\\
&\uvec{m}\left(\pi\right)\cdot\uvec{e}_2  = \dvec{m}_1\left(\pi\right)  =0\,. \label{eq:bc1end}
\end{align}
In addition, at the end the rod will twist by a quarter turn relative to the orientation in (\ref{eq:bc0end}). Thus the directors at $s=\pi$ have the form
\begin{align}
\label{eq:directors1}
\uvec{d}_1\left(\pi\right) & =- \uvec{e}_2\,, \\ 
\uvec{d}_2\left(\pi\right) & = \cos{\beta}\,\uvec{e}_1+\sin{\beta}\,\uvec{e}_3\,, \\ 
\uvec{d}_3\left(\pi\right) & =-\sin{\beta}\,\uvec{e}_1+\cos{\beta}\,\uvec{e}_3\,,\label{eq:directors2}
\end{align}
where $\beta$ is some unspecified angle. Comparing the rotation matrix specified by (\ref{eq:directors1})-(\ref{eq:directors2}) with the parameterization of $\tilde{R}\left(\mathbf{q}\right)$, cf. (\ref{eq:rqmatrix}), we may choose the two non-redundant conditions
 \begin{align}\label{eq:q1bc}
q_0^2\left(\pi\right)+q_2^2\left(\pi\right)-\frac{1}{2}&=0\,,\\
q_2\left(\pi\right)\,q_3\left(\pi\right)-q_0\left(\pi\right)\,q_1\left(\pi\right)&=0\,.
\end{align}
In addition, we impose the normalization
\begin{align}
 q_1^2\left(\pi\right)+q_2^2\left(\pi\right)+q_3^2\left(\pi\right)+q^2_0\left(\pi\right)=1\,,\label{eq:bcqend}
\end{align}
fulfilling  (\ref{eq:qconstr}) at the boundary point. This completes the required set of 14 boundary conditions.

\subsection{Initial Equilibrium Configuration}
As in \cite{Mahadevan1993}, the starting equilibrium configuration for our continuation scheme is a rod with $\lambda=1$ (i.e. circular cross-sections made of an isotropic material) and $\gamma=1.5$ deformed in a semi-circular configuration which is rotated clockwise (from the viewpoint of $s=0$) by a total angle of $\pi/2$ at $s=\pi$, (Figure \ref{fig:bcsetup}). The configuration for the half-rod defines the configuration of the full M\"{o}bius strip on $s\in \left[0,2\pi\right]$ via reflection about the $\uvec{e}_2$ axis.

\subsection{Full Rod Construction}\label{sec:fullrodrecon}
Once a numerical solution for (\ref{eq:sysstart})-(\ref{eq:sysend2}), (\ref{eq:bc0start})-(\ref{eq:bc0end}), (\ref{eq:bc1start})-(\ref{eq:bc1end}), (\ref{eq:q1bc})-(\ref{eq:bcqend})  is obtained for $s\in[0,\pi]$, the rest of the closed-loop for $s\in[\pi,2\pi]$ is constructed via a rotation by $180$ degrees about the $\uvec{e}_2$-axis.  The following procedure is rigorously detailed in \cite{Domokos2001}.  Denote the calculated solution to (\ref{eq:sysstart})-(\ref{eq:sysend}) for $s\in[0,\pi]$ by a superscript $``c"$, e.g. $\uvec{r}^c\left(s\right)$ for the calculated rod centerline position. The flip across the axis of symmetry is given by
\begin{align}\label{eq:operator}
\mathbf{E} = -\,\left(\uvec{e}_1\otimes\uvec{e}_1\right)\,+ \,\left(\uvec{e}_2\otimes\uvec{e}_2\right) -\,\left(\uvec{e}_3\otimes\uvec{e}_3\right)\,.
\end{align}
The position of the centerline for $s\in[0,2\pi]$ is given by
\begin{align}\label{eq:rflip}
\uvec{r}\left(s\right)=\begin{cases}
\uvec{r}^c\left(s\right)\quad& s\in[0,\pi]\\
\mathbf{E}\,\uvec{r}^c\left(2\pi-s\right)\quad& s\in[\pi,2\pi]\\
\end{cases}\,.
\end{align}
The continuity of $\uvec{r}\left(\cdot\right)$ at $s=\pi$ follows from (\ref{eq:bc1start}), (\ref{eq:bc1mid2}), and (\ref{eq:operator}). It follows similarly that $\uvec{r}\left(0\right)=\uvec{r}\left(2\pi\right)$. 

The extension of the rod's orientation on $\left[\pi,2\pi\right]$ is defined in terms of the director fields $\uvec{d}_i\left(s\right)$:
\begin{align}\label{eq:d1flip}
\uvec{d}_1\left(s\right)&=\begin{cases}
\uvec{d}_1^c\left(s\right)\quad& s\in[0,\pi]\\
\mathbf{E}\,\uvec{d}_1^c\left(2\pi-s\right)\quad& s\in[\pi,2\pi]\\
\end{cases}\,,\\\label{eq:d2flip}
\uvec{d}_2\left(s\right)&=\begin{cases}
\uvec{d}_2^c\left(s\right)\quad& s\in[0,\pi]\\
-\mathbf{E}\,\uvec{d}_2^c\left(2\pi-s\right)\quad& s\in[\pi,2\pi]
\end{cases}\,,\\\label{eq:d3flip}
\uvec{d}_3\left(s\right)&=\begin{cases}
\uvec{d}_3^c\left(s\right)\quad& s\in[0,\pi]\\
-\mathbf{E}\,\uvec{d}_3^c\left(2\pi-s\right)\quad& s\in[\pi,2\pi]
\end{cases}\,.
\end{align}
Observe that $\uvec{d}_i\left(\cdot\right)$, $i=1,2,3$ is continuous on $\left[0,2\pi\right]$.

While the rod position and orientation are enough to reproduce the equilibrium configuration for the closed loop, we also give the extensions of the contact force and contact couple fields, $\uvec{n}$ and $\uvec{m}$ respectively, mainly for use in the stability analysis presented in Section \ref{sec:kstab}. Following the results in \cite{Domokos2001}, the required extensions are given by:
\begin{align}\label{eq:nflip}
\uvec{n}\left(s\right)&=\begin{cases}
\uvec{n}^c\left(s\right)\quad& s\in[0,\pi]\\
-\mathbf{E}\,\uvec{n}^c\left(2\pi-s\right)\quad& s\in[\pi,2\pi]\\
\end{cases}\,,\\
\label{eq:mflip}
\uvec{m}\left(s\right)&=\begin{cases}
\uvec{m}^c\left(s\right)\quad& s\in[0,\pi]\\
-\mathbf{E}\,\uvec{m}^c\left(2\pi-s\right)\quad& s\in[\pi,2\pi]\\
\end{cases}\,.
\end{align}
In view of (\ref{eq:bc1mid})-(\ref{eq:bc1end}), we see that $\uvec{n}\left(\cdot\right)$ and  $\uvec{m}\left(\cdot\right)$ are each continuous at $s=\pi$. We further claim that $\uvec{n}\left(0\right)=\uvec{n}\left(2\pi\right)$ and $\uvec{m}\left(0\right)=\uvec{m}\left(2\pi\right)$. To see this, note that the global balance of forces and moments for the half rod on $\left[0,\pi\right]$, together with (\ref{eq:bc1mid})-(\ref{eq:bc1end}) reveal that $\uvec{n}\left(0\right)\cdot\uvec{e}_2=\uvec{m}\left(0\right)\cdot\uvec{e}_2=0$. The claim now follows directly from (\ref{eq:nflip})-(\ref{eq:mflip}).
\section{Kirchhoff Rod Theory Results}\label{sec:kirchoff}
The system (\ref{eq:sysstart})-(\ref{eq:sysend2}), subject to boundary conditions (\ref{eq:bc0start})-(\ref{eq:bc0end}), (\ref{eq:bc1start})-(\ref{eq:bc1end}), and (\ref{eq:q1bc})-(\ref{eq:bcqend}) is solved via collocation methods by AUTO-07p Continuation and Bifurcation software \cite{Doedel2007,Doedel2009}. The rod is divided into 30 mesh intervals with 4 collocation points per interval for a total of 121 nodes along the interval $[0,\pi]$, and the mesh is updated every three continuation steps. These numbers were chosen based on recommended values given in \cite{Doedel2007}, and increasing the number of nodes further did not lead to a significant quantitative difference in the numerical results. 

\begin{figure}
\centering
\subfigure[$\lambda=1$]{\includegraphics[width=.45\textwidth] {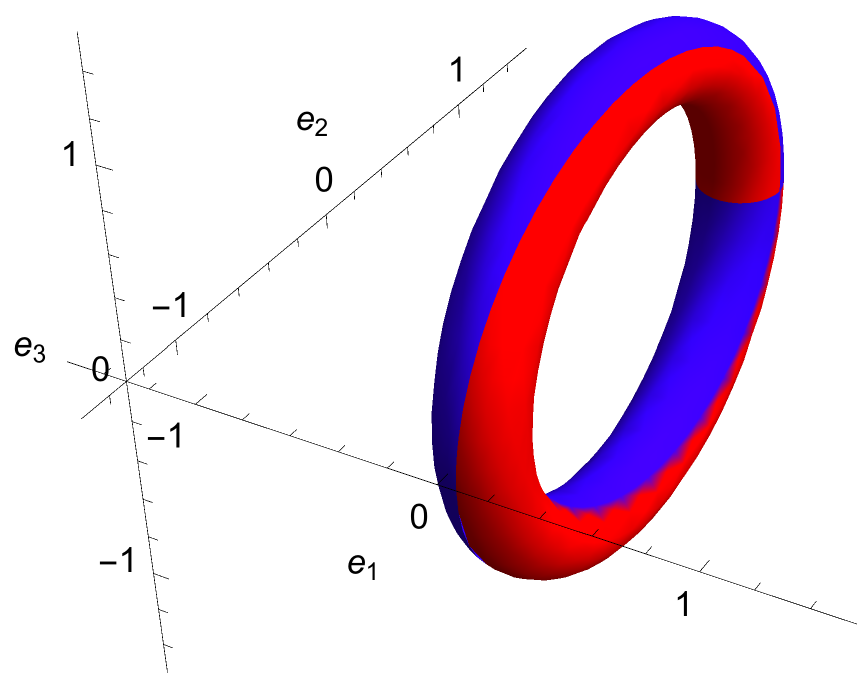}
}\hspace{.05\textwidth}
\subfigure[$\lambda=2$]{\includegraphics[width=.45\textwidth] {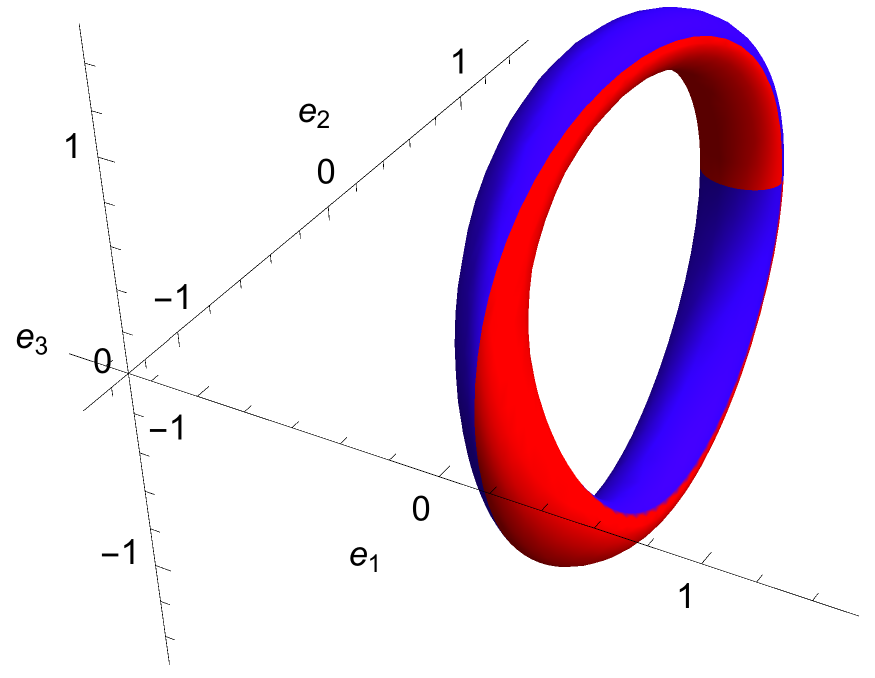}}\\
\subfigure[$\lambda=10$]{\includegraphics[width=.45\textwidth]{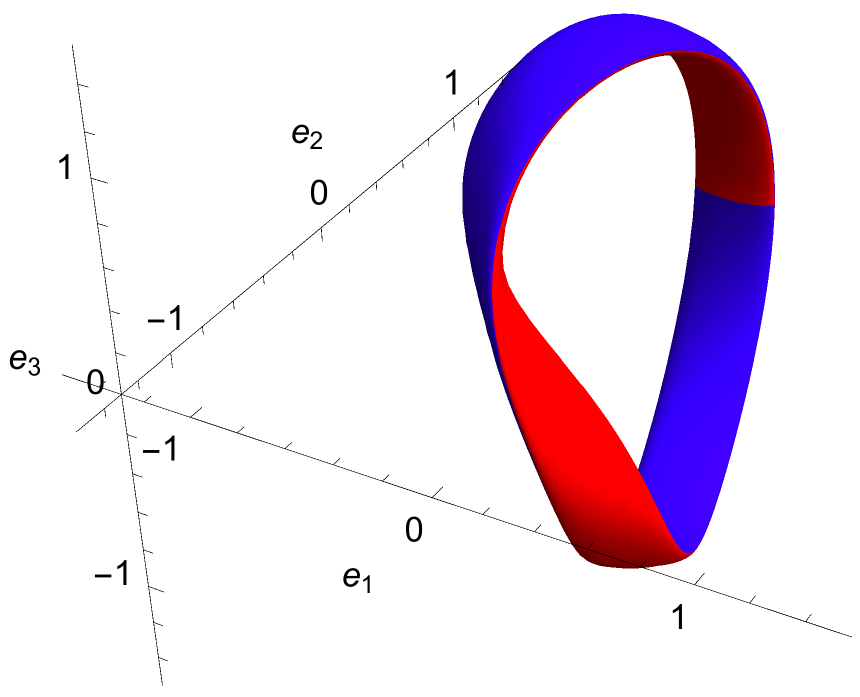}}\hspace{.05\textwidth}
\subfigure[$\lambda=1000$]{\includegraphics[width=.45\textwidth]{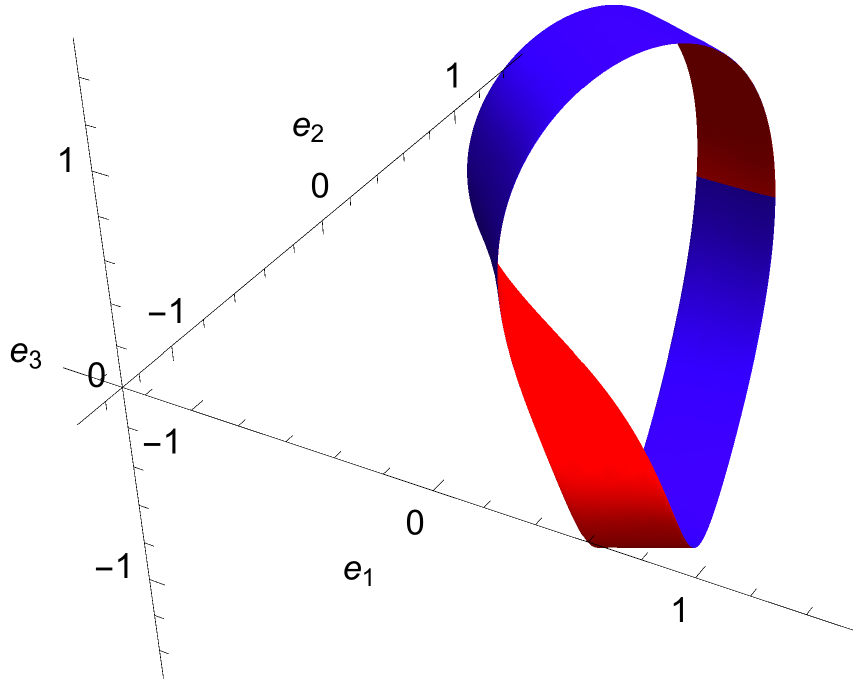}}
\caption{\small Kirchhoff rod theory results. All plots have $\gamma=1.5$. Note that the origin of the axes is at the point  $\left(-1.3,-1.3,0\right)$. Coloring is used to indicate the orientation of the strip.}
\label{fig:mahatable}
\end{figure}

 Starting from the equilibrium configuration with $\lambda=1$ and $\gamma=1.5$,  new equilibrium configurations are found as $\lambda$ is increased and $\gamma$ is held fixed, cf. (\ref{eq:sysend2}). The calculated equilibrium configurations in Figure \ref{fig:mahatable} are in qualitative agreement with the smoothly varying configurations found in \cite{Mahadevan1993}. We note that different values of $\gamma$ in the allowed range do not produce qualitatively different equilibrium configurations. These updated results confirm that Kirchhoff theory does not capture the sharp localized bending and twisting seen in \cite{Starostin2007,Starostin2014}. 
\section{Local Stability of Kirchhoff-Model Configurations}\label{sec:kstab}
   The local stability of the equilibrium configurations of the entire closed M\"{o}bius strip on $[0,2\pi]$ is investigated in this section, for small but arbitrary perturbations -- in particular, perturbations that break the flip symmetry of the equilibrium configuration.  For any computed solution of the half-rod, the first step is to generate the entire solution on $[0,2\pi]$ via (\ref{eq:operator})-(\ref{eq:mflip}).  We then employ the methodology of \cite{Kumar2010}.  The latter is quite general and can accommodate an accurate, discrete-numerical representation of a rod equilibrium in the presence of constraints -- regardless of the numerical discretization method used, e.g., finite differences, finite elements, collocation, shooting methods, etc.  Here we have the point-wise constraints inherent in (\ref{eq:unshear_inext}).  We now summarize the methodology.
\subsection{Formulation}\label{sec:kstabform}
We first introduce the spatially weak forms associated with the dynamics of unshearable, inextensible rods of length $2\pi$ without the presence of body forces or body couples: 
\begin{align}
G_{dynamic} & = \int\limits_0^{2\pi}\,\left[\, p\,A\,\uvec{\ddot{r}}\,\cdot\boldsymbol{\rho}+p\,\dot{\left(\mathbf{I}\,\mathbf{w}\right)}\cdot\boldsymbol{\psi}\right]\,\mathrm{d}s\label{eq:Gdym}\,,\\
G_{static} &=\int\limits_0^{2\pi}\,\left[\, \uvec{n}\cdot\left(\boldsymbol{\rho}^\prime-\boldsymbol{\psi}\times\uvec{r}^\prime\right)+\uvec{m}\cdot\boldsymbol{\psi}^\prime+...\right.\nonumber\\&\hspace{.05\textwidth}\left....+\boldsymbol{\xi}_\alpha\uvec{r}^\prime\cdot\uvec{R}\,\uvec{e}_\alpha+\boldsymbol{\xi}_3\,\left(\uvec{r}^\prime\cdot\uvec{R}\,\uvec{e}_3-1\right)\,\right]\,\mathrm{d}s+ \left[ \uvec{n}\cdot\boldsymbol{\rho}+\uvec{m}\cdot\boldsymbol{\psi}\right|_0^{2\pi}\label{eq:Gstat}\,,
\end{align}
 where $p$ is the density of the rod, $A$ is the cross-sectional area, $\mathbf{I}$ is the moment of area tensor, $\mathbf{w}$ is the angular velocity of the cross-sections, the over dot $(\,\dot{\,}\,)$ indicates a derivative with respect to time, and $\boldsymbol{\rho},\boldsymbol{\psi}$, and $\boldsymbol{\xi}$ correspond to smooth variations in $\uvec{r},\uvec{R}$, and $\uvec{n}$ respectively. 
The spatially weak form of the governing partial differential equations governing the dynamics of the rod is represented by
\begin{align}
G\left(\uvec{r},\uvec{R},\uvec{n}\,;\,\boldsymbol{\rho},\boldsymbol{\psi},\boldsymbol{\xi}\right) = G_{dynamic}-G_{static}=0\,,
\end{align}
is satisfied identically at an equilibrium $\left(\mathbf{r},\mathbf{R},\mathbf{n}\right)$, for all smooth variations $\left(\boldsymbol{\rho},\boldsymbol{\psi},\boldsymbol{\xi}\right)$.
We now consider small perturbations from an equilibrium $\left(\mathbf{r},\mathbf{R},\mathbf{n}\right)$ via
\begin{align}
\uvec{r}_\epsilon & = \uvec{r} + \epsilon\,\Delta\uvec{r}\,,\\
\uvec{R}_\epsilon & = \text{exp}\left(\epsilon\,\Delta \Theta\right)\uvec{R}\,,\\
\uvec{n}_\epsilon & = \uvec{n} + \epsilon\,\Delta\uvec{n}\,,
\end{align}
where $\Delta\uvec{r}$, $\Delta\uvec{n}$ are smooth admissible variations, $\text{exp}\left(\cdot\right)$ denotes the matrix exponential, $\Delta\Theta$ is a smooth admissible skew-symmetric matrix, and $\epsilon$ is a small parameter. We define $\Delta\boldsymbol{\theta}=\text{axial}\left(\Delta\Theta\right)$ along with the vector
\begin{align}\label{eq:kirchzeta}
\Delta\boldsymbol{\zeta}_0 = \begin{pmatrix} \left[ \Delta\uvec{r} \right]& \left[ \Delta\boldsymbol{\theta} \right] &\left[ \Delta\uvec{n} \right]\end{pmatrix}^T\,.
\end{align}
The time dependent perturbations take the form
\begin{align}
\Delta\boldsymbol{\zeta}=\Delta\boldsymbol{\zeta}_0\,e^{\,\sigma\,t}\label{eq:pertb}\,.
\end{align}
Taylor's expansion about an equilibrium point generates
\begin{align}
G\left(\uvec{r}_\epsilon,\uvec{R}_\epsilon,\uvec{n}_\epsilon\right) = \epsilon\,DG\left(\uvec{r},\uvec{R},\uvec{n}\right)\,\Delta\boldsymbol{\zeta} + o\left(\epsilon\abs{\Delta\boldsymbol{\zeta}}\right)\label{eq:taylor}\,.
\end{align}
Substituting (\ref{eq:pertb}) into the linear part of (\ref{eq:taylor}), we obtain the generalized eigenvalue problem
\begin{align}
DG_{static}\,\Delta\boldsymbol{\zeta}_0 =\mu\,DG_{dynamic}\,\Delta\boldsymbol{\zeta}_0 \label{eq:eigenstab}\,.
\end{align}
where $\mu:=-\sigma^2$ is the eigenvalue. As discussed in \cite{Kumar2010}, the structure of (\ref{eq:eigenstab}) is nonstandard, due to the presence of the linearized constraints, e.g. (\ref{eq:unshear_inext}), on the left side, which are equated to zero on the right side.  Moreover, for conservative problems, like the one at hand, the eigenvalues are necessarily real: A negative eigenvalue, $\mu<0$, indicates instability, since $\sigma=\sqrt{-\mu}$ in (\ref{eq:pertb}) engenders exponential growth; a positive eigenvalue implies that $\sigma$ is purely imaginary, showing that (\ref{eq:pertb}) is oscillatory.  Accordingly, the solution is stable if all eigenvalues are positive.

Explicit forms of $DG_{static}$ and $DG_{dynamic}$ for an unshearable-inextensible rod are derived in \cite{Kumar2010}.
\subsection{Numerical Implementation}
To calculate the eigenvalues in (\ref{eq:eigenstab}), we employ the finite-element method, as implemented in \cite{Kumar2010,Simo1986}. We approximate the smooth test functions $\left(\boldsymbol{\rho},\boldsymbol{\psi},\boldsymbol{\xi}\right)$ and spatial perturbations $(\Delta\uvec{r}, \Delta\boldsymbol{\theta},\Delta\uvec{n})$ with piecewise linear functions. Nodal values for the variables $(\uvec{r},\uvec{R},\uvec{n},\uvec{m})$ on $[0,2\pi]$ are obtained from the continuation results on $[0,\pi]$ and symmetry transformations in (\ref{eq:rflip})-(\ref{eq:nflip}) on $[\pi,2\pi]$. For $N$ elements, this discretization transforms (\ref{eq:eigenstab}) into the matrix eigenvalue problem
\begin{align}\label{eq:eigenstabM}
\left[\begin{array}{cc}\mathbf{K}_{m\times m}&\mathbf{C}_{m\times p}\\\mathbf{C}^T_{p\times m}&\mathbf{0}_{p\times p} \end{array}\right]\left[\begin{array}{c}\left[\begin{array}{c}\Delta\uvec{r}\\\Delta\boldsymbol{\theta}\end{array}\right]\\\Delta\dvec{n}_0\end{array}\right]=\mu\left[\begin{array}{cc}\mathbf{M}_{m\times m}&\mathbf{0}\\\mathbf{0}&\mathbf{0}\end{array}\right]\left[\begin{array}{c}\left[\begin{array}{c}\Delta\uvec{r}\\\Delta\boldsymbol{\theta}\end{array}\right]\\\Delta\dvec{n}_0\end{array}\right]
\end{align}
where $\mathbf{K}$ is the global stiffness matrix, $\mathbf{C}$ is the global constraint matrix, $\mathbf{M}$ is the global mass matrix, $m=6N$, and $p=3N$. Note that $p$ represents the total number of point-wise constraints on the discretized rod. Also note that $\mathbf{K}$ is block tri-diagonal and symmetric for conservative loadings at equilibrium \cite{Simo1986}.

As mentioned before, the constraint terms cause (\ref{eq:eigenstab}) and (\ref{eq:eigenstabM}) to be singular. The Q-R factorization of of $\mathbf{C}$ has the form
\begin{align}
\mathbf{C}=\left[\begin{array}{cc}\mathbf{Q1}&\mathbf{Q2}\end{array}\right]\left[\begin{array}{c}\mathbf{R1}\\\mathbf{0}\end{array}\right]=\mathbf{Q1}\,\mathbf{R1}\,,\label{eq:q2eq}
\end{align}
Following the procedure in \cite{Kumar2010}, we generate a non-singular reduced version of (\ref{eq:eigenstabM}) via
\begin{align}
\left(\mathbf{Q2}^T\,\mathbf{K}\,\mathbf{Q2}\right)\,\Delta\zeta_0&=\mu\,\left(\mathbf{Q2}^T\,\mathbf{M}\,\mathbf{Q2}\right)\,\Delta\zeta_0\,,\\
\tilde{\mathbf{K}}\,\Delta\zeta_0&=\mu\tilde{\mathbf{M}}\,\Delta\zeta_0\,,\label{eq:methend}
\end{align}
where $\tilde{\mathbf{K}}$ and $\tilde{\mathbf{M}}$ are the projected stiffness and mass matrices. This eliminates all the spurious eigenvalues and reduces the total dimension of the problem from $m+p$ to $m-p$.

The M\"{o}bius strip problem considered here is conservative, and the projected mass matrix $\tilde{\mathbf{M}}$ is positive definite.  Accordingly, the latter may be replaced by the identity matrix without impacting the signs of the eigenvalues in (\ref{eq:q2eq}), and our final form of the eigenvalue problem is
\begin{align}
\tilde{\mathbf{K}}\,\Delta\zeta_0&=\mu\,\mathbf{I}\,\Delta\zeta_0\,,\label{eq:methend}
\end{align}
where positive eigenvalues of $\tilde{\mathbf{K}}$ indicate stability and negative eigenvalues indicate unstable perturbations. Since the problem at hand is conservative, $\tilde{\mathbf{K}}$ is the discreteHessian corresponding to the constrained potential energy. Thus, (\ref{eq:methend}), while derived as part of a linearized stability method, the positivity of $\tilde{\mathbf{K}}$ is equivalent to the minimum-potential-energy criterion.
\subsection{Boundary Conditions}\label{sec:bcz}
For the closed loop, both the position and the orientation of the rod at $s=0$ and $s=2\pi$ are clamped. Assuming the rod is divided into $N$ elements with $N+1$ nodes, the boundary conditions are
\begin{align}\label{eq:stabbc1}
\Delta\uvec{r}^{\left(0\right)}& = 0\,,\hspace{.13\textwidth}
\Delta\uvec{\boldsymbol{\theta}}^{\left(0\right)} = 0\,,\\
\Delta\uvec{r}^{\left(N+1\right)}& = 0\,,\hspace{.1\textwidth}
\Delta\uvec{\boldsymbol{\theta}}^{\left(N+1\right)} = 0\,.\label{eq:stabbc2}
\end{align}
These ensure that the $s=0$ and $s=2\pi$ ends of the rod will remain smoothly connected under any perturbation. In addition, (\ref{eq:stabbc1})-(\ref{eq:stabbc2}) eliminate the six neutrally stable rigid-body modes corresponding to uniform translation and rotation of the closed rod and also one additional neutral degeneracy associated with the axial motion of the strip acting through its own fixed configuration \cite{Domokos1995,Domokos2001}. For the initial isotropic configuration ($\lambda=1$), there is one remaining zero eigenvalue, due to a one-parameter family of equilibria corresponding to continuous precession of the centerline configuration accompanied by rolling of the cross sections in the opposite sense, cf. \cite{Domokos2001, Domokos1995, Healey1992}. This degeneracy disappears for $\lambda>1$.
\subsection{Results}\label{sec:stabilityk}
\begin{table}
\caption{The four smallest eigenvalues of $DG_{static}$ for a Kirchhoff rod with $N=240$ elements}
\centering\label{table:kirchhoffeigs}
\vspace{.1in}
\begin{tabular}{c|c|c|c|c}
 &Smallest&2nd smallest &3rd smallest&4th smallest\\
\hline\hline
$\lambda=10$ &1.07e-4 &0.0449&0.06895&0.141\\
$\lambda=100$ &0.00126&0.0490&0.0730&0.163\\
$\lambda=1000$ &0.0118	& 0.0510	&0.0794&0.169
\end{tabular}
\end{table}
The numerical equilibrium solutions from AUTO calculated in Section \ref{sec:kirchoff} are extended to the full M\"{o}bius strip on $[0,2\pi]$ and used for the finite element calculation. This results in a mesh resolution of 240 elements for the full rod. We find the eigenvalues of $\tilde{\mathbf{K}}$ using the eigs() function in Matlab for each equilibrium configuration.

We list the three smallest eigenvalues of the Hessian $\tilde{\mathbf{K}}$ in Table \ref{table:kirchhoffeigs} for several values of the the bending stiffness ratio ``$\lambda$".  In all cases the computed eigenvalues are positive, and we conclude that the Kirchhoff-rod equilibria for the M\"{o}bius strip are stable with respect to all local perturbations - symmetric and anti-symmetric.

\section{Developable Rod Model}\label{sec:devrod}
In this section, we model the M\"{o}bius strip as a developable surface as in \cite{Wunderlich1962, Starostin2007,Starostin2014}. We employ the developable-strip plate model of Wunderlich \cite{Wunderlich1962} as derived in \cite{Dias2014}. Once equilibria are calculated, their stability can be assessed by adapting the approach of \cite{Kumar2010} as employed in section \ref{sec:kstab}.

Using the same notation introduced in section \ref{sec:kform}, the developable strip is defined with centerline coordinate $s\in\left[0,L\right]$ in a straight, stress free reference configuration. The position of the strip is defined by the vector-valued function $\uvec{r}\left(s\right)$ with the reference configuration's centerline given by $\mathbf{r}_0\left(s\right)=s\uvec{e}_3$. The directors are again defined in (\ref{eq:director})-(\ref{eq:kappaaxial}), and the definitions (\ref{eq:rprime})-(\ref{eq:mveckinematics}) remain valid.

Departing from Cosserat rod theory, we assume that the strip has an instantaneous axis of bending given by the vector, $\uvec{b}\left(s\right)\in\text{span}\left\{\uvec{d}_1,\uvec{d}_3\right\}$. Let $\phi$ denote the angle between the centerline tangent vector, $\uvec{d}_3$, and the instantaneous axis of bending. Define the quantity $\eta=\cot{\phi}$. Note that $\eta\equiv0$ corresponds to the usual Cosserat rod theory.

Inherent in the approach of \cite{Wunderlich1962}, echoed in \cite{Dias2014} and \cite{Starostin2014}, is the tacit assumption that the flat, stress-free reference configuration admits the representation
\begin{align}\label{eq:devstripans}
\mathbf{X}=s\,\uvec{e}_3+v\,\left[\uvec{e}_1+\eta\left(s\right)\,\uvec{e}_3\right]\,,
\end{align}
where $s\in\left[0,L\right]$, $v\in\left[-w/2,w/2\right]$. That is, the mapping $\left(v,s\right)\rightarrow \left(v,s+v\eta\left(s\right)\right)$ should be locally injective on $\Omega:=\left[0,L\right]\times\left[-w/2,w/2\right]$, viz., $1+v\,\eta^\prime\left(s\right)>0$ on $\Omega$. Assuming this is the case, then the deformation of the strip, $\uvec{f}:\Omega\rightarrow\mathbb{E}^3$, is given by
\begin{align}\label{eq:devsurf}
\mathbf{x}=\mathbf{f}\left(\mathbf{X}\right)&=\uvec{r}\left(s\right)+v\,\mathbf{b}\left(s\right)\,,
\end{align}
with
\begin{align}\label{eq:devsurf2}
\mathbf{b}\left(s\right)=\uvec{d}_1\left(s\right)+\eta\left(s\right)\,\uvec{d}_3\left(s\right)\,,
\end{align}
Note that (\ref{eq:devsurf}) defines a ruled surface with normal vector $\mathbf{N}=\uvec{d}_2\left(s\right)$.

The strip is presumed inextensible and the centerline in (\ref{eq:devsurf}) is constrained to be inextensible and unshearable via (\ref{eq:unshear_inext}). In addition, the constraint
\begin{align}\label{eq:k2contr}
\kappa_2\equiv0\,,
\end{align}
precludes bending along the stiff axis. The ruled surface in (\ref{eq:devsurf}), (\ref{eq:devsurf2}) is developable if (cf. \cite{DoCarmo1976})
\begin{align}
\kappa_3-\eta\,\kappa_1=0\label{eq:develconstraint}\,.
\end{align}
The strip (\ref{eq:devsurf}) is a tangent developable with one generator of curvature, given by $\uvec{b}\left(s\right)$. The constraints (\ref{eq:unshear_inext}), (\ref{eq:k2contr}), and (\ref{eq:develconstraint}) enforce developability.

Following \cite{Wunderlich1962}, the stored energy for the thin strip is derived from a constrained St Venant-Kirchhoff plate. In particular, the only contribution to the stored energy is that due to pure bending about the instantaneous axis (\ref{eq:devsurf2}); an integration across the width yields the total stored energy expression due to Wunderlich:
\begin{align}\label{eq:devenergy}
V=\frac{D\,w}{2}\int\limits_0^{L}\,\kappa_1^2\left[1+\eta^2\right]^2\,g\left(w\,\eta^\prime \right) \,\mathrm{d}s\,.
\end{align}
where
\begin{align}
D&:=\frac{Eh^3}{12\left(1-\nu^2\right)}\,,\quad \quad
g\left(w\,\eta^\prime\right):=\frac{1}{\eta^\prime w}\log{\left(\frac{1+\eta^\prime w/2}{1-\eta^\prime w/2}\right)}\,,\label{eq:gpar}
\end{align} 
and where $E$ is Young's modulus, $\nu$ is Poisson's ratio, and $\kappa_1$ is defined in (\ref{eq:rprime}).  Following \cite{Dias2014}, (\ref{eq:devenergy}) is now amended by integral terms involving the constraints (\ref{eq:unshear_inext}), (\ref{eq:k2contr}), and (\ref{eq:develconstraint}) and the appropriate Lagrange multipliers:
\begin{align}\label{eq:fullenergy}
U=V+\int\limits_0^{L}\dvec{m}_3\left(\kappa_3-\eta\,\kappa_1\right)\,\mathrm{d}s+\int\limits_0^{L}\dvec{m}_2\,\kappa_2\,\mathrm{d}s+\int\limits_0^{L}\uvec{n}\cdot\left(\uvec{r}^\prime-\uvec{d}_3\right)\,\mathrm{d}s
\end{align}

In order to derive the Euler-Lagrange equilibrium equations, we consider smooth, L-periodic variations $\hat{\mathbf{r}}$, $\hat{\boldsymbol{\Theta}}$, $\hat{\eta}$, where $\hat{\boldsymbol{\Theta}}$ is skew-symmetric valued, as follows:
\begin{align}\label{eq:vars}
\uvec{r}\rightarrow\uvec{r}+\alpha\hat{\uvec{r}}\,,\quad
\mathbf{R}\rightarrow\mathbf{R}+\text{exp}\left(\alpha\hat{\boldsymbol{\Theta}}\right)\mathbf{R}\,,\quad
\eta\rightarrow\eta+\alpha\hat{\eta}
\end{align}
where $\alpha$ is a small parameter. We then find
\begin{align}
\mathbf{r}^\prime\,&\rightarrow\,\mathbf{r}^\prime+\alpha\,\left(\hat{\mathbf{r}}^\prime+\mathbf{r}^\prime\times\hat{\boldsymbol{\theta}}\right)+o\left(\alpha\right)\,,\\
\boldsymbol{\kappa}\,&\rightarrow\,\boldsymbol{\kappa}+\alpha\,\hat{\boldsymbol{\theta}}^\prime+o\left(\alpha\right)\,,\quad\text{as}\quad \alpha\rightarrow0\,,\label{eq:axvars}
\end{align}
where $\hat{\boldsymbol{\theta}}:=\text{axial}\left(\hat{\boldsymbol{\Theta}}\right)$. We substitute (\ref{eq:vars})-(\ref{eq:axvars}) into (\ref{eq:fullenergy}), take the derivative of the resulting expression with respect to $\alpha$, and then evaluate it at $\alpha=0$. A formal integration by parts then delivers the first variation condition:
\begin{align}
\delta U&= \int\limits_0^{L}\Bigg\{-\frac{\mathrm{d}\,}{\mathrm{d}s}\left[ \frac{Dw^2}{2}\kappa_1^2\left[1+\eta^2\right]^2\,\dot{g}\left(w\eta^\prime \right)\right]+2Dw\kappa_1^2\eta\left[1+\eta^2\right]g\left(w\eta^\prime\right)+...\nonumber\\
&\hspace{.1\textwidth}...-\dvec{m}_3\kappa_1\Bigg\}\,\hat{\eta}\,\mathrm{d}s \label{eq:bigvar}-\int\limits_0^{L} \uvec{n}^\prime\cdot\hat{\uvec{r}}\,\mathrm{d}s\,-\,\int\limits_0^{L} \left\{\uvec{m}^\prime+\left(\uvec{d}_3\times\uvec{n}\right)\right\}\cdot\hat{\boldsymbol{\theta}}\,\mathrm{d}s =0\, ,
\end{align}
for all smooth variations $\hat{\mathbf{r}}$, $\hat{\boldsymbol{\theta}}$, $\hat{\eta}$, where the overdot, $\dot{\left(\,\,\right)}$ indicates a derivative of a function with respect to the whole argument and prime, $\left(\,\right)^\prime$ denotes a derivative with respect to centerline arclength, $s$, and 
\begin{align}
\dvec{m}_1=\mathbf{d}_1\cdot\mathbf{m}:=D\,w\,\kappa_1\left[1+\eta^2\right]^2\,g\left(w\,\eta^\prime\right)-\eta\,\dvec{m}_3 \,,\label{eq:m1const}
\end{align}
Choosing $\hat{\eta}\equiv 0$ and $\hat{\boldsymbol{\theta}}\equiv0$ for all variations $\hat{\mathbf{r}}$ in (\ref{eq:bigvar}) yields (\ref{eq:govfixed1}), and the $\hat{\eta}\equiv0$ for all variations $\hat{\boldsymbol{\theta}}$ delivers (\ref{eq:govfixed2}). A new, third Euler-Lagrange equations, corresponding to all variations $\hat{\eta}$ in (\ref{eq:bigvar}), reads
\begin{align}\label{eq:etaeuler}
-\frac{\mathrm{d}\,}{\mathrm{d}s}\left[ \frac{Dw^2}{2}\kappa_1^2\left[1+\eta^2\right]^2\,\dot{g}\left(w\eta^\prime \right)\right]+2Dw\kappa_1^2\eta\left[1+\eta^2\right]g\left(w\eta^\prime\right)-\dvec{m}_3\kappa_1=0\,. 
\end{align}
We observe that (\ref{eq:etaeuler}) is singular, corresponding to the pointwise vanishing of $\boldsymbol{\kappa}_1$. The latter, in view of (\ref{eq:k2contr}), is the total curvature of the center-line curve, $s\to\mathbf{r}\left(s\right)$ and our director basis here coincides with the usual Frenet-Serret frame.  As such, the validity of (\ref{eq:devsurf}), (\ref{eq:devsurf2}), and (\ref{eq:devenergy}) requires $\boldsymbol{\kappa}_1\neq0$,cf. \cite{Starostin2014, Freddi}.

Our intended strategy here is the same used before in section \ref{sec:solutionmethod}, viz., solve the governing equations on $\left[0,\pi\right]$ and then generate the rest by rotation (flip symmetry).  In particular, the latter, purely kinematical requirement (flip symmetry of the configuration) implies that  $\boldsymbol{\kappa}_1\left(\pi\right)=0$. In order to overcome the otherwise certain numerical difficulties associated with that, we introduce the following ``elliptic regularization" into the energy functional:
\begin{align}\label{eq:regenergy}
\hat{U}=U+\int\limits_0^{L}\frac{\epsilon}{2}\,\left(\eta^\prime\right)^2\,\mathrm{d}s
\end{align}
where $\epsilon>0$ is a very small parameter and $\hat{U}$ is the new regularized energy. This, in turn, modifies the principal part of (\ref{eq:etaeuler}) as follows
\begin{align}
&\left\{\epsilon+\frac{D\,w^3}{2} \kappa_1^2 \left[1+\eta^2\right]^2 \,\ddot{g}\left(w \eta '\right)\,\right\}\,\eta ''+D\,w^2 \,\kappa_1\left[1+\eta^2\right]^2\dot{g}\left(w \eta '\right)\, \kappa_1'+...\label{eq:ELeta1}\\
&\hspace{.05\textwidth}...+2\,D\,w^2\,\kappa_1\left[1+\eta^2\right]\dot{g}\left(w \eta '\right) \,\eta \, \eta ' -2 D\,w\kappa_1^2\,\left[1+\eta^2\right] \, g\left(w \eta '\right)\,\eta +\kappa_1\,\dvec{m}_3 =0\nonumber
\end{align}
We normalize all variables and the strip length to $2\pi$ according to (\ref{eq:norm1k}) and
\begin{align}
\bar{w}=\frac{2\pi\,w}{L}\hspace{.1\textwidth}
\bar{\dvec{n}}_i=\frac{\dvec{n}_i}{D\,w} \hspace{.05\textwidth} \hspace{.05\textwidth}
\bar{\dvec{m}}_i=\frac{2\pi\,\dvec{m}_i}{D\,w\,L}\,.
\end{align}
Note that $\eta$ is already a unitless parameter on $\left[-1,1\right]$). The complete system of differential equations for the developable rod (dropping all overbars) is given by
\begin{align}\label{eq:sysstartdev}
\dvec{n}^\prime +\tilde{\dvec{k}}\left(\dvec{m}\right)\times\dvec{n} &=0\,,\\
\dvec{m}^\prime+\tilde{\dvec{k}}\left(\dvec{m}\right)\times\dvec{m}+\hat{\dvec{d}}\times\dvec{n}&=0\,,\\
 \bar{r}^\prime - \bar{R}\left(\uvec{q}\right)\,\hat{\dvec{d}} & = 0\,,\\
\uvec{q}^\prime-\bar{A}\left(\uvec{q}\right)\,\tilde{\dvec{k}}\left(\dvec{m}\right)-\mu\,\uvec{q} & = 0\,,\\
a_{2}\left[\eta,\eta^\prime\right]\,\eta^{\prime\prime}+a_1\left[\eta,\eta^\prime\right]\,\eta^\prime+a_0\left[\eta,\eta^\prime\right]&=0\,,\label{eq:sysenddev}
\end{align}
where
\begin{align}\label{eq:sysend2dev}
\hat{\dvec{d}}:=&\left(0,0,1\right)\\
\tilde{\dvec{k}}\left(\dvec{m}\right):=&\left(\frac{\left(\dvec{m}_1+\eta\,\dvec{m}_3\right)L^2}{4\pi^2\left[1+\eta^2\right]^2\,g\left(w\,\eta^\prime\right)},\,\,\,0,\,\,\,  \frac{\eta\,\left(\dvec{m}_1+\eta\,\dvec{m}_3\right)L^2}{4\pi^2\left[1+\eta^2\right]^2\,g\left(w\,\eta^\prime\right)} \right)\label{eq:kvec}\\
a_{2}\left[\eta,\eta^\prime\right]=&\, \epsilon+\frac{w^2}{2} \left(\frac{\dvec{m}_1+\eta\,\dvec{m}_3}{\left[1+\eta^2\right]\,g\left(w\eta^\prime\right)}\right)^2 \left(\frac{\ddot{g}\left(w\eta^\prime\right)\,g\left(w\eta^\prime\right)-2\,g^{\prime}\left(w\eta^\prime\right)^2}{g\left(w\eta^\prime\right)}\right)\,,\\
a_{1}\left[\eta,\eta^\prime\right]=& \, w\,\frac{\left(\dvec{m}_3\left[1-\eta^2\right]-2\,\dvec{m}_1\,\eta\right)\,\left(\dvec{m}_1+\eta\,\dvec{m}_3\right)g^{\prime}\left(w\eta^\prime\right)\,}{\left[1+\eta^2\right]^3\,g\left(w\eta^\prime\right)^2}\,,\\
a_{0}\left[\eta,\eta^\prime\right]=&\,\frac{\left(\dvec{m}_1+\eta\,\dvec{m}_3\right)}{\left[1+\eta^2\right]^2\,g\left(w\eta^\prime\right)} \left[  w\,\dvec{n}_2\,\frac{\dot{g}\left(w\eta^\prime\right)}{g\left(w\eta^\prime\right)}+ \frac{\dvec{m}_3\left[1-\eta^2\right]-2\eta\dvec{m}_1}{\left[1+\eta^2\right]}\right]\,,\label{eq:sysend2devend}
\end{align}
and where the overdot, $\dot{\left(\,\,\right)}$, indicates a derivative of a function with respect to the whole argument and the prime, $\left(\,\right)^\prime$, denotes a derivative with respect to centerline arclength, $s$.

From (\ref{eq:gpar}), we see that the stored energy in (\ref{eq:devenergy}) has a diverging integrand when $\eta^\prime\,w=0$. This is a removable singularity which will occur at the boundary point $s=0$ due to (\ref{eq:etabc1}). To avoid division by zero, $g$ is expanded in the following Taylor Series about $w\eta^\prime=0$:
\begin{align}
g\left(w\eta^\prime\right)&=1+\frac{\left(w\eta^\prime\right)^2}{12}+\frac{\left(w\eta^\prime\right)^4}{80}+\frac{\left(w\eta^\prime\right)^6}{448}+\frac{\left(w\eta^\prime\right)^8}{2304}+O\left(\left[w\eta^\prime\right]^{10}\right)\,. \label{eq:g}
\end{align}
This Taylor series is used instead of the exact expression in (\ref{eq:gpar}) in the neighborhood of points where $\eta^\prime=0$.
\section{Developable Rod Solution Method}\label{sec:developablesol}
The system of ordinary differential equations in (\ref{eq:sysstartdev})-(\ref{eq:sysenddev}) is solved using the same procedure outlined in section \ref{sec:solutionmethod}: AUTO is again used to produce a solution to the half problem on $\left[0,\pi\right]$ and the full M\"{o}bius strip on $\left[0,2\pi\right]$ is constructed through a flip about an axis of symmetry. However, a major point of departure from section \ref{sec:solutionmethod} is that there is no explicit (twisted) solution from which to initiate continuation. Instead an appropriate initial configuration is found utilizing a series of intermediate continuation calculations.

\subsection{Initial Equilibrium Configuration}\label{sec:devstartconfig}
To form a twisted M\"{o}bius strip, the centerline must undergo a non-zero twist, $\kappa_3$, which induces non-zero curvature, $\kappa_1$, and a non-zero scalar parameter, $\eta$, via the constraint (\ref{eq:develconstraint}). Thus, the most direct closed form equilibrium configuration is an untwisted strip  with $\eta\equiv0$ which satisfies (\ref{eq:ELeta1}) trivially. Only a strip in the shape of a cylinder satisfies $\eta\equiv0$ while being developable \cite{Audoly2010}, so the initial configuration is a strip with a semi-circular centerline on $\left[0,\pi\right]$ with uniform curvature $\kappa_1\equiv\pi$.

In order to sustain the uniform curvature in this configuration, a moment on the $s=\pi$ boundary will be applied. This will initially violate the zero-moment condition for the hinge at $s=\pi$, so we use continuation to follow the path of equilibria to a non-uniformly curved strip with zero moment at the $s=\pi$ boundary (detailed in Figure \ref{fig:momentrelax}(a)-(b)). Once this is completed, the hinge boundary condition is fixed by setting $\dvec{m}_1\left(\pi\right)=0$. Next, another continuation step is used to twist the strip by $\pi/2$ and align the hinge with the axis of symmetry, say $\uvec{e}_2$ (detailed in Figure \ref{fig:momentrelax}(c)-(f)).

After both of these continuation steps are completed, a M\"{o}bius strip is formed via a 180 degree rotation about the axis of symmetry. Continuation is performed in $\epsilon$ to reduce the regularizing term followed by continuation in the width $w$ to find the M\"{o}bius-strip equilibria. 
In summary:
\begin{enumerate}
\item Start with an untwisted strip with constant curvature. This requires an applied moment at $s=\pi$ to sustain the curvature.
\item Use continuation to relax the moment until there is no applied moment at $s=\pi$.
\item Fix the moment at zero, and use continuation to twist the orientation at $s=\pi$ until the strip will form a M\"{o}bius half-strip.
\item Perform continuation in the width and parameter $\epsilon$.
\end{enumerate}
\subsection{Boundary Conditions}\label{sec:devbc}
As in Section \ref{sec:bc}, the half-strip satisfies (\ref{eq:bc0start})-(\ref{eq:bc1end}) and (\ref{eq:q1bc})-(\ref{eq:bcqend}).
In addition, we require
\begin{align}\label{eq:devbcalmostend}
\eta^\prime\left(0\right)&=0\,,\\
\eta\left(\pi\right)&=0\,,\label{eq:devbcend}
\end{align}
which together with the other boundary conditions ensure flip symmetry of the complete M\"{o}bius band with the generator of curvature at $s=\pi$, $\mathbf{b}\left(\pi\right)$, is aligned with the axis of symmetry $\uvec{e}_2$.

As explained in Section \ref{sec:devstartconfig}, we require some preliminary continuations steps in order to arrive at a starting M\"{o}bius strip configuration. Each stage of the preliminary continuation calculations requires its own set of 16 boundary conditions as detailed below.
\subsubsection{Initial Configuration}
When starting from the uniformly curved configuration (see Figure \ref{fig:momentrelax}(a)), the boundary conditions for the fixed end are defined as above in (\ref{eq:bc0start})-(\ref{eq:bc1end}). As in the Kirchhoff rod case in section \ref{sec:solutionmethod}, only the displacements in the $\uvec{e}_1$ and $\uvec{e}_3$ directions are constrained at $s=\pi$
\begin{align}
&\mathbf{r}\left(\pi\right)\cdot\uvec{e}_1 =\bar{r}_1\left(\pi\right) = 0\,,\\
&\mathbf{r}\left(\pi\right)\cdot\uvec{e}_3 =\bar{r}_3\left(\pi\right) = 0\,,
\end{align}
At $s=\pi$ the positions are held fixed via
\begin{align}
&\mathbf{r}\left(\pi\right)\cdot\uvec{e}_1 =\bar{r}_1\left(\pi\right) = 0\,,\\
&\mathbf{r}\left(\pi\right)\cdot\uvec{e}_2 =\bar{r}_2\left(\pi\right) = -1\,,\\
&\mathbf{r}\left(\pi\right)\cdot\uvec{e}_3 =\bar{r}_3\left(\pi\right) = 0\,,
\end{align}
while the hinge with variable end moment is given by
\begin{align}
\dvec{n}_1\left(\pi\right)&=0\,,\\
\dvec{m}_1\left(\pi\right)&=\pi\left(1-\Xi_1\right)\,,
\end{align}
where the continuation parameter $\Xi_1\in\left[0,1\right]$ begins at $0$ and is continued to $1$. For the strip orientation, the director $\uvec{d}_1$ is constrained to point along the hinge axis via
\begin{align}
q_1\left(\pi\right)\,q_2\left(\pi\right)+q_0\left(\pi\right)\,q_3\left(\pi\right)=0\,,
\end{align}
and the the quaternions are normalization through
\begin{align}
 q_1^2\left(\pi\right)+q_2^2\left(\pi\right)+q_3^2\left(\pi\right)+q^2_0\left(\pi\right)=1\,.
\end{align}
The final two boundary conditions are given by (\ref{eq:devbcalmostend})-(\ref{eq:devbcend}).
\subsubsection{Moment Relaxing Continuation}
At $s=\pi$ the positions are held fixed via
\begin{align}
&\mathbf{r}\left(\pi\right)\cdot\uvec{e}_1 =\bar{r}_1\left(\pi\right) = 0\,,\\
&\mathbf{r}\left(\pi\right)\cdot\uvec{e}_2 =\bar{r}_2\left(\pi\right) = -1\,,\\
&\mathbf{r}\left(\pi\right)\cdot\uvec{e}_3 =\bar{r}_3\left(\pi\right) = 0\,,
\end{align}
while the hinge with variable end moment is given by
\begin{align}
\dvec{n}_1\left(\pi\right)&=0\,,\\
\dvec{m}_1\left(\pi\right)&=\pi\left(1-\Xi_1\right)\,,
\end{align}
where the continuation parameter $\Xi_1\in\left[0,1\right]$ begins at $0$ and is continued to $1$. For the strip orientation, the director $\uvec{d}_1$ is constrained to point along the hinge axis via
\begin{align}
q_1\left(\pi\right)\,q_2\left(\pi\right)+q_0\left(\pi\right)\,q_3\left(\pi\right)=0\,,
\end{align}
and the the quaternions are normalized through
\begin{align}
 q_1^2\left(\pi\right)+q_2^2\left(\pi\right)+q_3^2\left(\pi\right)+q^2_0\left(\pi\right)=1\,.
\end{align}
The final two boundary conditions are
\begin{align}
\eta^\prime\left(0\right)&=0\label{eq:etabc1}\,,\quad \quad
\eta\left(\pi\right)=0\,,
\end{align}
which ensure the $s=\pi$ end of the strip is aligned with the axis of symmetry $\uvec{e}_2$.
\subsubsection{Twisting Continuation}\label{sec:twistingcont}
For the twisting continuation, the $s=0$ end remains via (\ref{eq:bc0start})-(\ref{eq:bc1end}).
As in the Kirchhoff rod case in section \ref{sec:solutionmethod}, only the displacements in the $\uvec{e}_1$ and $\uvec{e}_3$ directions are constrained at $s=\pi$
\begin{align}
&\mathbf{r}\left(\pi\right)\cdot\uvec{e}_1 =\bar{r}_1\left(\pi\right) = 0\,,\\
&\mathbf{r}\left(\pi\right)\cdot\uvec{e}_3 =\bar{r}_3\left(\pi\right) = 0\,,
\end{align}
while the hinge conditions are enforced via
\begin{align}
\dvec{n}_1\left(\pi\right)=0\,,\quad \quad
\dvec{m}_1\left(\pi\right)=0\,,\label{eq:m1fixeddev}
\end{align}
where (\ref{eq:m1fixeddev}) is equivalent to fixing $\Xi_1=1$. The strip needs to be twisted by $\pi/2$ and the end oriented along the $\uvec{e}_2$ axis of symmetry. This requires the transformations
 \begin{align}\label{eq:devrot1}
 \uvec{d}_1\left(\pi\right)&\rightarrow-\uvec{e}_2\,,\quad \quad  \uvec{d}_3\left(\pi\right)\cdot\uvec{e}_2\rightarrow \,0\,.
  \end{align}
The rotation that executes (\ref{eq:devrot1}) is facilitated by the dummy continuation parameter $\Xi_2$, which starts at $0$ and is continued to $1$. Define
\begin{align}
\hat{R}_{23}=\left[\,\uvec{d}_3\left(\pi\right)\cdot\uvec{e}_2\,\right]_{\left(\Xi_1=1\,,\,\Xi_2=0\right)}\,,
\end{align}
which corresponds to the $\uvec{e}_2$ component of the $\uvec{d}_3$ vector when the hinge is fully relaxed and the strip is still untwisted. The boundary conditions
\begin{align}\label{eq:twistbc1}
2\,\left(q_1\left(\pi\right)\,q_2\left(\pi\right)+q_0\left(\pi\right)\,q_3\left(\pi\right)\right)+\Xi_2&=0\,,\\
2\,\left(q_2\left(\pi\right)\,q_3\left(\pi\right)-q_1\left(\pi\right)\,q_0\left(\pi\right)\right)+\hat{R}_{23}\,\Xi_2&=0\,,\label{eq:twistbc2}\\
 q_1^2\left(\pi\right)+q_2^2\left(\pi\right)+q_3^2\left(\pi\right)+q^2_0\left(\pi\right)&=1\,,\label{eq:twistbc3}
\end{align}
execute the transformation (\ref{eq:devrot1}) and twist the strip about its centerline by $\pi/2$ when $\Xi_2$ is continued from zero to one. The final two boundary conditions are again (\ref{eq:etabc1}). Once $\Xi_1=\Xi_2=1$, the strip is a half-M\"{o}bius strip on $s\in\left[0,\pi\right]$ and the boundary conditions are held fixed for continuation in the width, $w$ and regularizing term $\epsilon$.
\subsection{Full Strip Construction}
Following the procedure outlined in section \ref{sec:fullrodrecon} and \cite{Domokos2001}, the numerical solution is obtained for $s\in\left[0,\pi\right]$, and the rest of the closed strip is constructed via a flip rotation by $180$ degrees about the $\uvec{e}_2$-axis. The transformations (\ref{eq:operator})-(\ref{eq:mflip}) are applied along with the transformation
\begin{align}\label{eq:etaflip}
\eta\left(s\right)=\begin{cases}
\eta^c\left(s\right)\quad& s\in[0,\pi]\\
-\eta^c\left(2\pi-s\right)\quad& s\in[\pi,2\pi]
\end{cases}\,,
\end{align}
where the superscript $c$ indicates the calculated value on the domain $\left[0,\pi\right]$. Note that at $s=2\pi$, the strip is twisted by $\pi$ about the centerline, so $\uvec{d}_1\left(2\pi\right)$ is in the opposite direction of $\uvec{d}_1\left(0\right)$. The negative sign in (\ref{eq:etaflip}) ensures that the strip forms a smooth closed loop after this orientation change.

\section{Developable Rod Results} \label{sec:devrodresults}
Due to the presence of constraints (\ref{eq:k2contr}) and (\ref{eq:develconstraint}), a finer mesh is needed to converge to equilibrium configurations for the developable rod than the one used in Section \ref{sec:kirchoff} for the Kirchhoff rod. In particular, this ensures convergence during the twisting continuations steps detailed in Section \ref{sec:twistingcont} when $\eta,\eta^\prime \neq 0$. The rod is divided into $100$ mesh intervals with $5$ collocation points on each element for a total of $501$ points for the rod on $s\in\left[0,\pi\right]$. The same continuation step size, tolerances and plotting routine from section \ref{sec:kirchoff} are used.
\begin{figure} 
\centering
\subfigure[Initial config., $\Xi_1=0$, $\phi=\pi/2$]{\includegraphics[width=.32\textwidth] {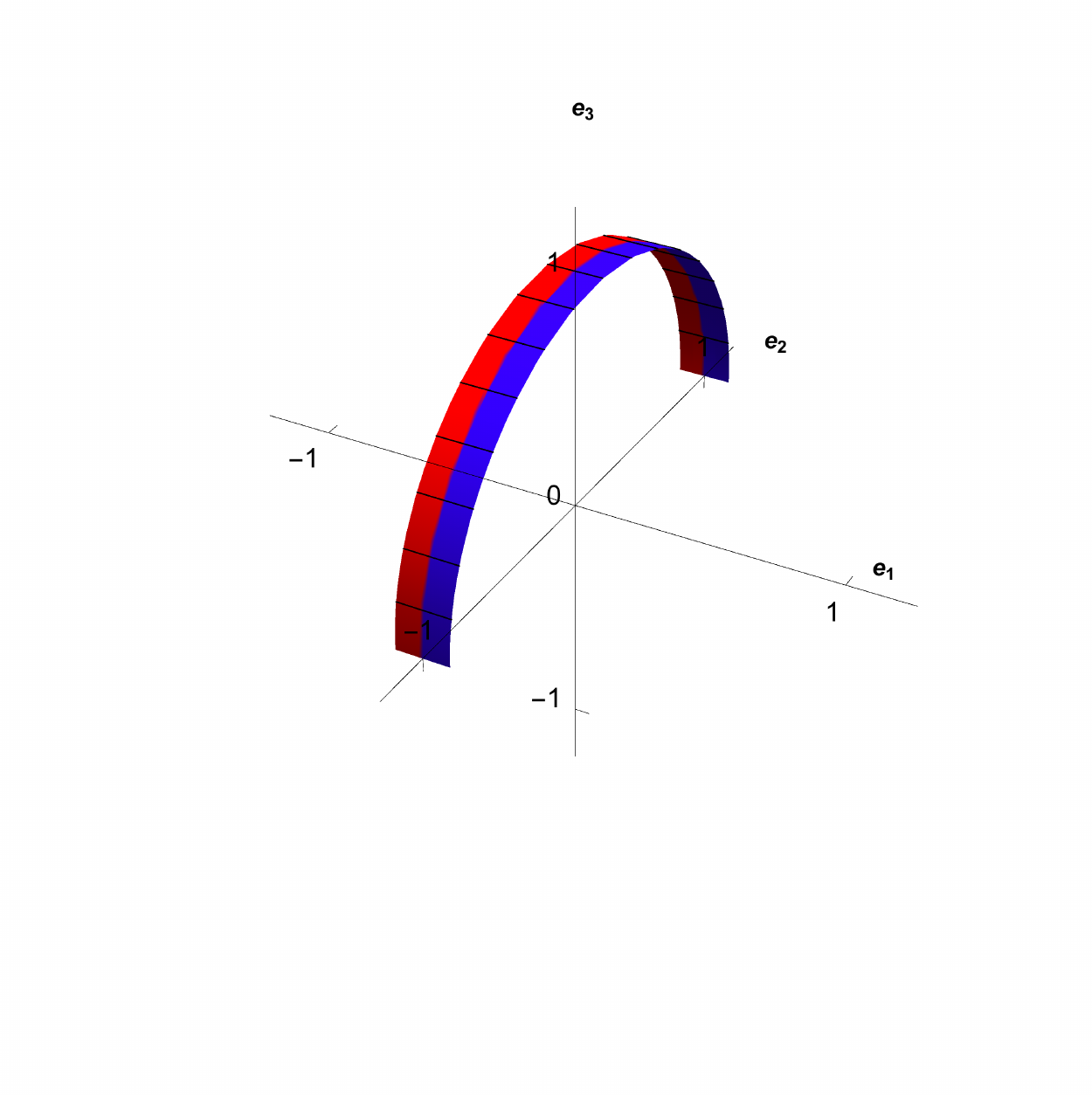}}
\subfigure[Relaxed hinge, $\Xi_1=1$]{\includegraphics[width=.32\textwidth] {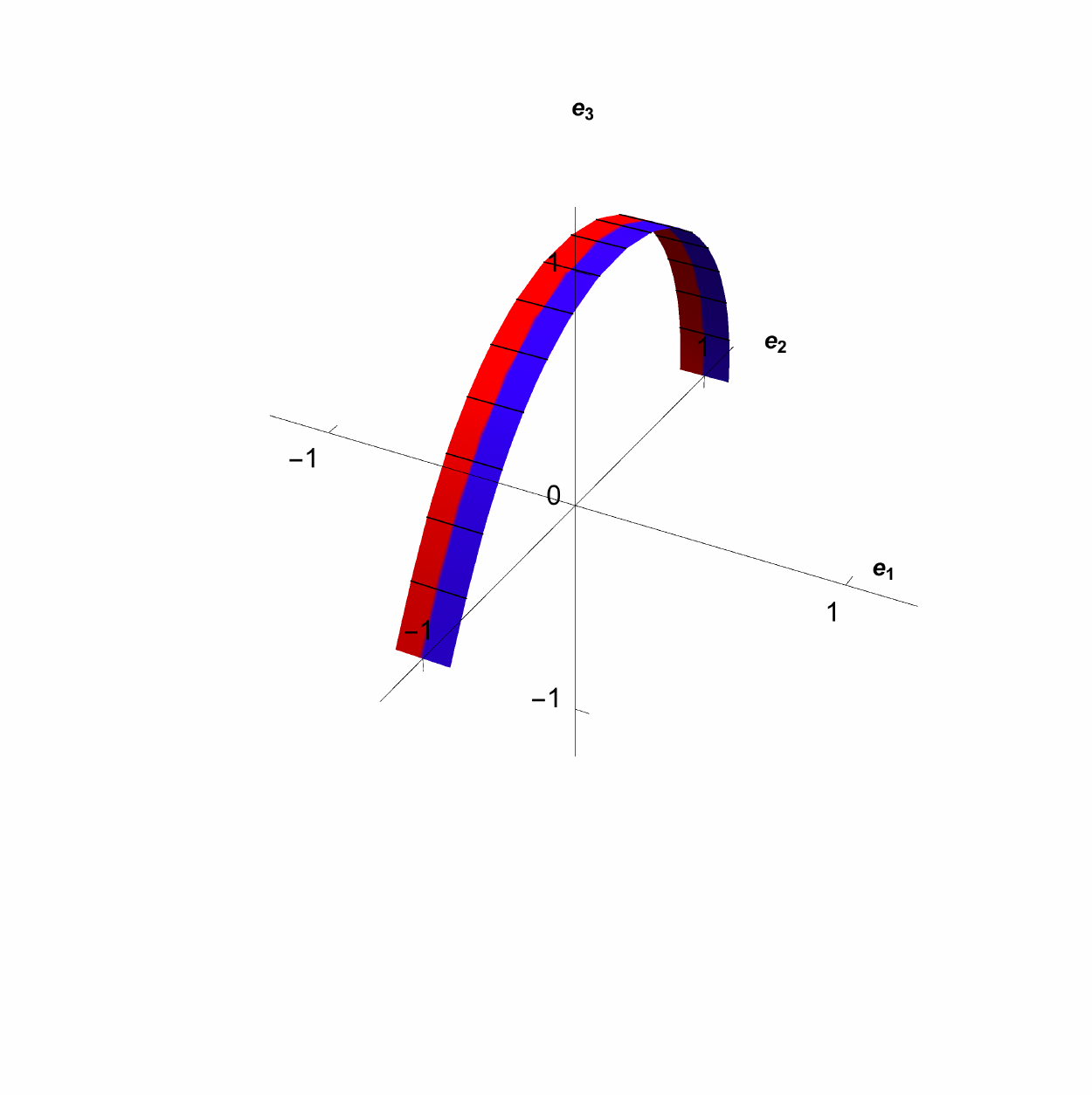}}
\subfigure[Twist continuation, $\phi=5\pi/12$]{\includegraphics[width=.32\textwidth] {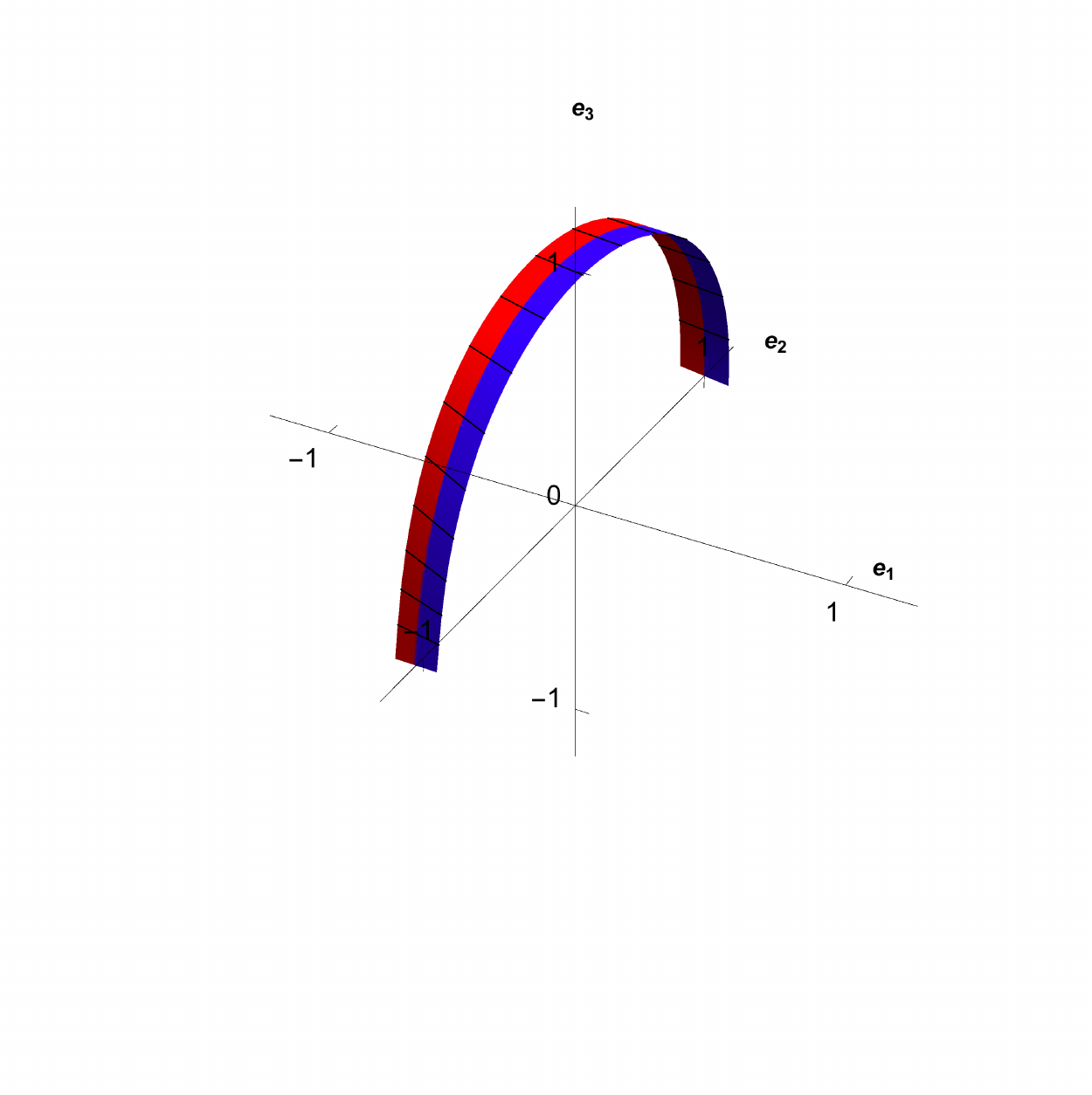}}
\subfigure[Twist continuation, $\phi=5\pi/6$]{\includegraphics[width=.32\textwidth] {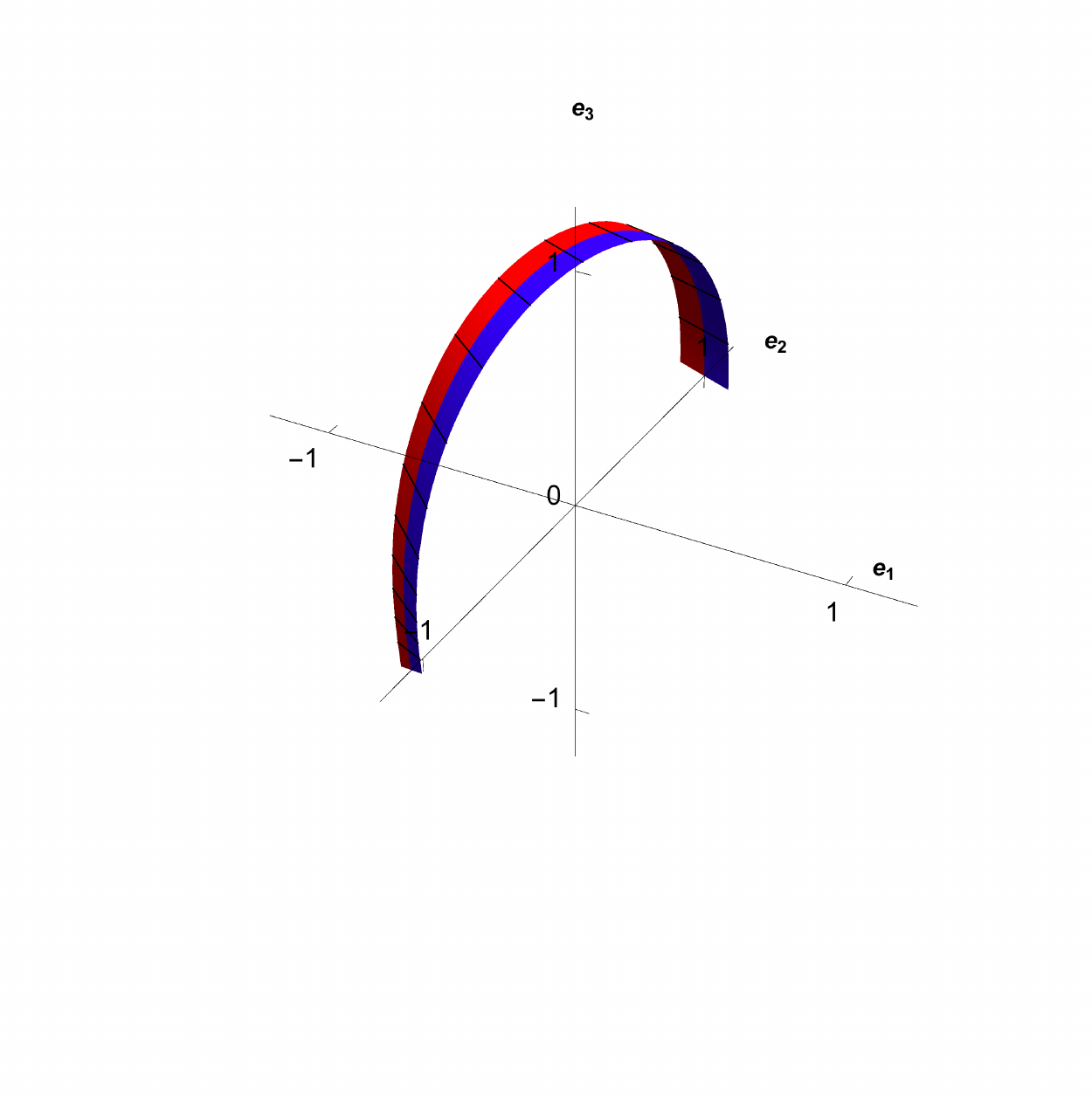}}
\subfigure[Twist continuation, $\phi=2\pi/3$]{\includegraphics[width=.32\textwidth] {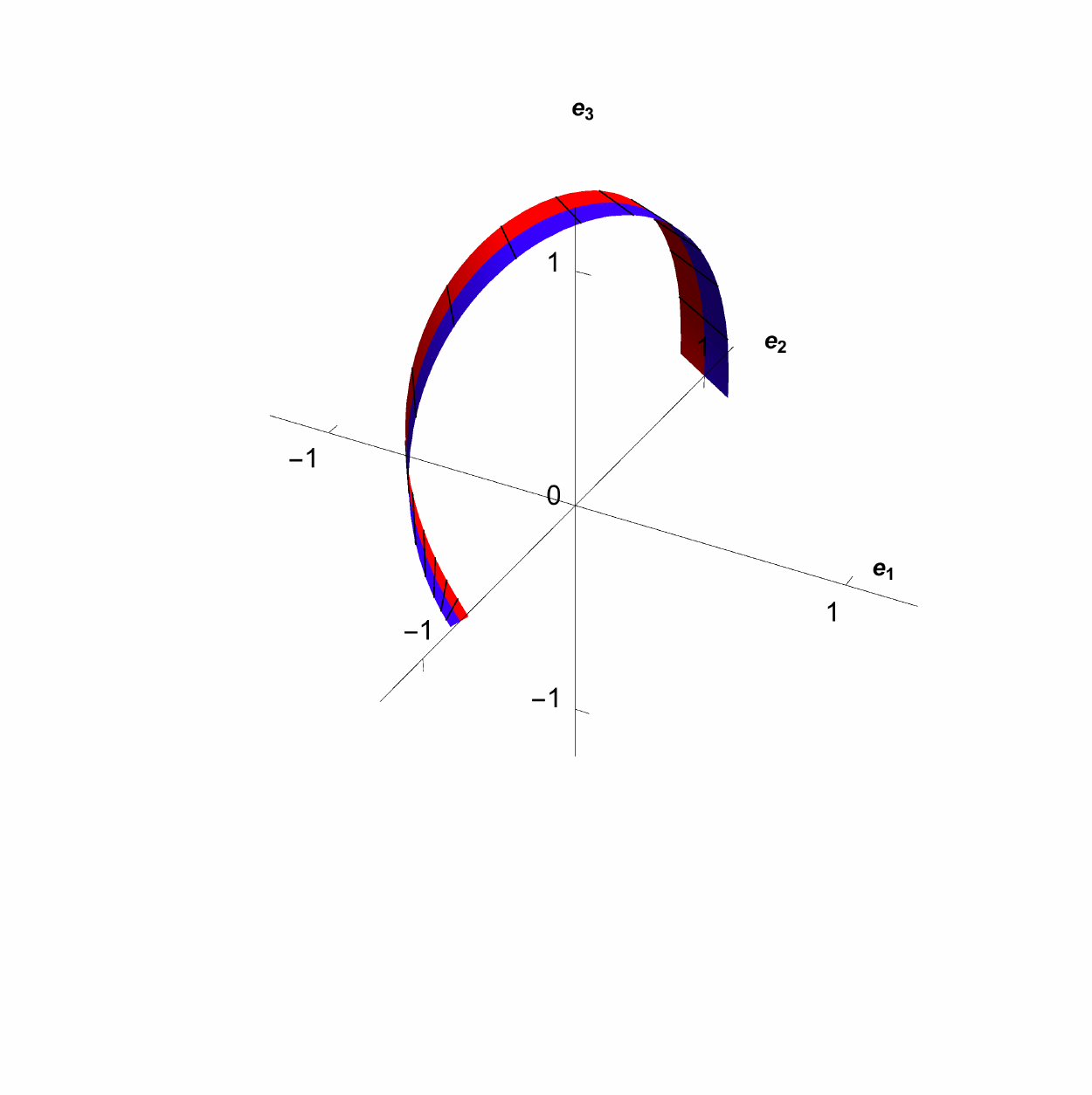}}
\subfigure[End of continuation, $\phi=0$, forming half of a M\"{o}bius strip]{\includegraphics[width=.32\textwidth] {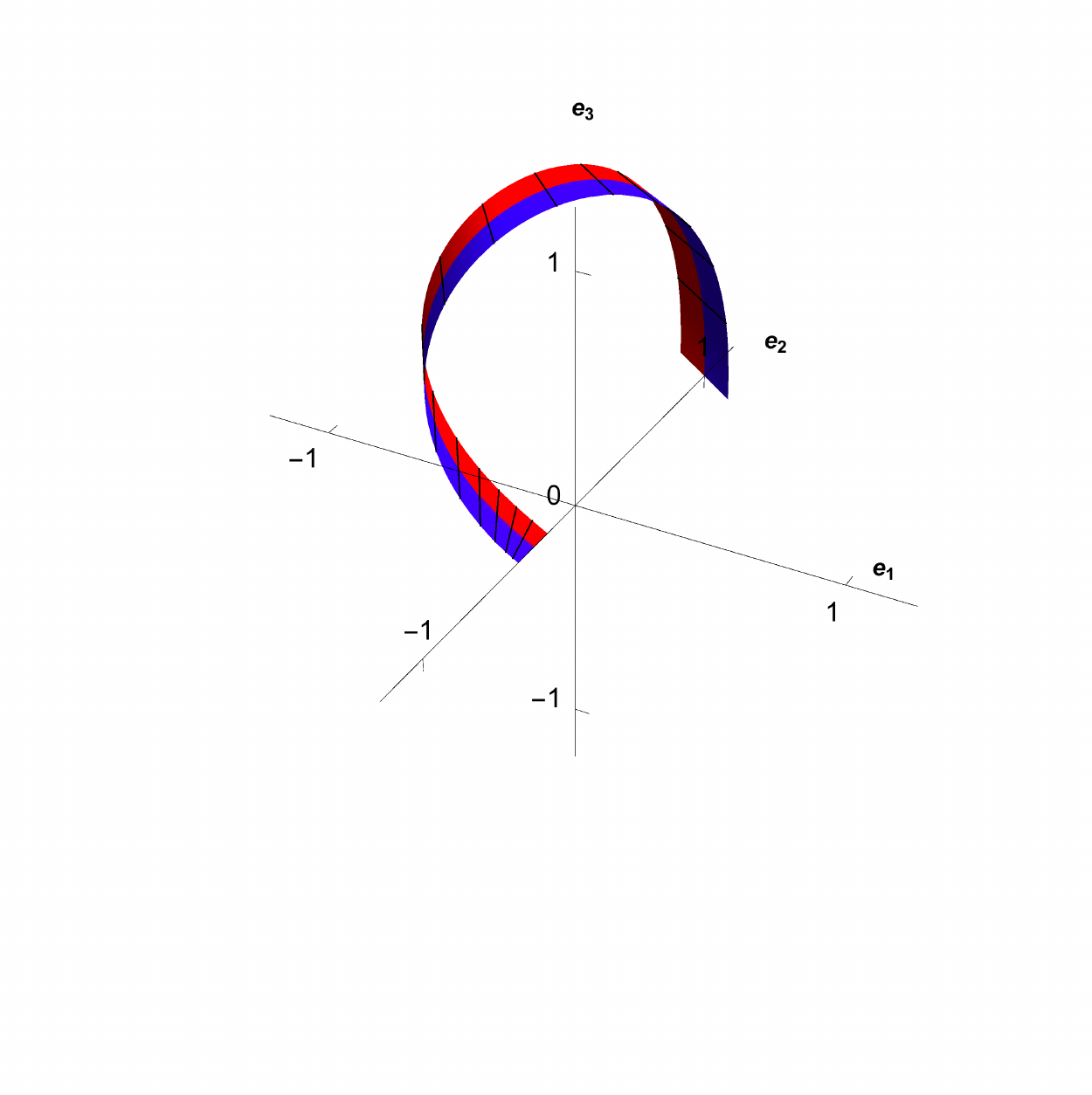}}
\caption{\footnotesize Initial and intermediate configurations to set up the M\"{o}bius strip.  $\phi$ is the angle between the $\uvec{d}_1\left(\pi\right)$ and the axis of symmetry $\uvec{e}_2$. (a)-(b) Continuation in $\Xi_1$ to relax the $\dvec{m}_1$ moment about the hinge to zero. (c)-(f) Twisting continuation in $\Xi_2$ to align the strip end with the axis of symmetry. Black lines indicate the direction of the generator of curvature, $\mathbf{b}\left(s\right)$.}
\label{fig:momentrelax}
\end{figure}

\begin{figure}
\centering
\subfigure[$w=0.2$\quad$\epsilon=0.033$]{\includegraphics[width=.4\textwidth] {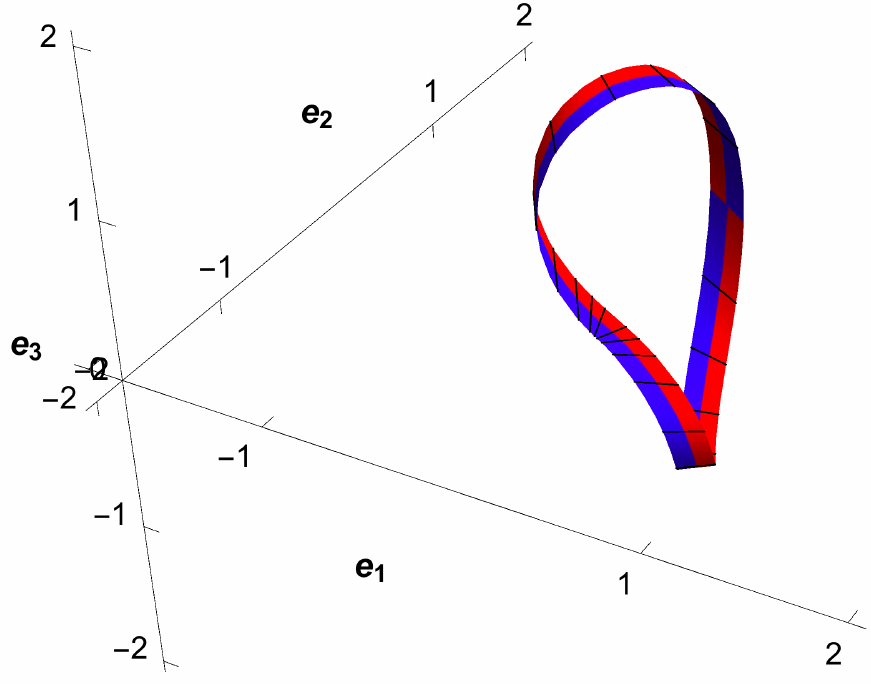}}\hspace{.05\textwidth}
\subfigure[$w=0.4$\quad$\epsilon=0.11$]{\includegraphics[width=.4\textwidth] {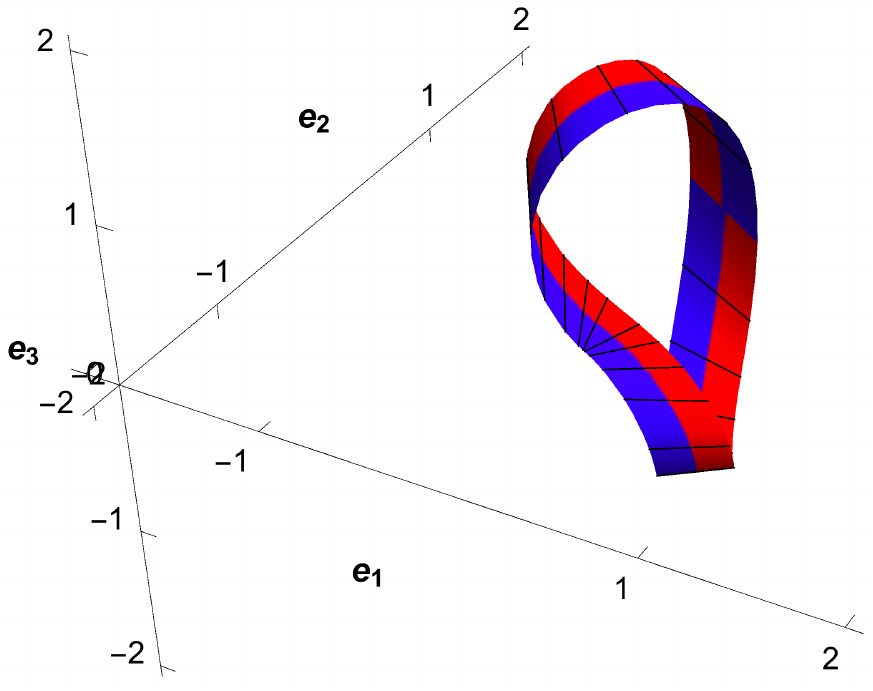}}\vspace{.05\textheight}
\subfigure[$w=1$\quad$\epsilon=1.0$]{\includegraphics[width=.4\textwidth] {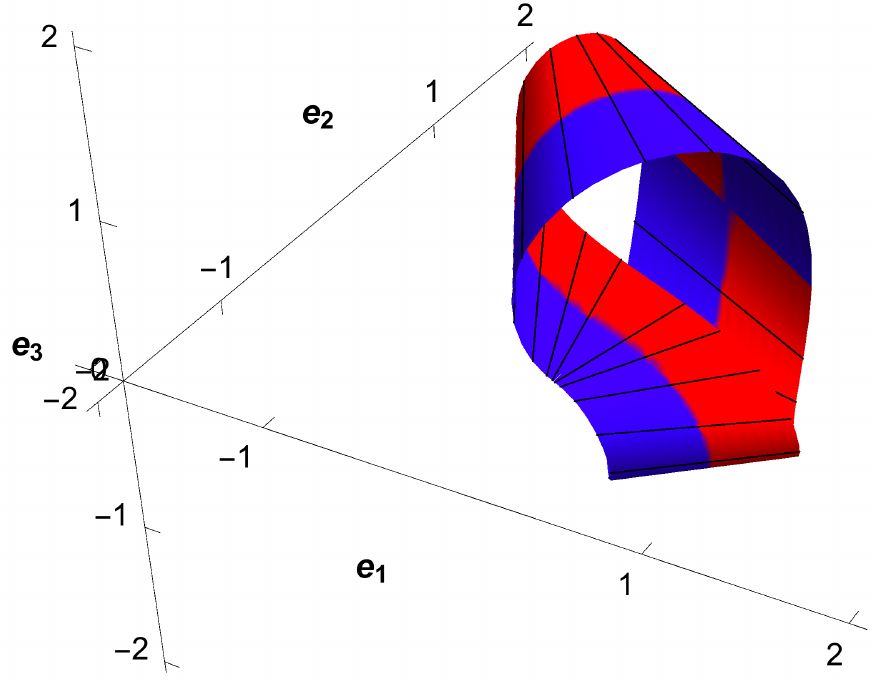}}\hspace{.05\textwidth}\subfigure[$w=1.6$\quad$\epsilon=4.08$]{\includegraphics[width=.4\textwidth] {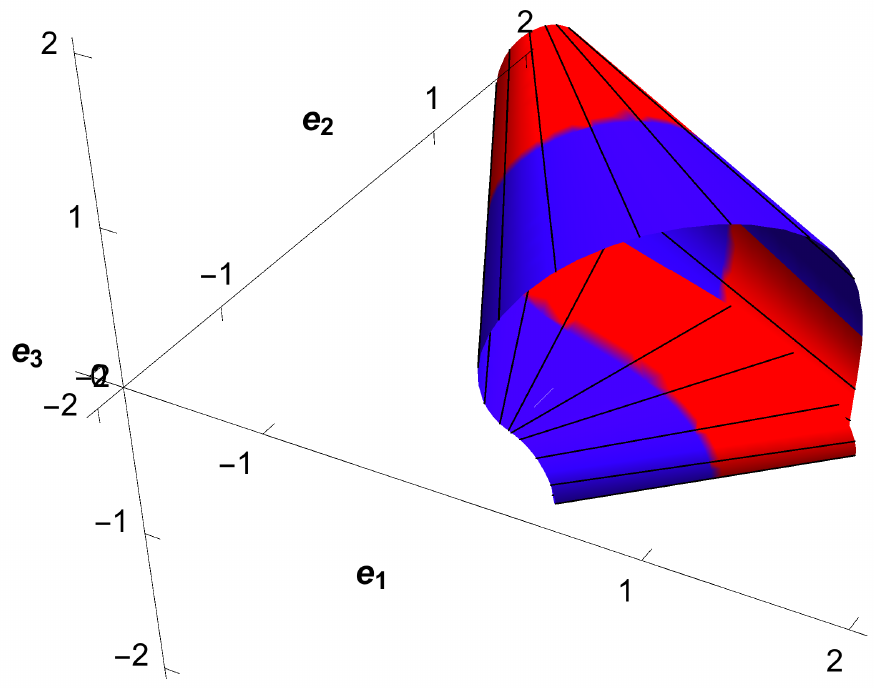}}\vspace{.05\textheight}
\subfigure[$w=2.0$\quad$\epsilon=10$]{\includegraphics[width=.4\textwidth] {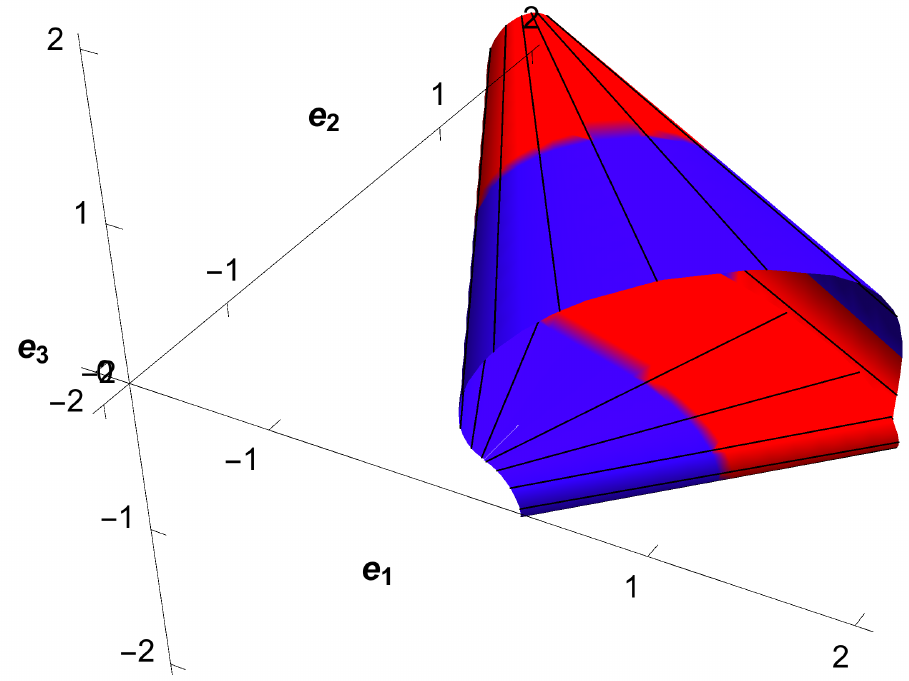}}
\caption{\footnotesize M\"{o}bius strip configurations for different widths. For each $w$ value, the configuration with the smallest $\epsilon$ value is shown. Note that the width in \cite{Starostin2014} corresponds to one-half the width in this formulation.}
\label{fig:o8tableresults}
\end{figure}

\begin{figure}
\centering
\subfigure{\includegraphics[width=.47\textwidth] {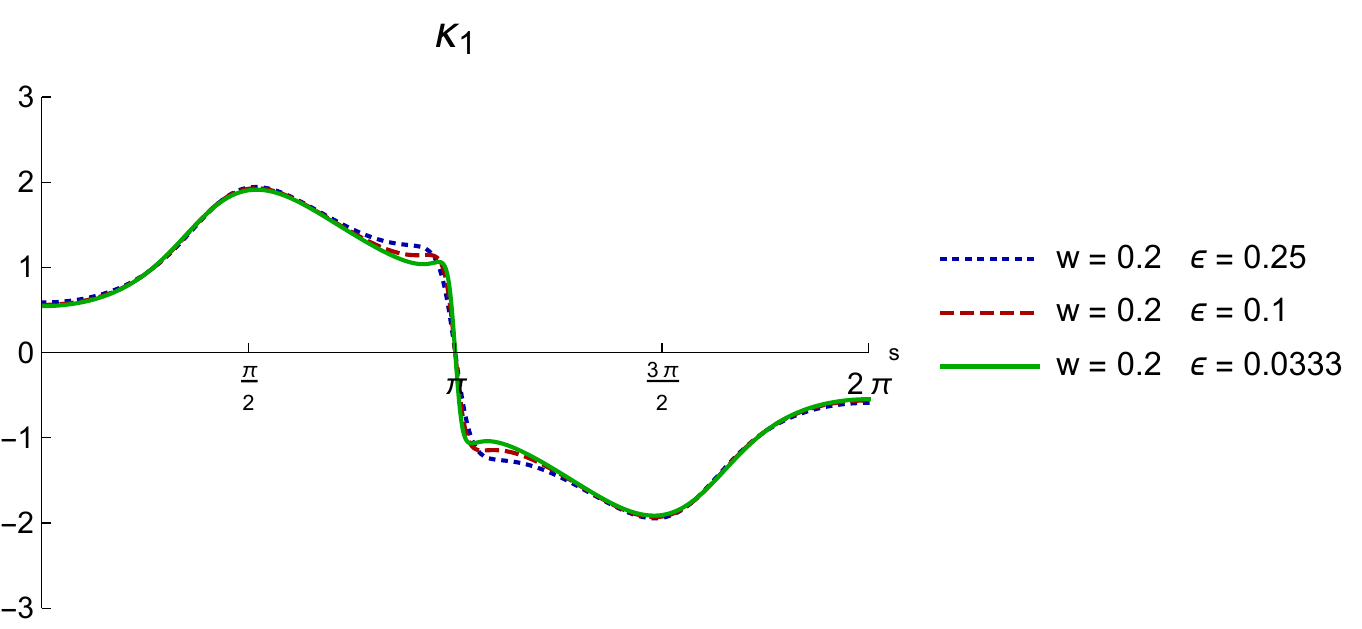}}\quad
\subfigure{\includegraphics[width=.47\textwidth] {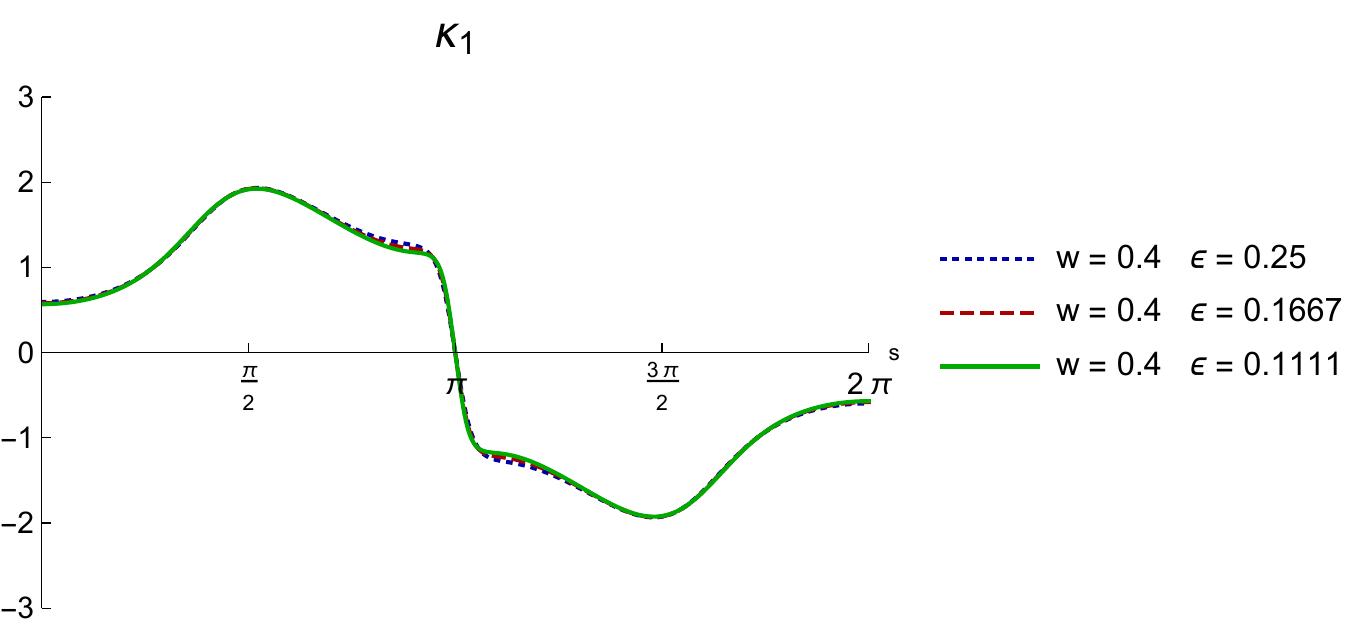}}
\subfigure{\includegraphics[width=.47\textwidth] {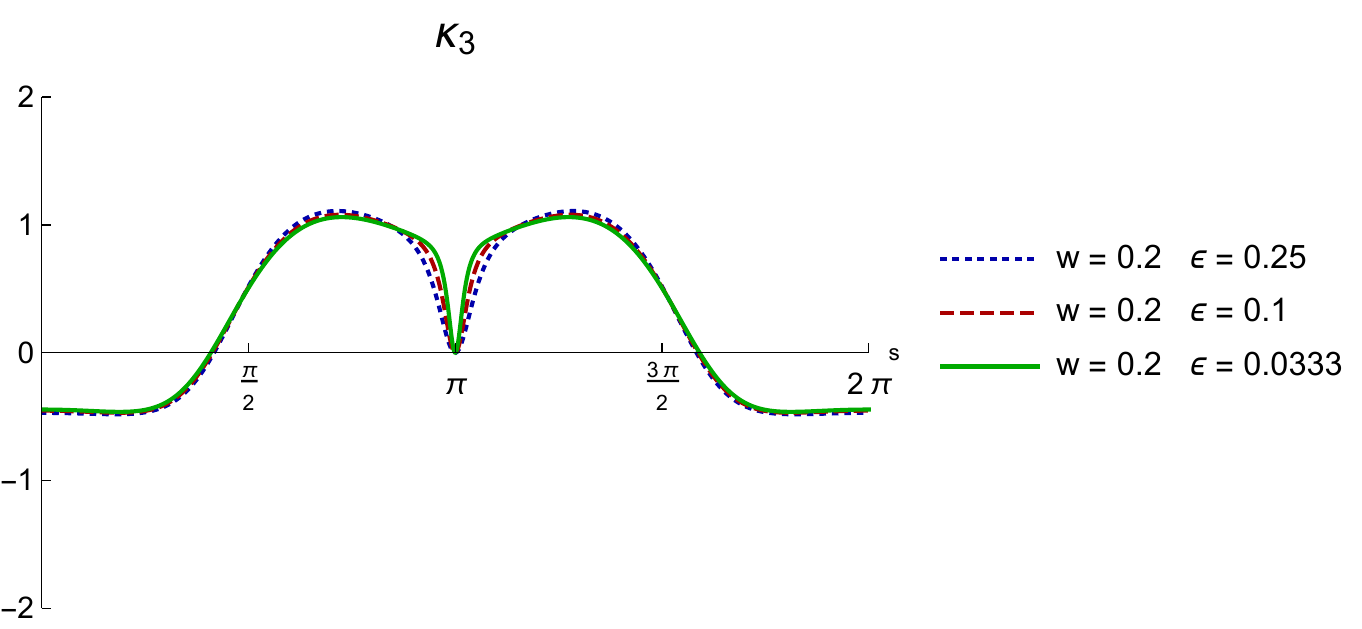}}\quad
\subfigure{\includegraphics[width=.47\textwidth] {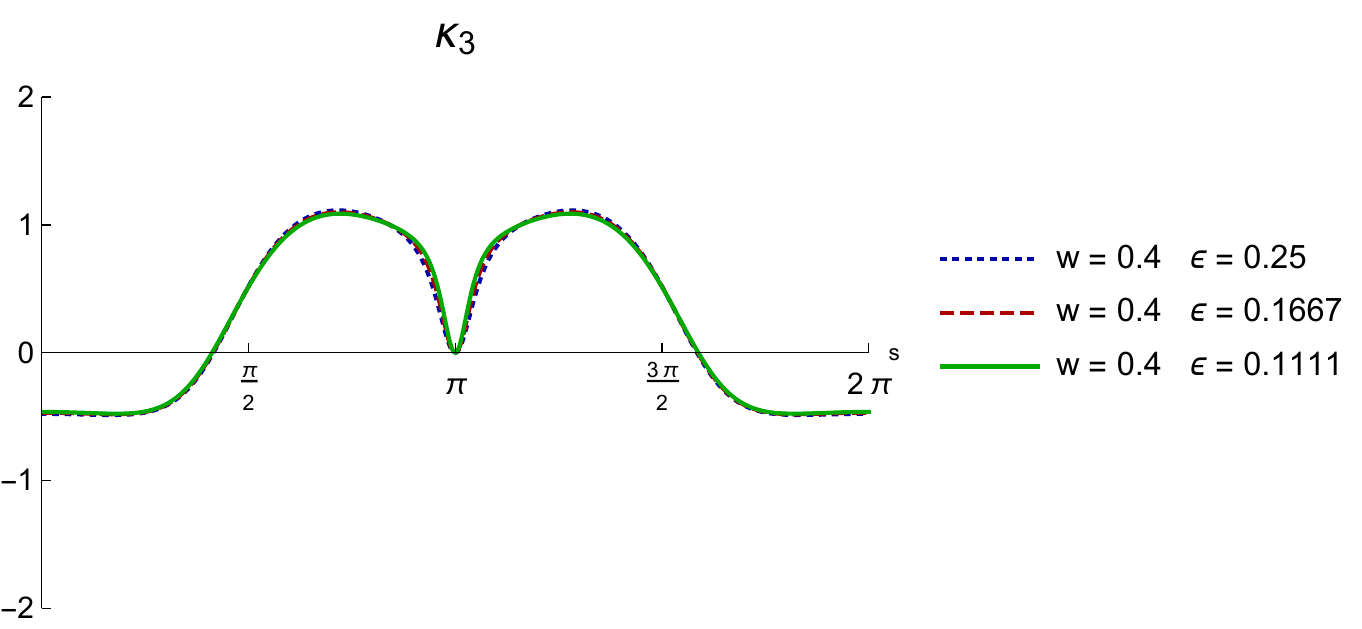}}
\subfigure{\includegraphics[width=.47\textwidth] {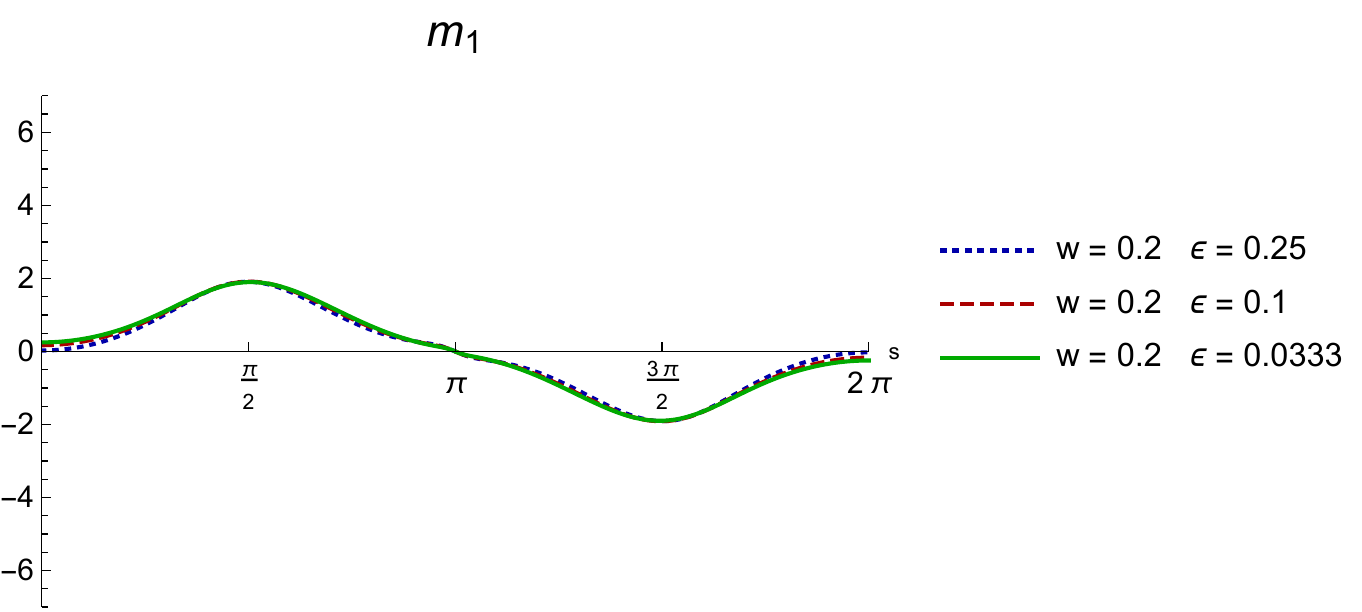}}\quad
\subfigure{\includegraphics[width=.47\textwidth] {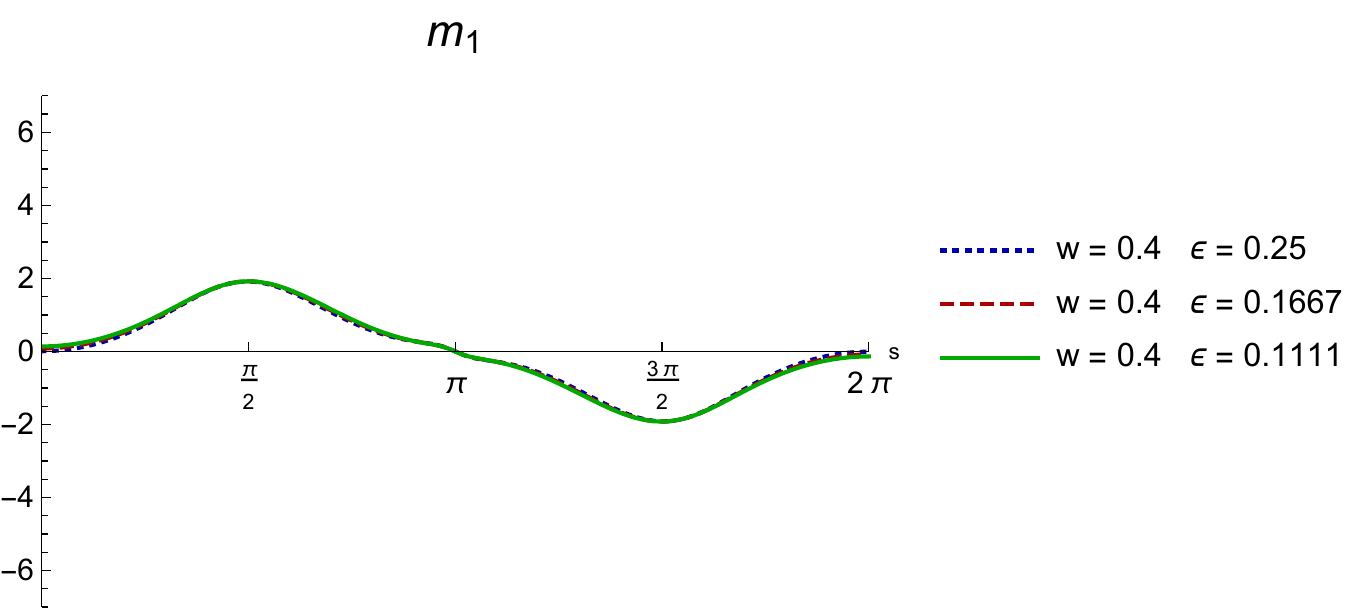}}
\subfigure{\includegraphics[width=.47\textwidth] {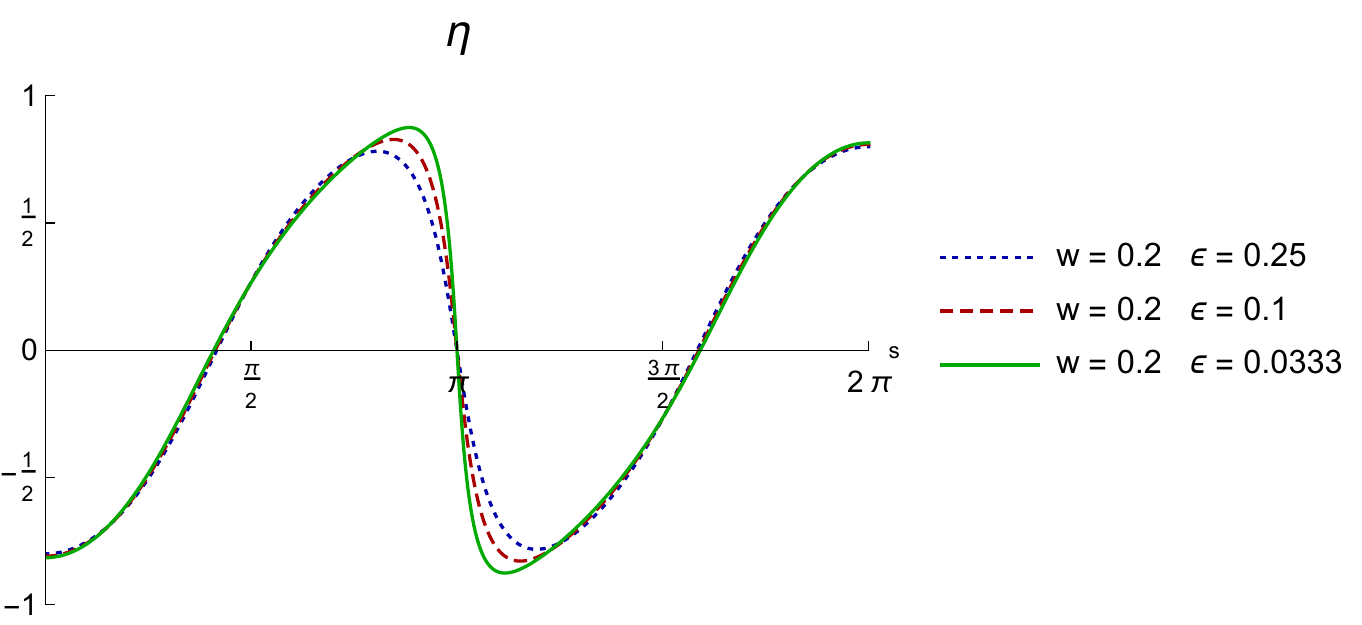}}\quad
\subfigure{\includegraphics[width=.47\textwidth] {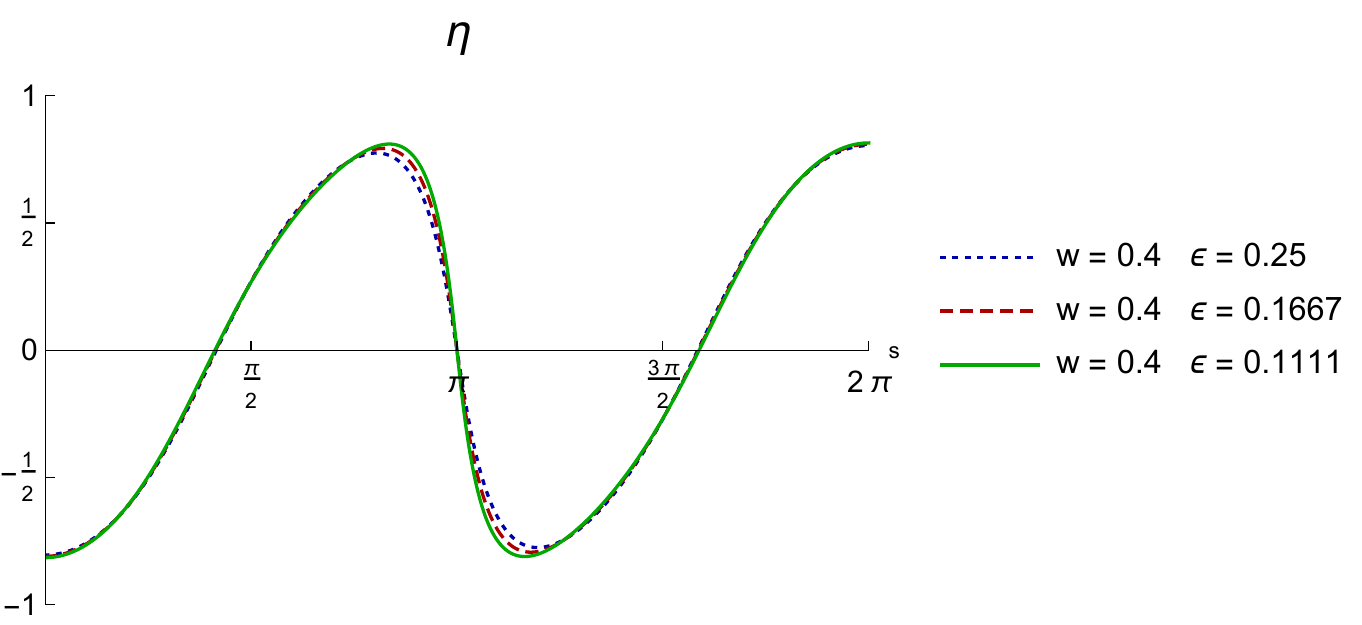}}
\subfigure{\includegraphics[width=.47\textwidth] {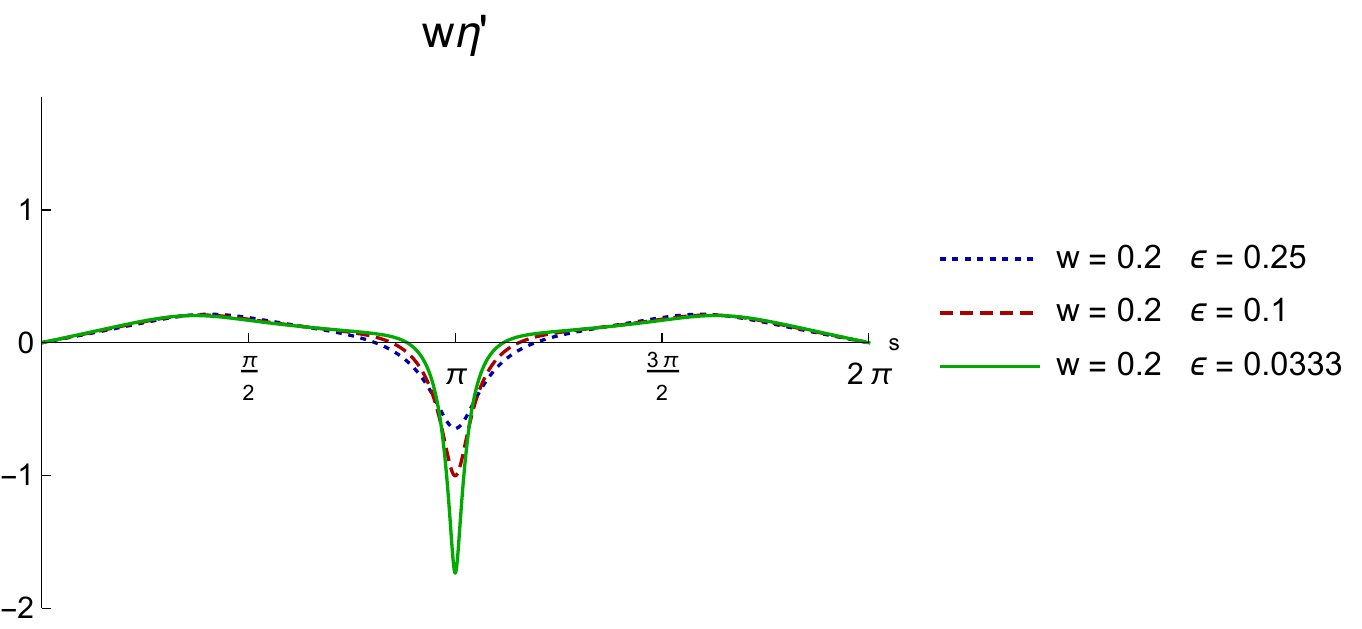}}\quad
\subfigure{\includegraphics[width=.47\textwidth] {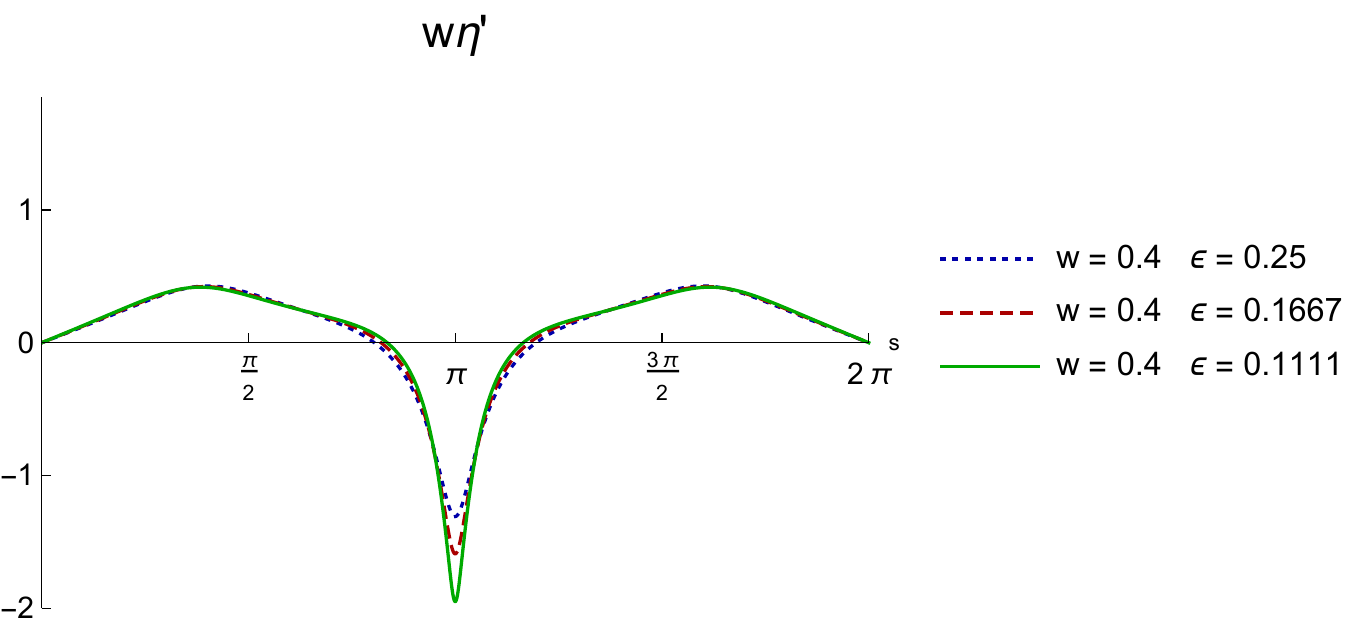}}
\caption{\footnotesize The curvature, twist, moment, $\eta$, and $\eta\prime w$ values for different continuations configurations for $w=0.2$ and $w=0.4$. Note that $\eta$ shows a boundary layer around the point $s=\pi$. Also note that the energy density in (\ref{eq:gpar}) becomes complex for $w=\pm2$. }
\label{fig:append1}
\end{figure}

\begin{figure}
\centering
\subfigure{\includegraphics[width=.47\textwidth] {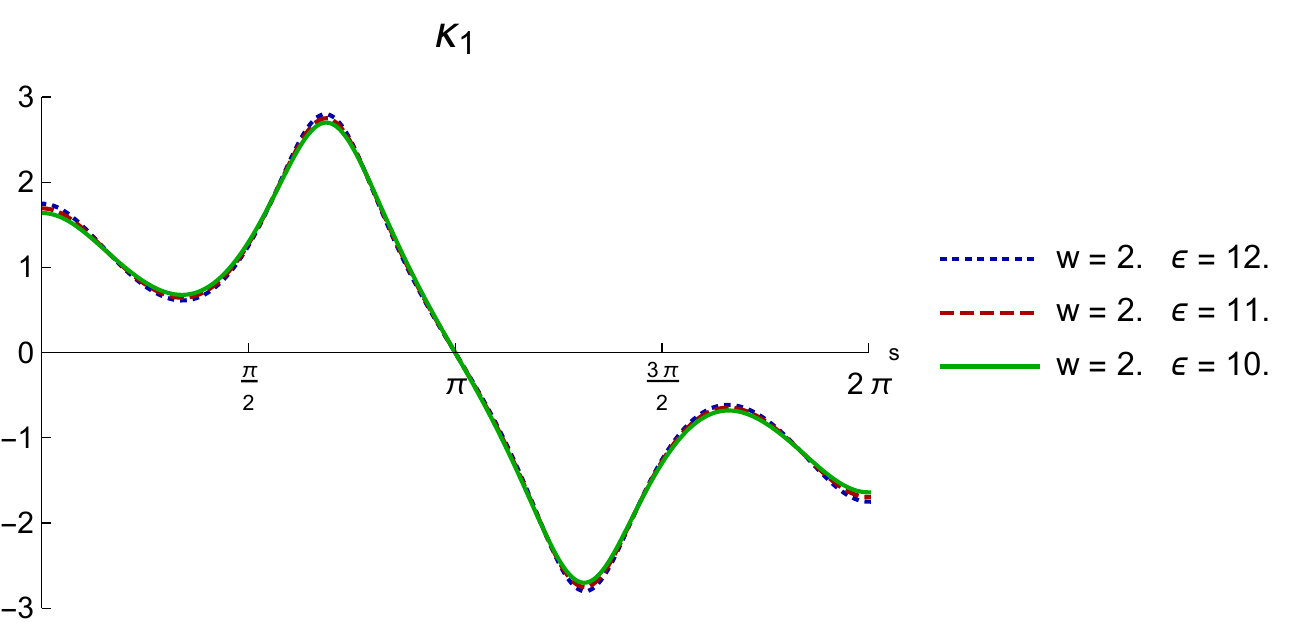}}
\subfigure{\includegraphics[width=.47\textwidth] {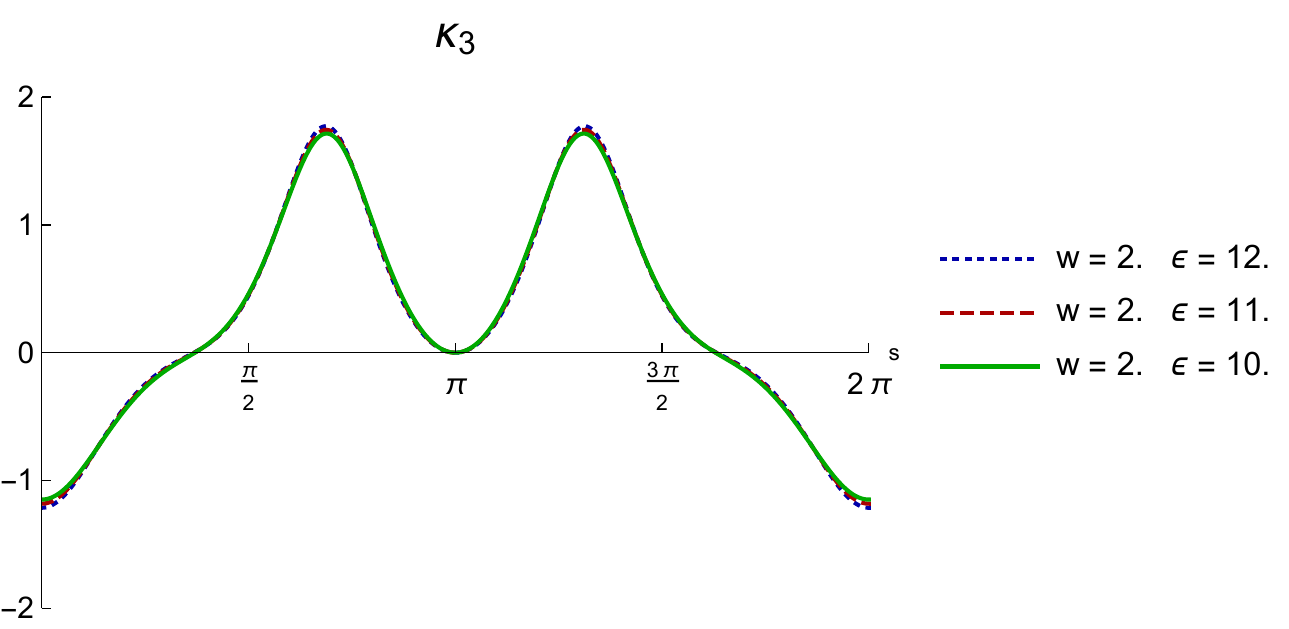}}
\subfigure{\includegraphics[width=.47\textwidth] {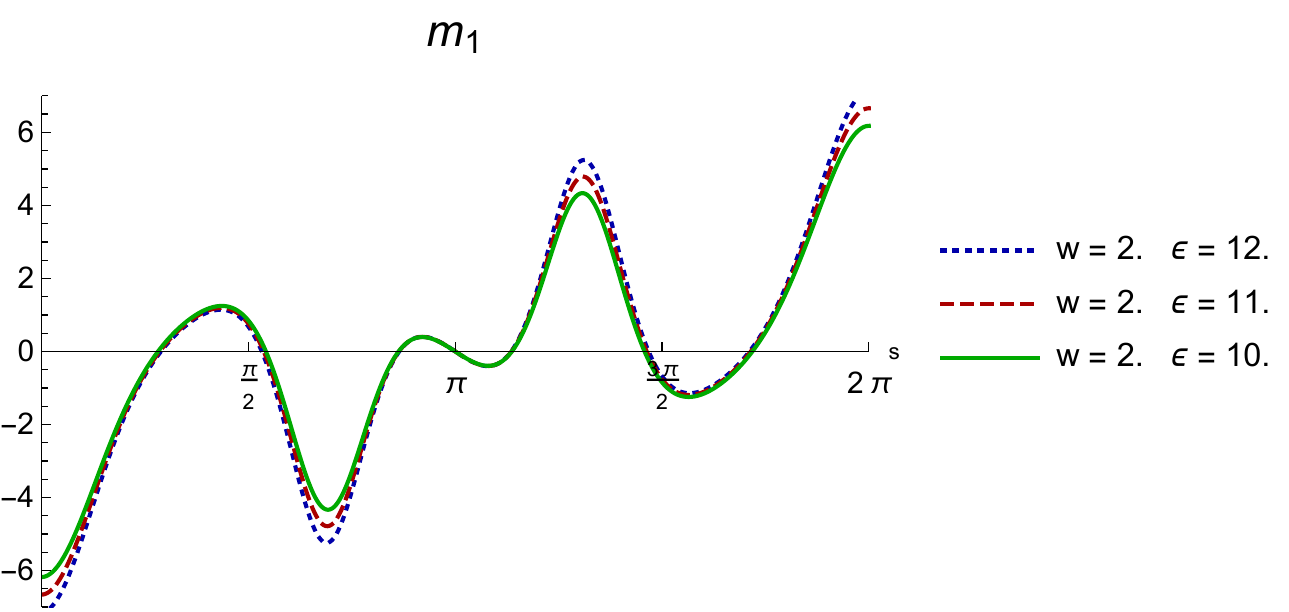}}
\subfigure{\includegraphics[width=.47\textwidth] {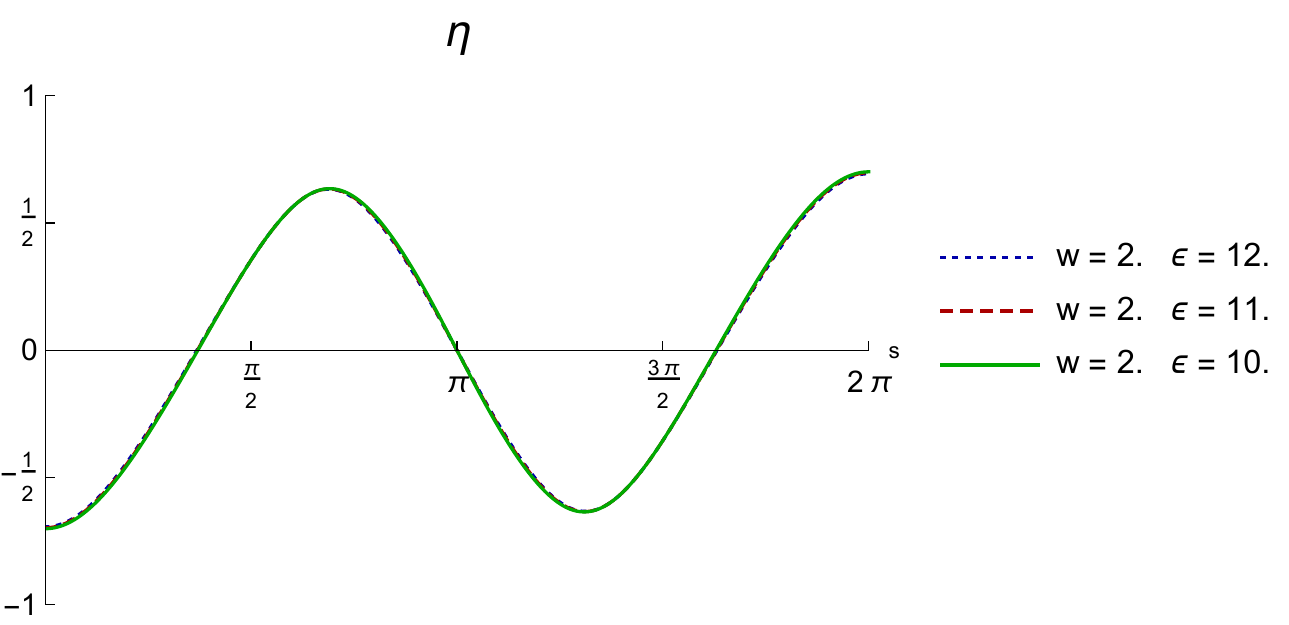}}
\subfigure{\includegraphics[width=.47\textwidth] {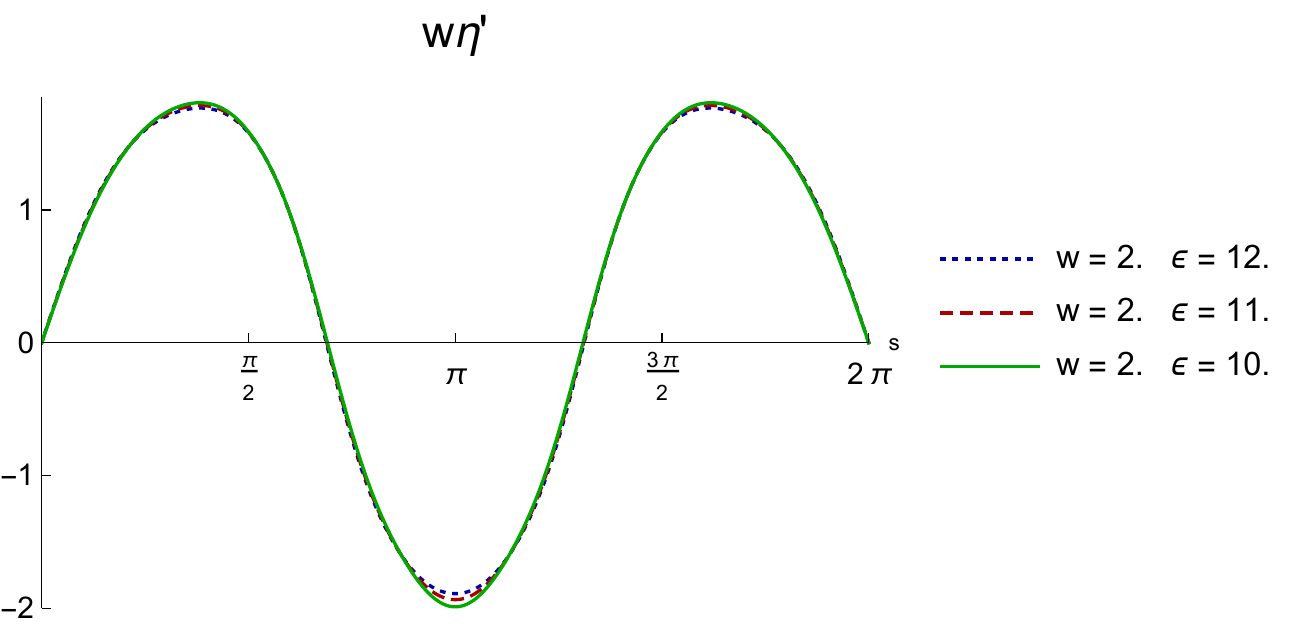}}
\caption{\footnotesize The curvature, twist, moment, $\eta$, and $\eta\prime w$ values for different continuations configurations for $w=2.0$. Note that $\eta$ shows a boundary layer around the point $s=\pi$. Also note that the energy density in (\ref{eq:gpar}) becomes complex for $w=\pm2$. }
\label{fig:append2}
\end{figure}

Starting with a fixed width, $w$, and regularizing term, $\epsilon$, continuation in the auxiliary variables, $\Xi_1$ and $\Xi_2$, is used to arrive at the M\"{o}bius strip configuration, as depicted in Figure \ref{fig:momentrelax}. Once the M\"{o}bius configuration is reached, continuation in the width $w$ and the regularizing coefficient $\epsilon$ are used to obtain the configurations shown in Figure \ref{fig:o8tableresults}.  In particular, for a given fixed width, $\epsilon$ is reduced as small as possible.  However, it cannot be decreased indefinitely:  As shown in Figures \ref{fig:append1}-\ref{fig:append2}, as $\epsilon$ is decreased for given fixed width, the product $w\eta^\prime$ approaches the value $-2$ at the hinge location $s=\pi$ at which the density function (\ref{eq:gpar}) blows up.  As discussed after equation (\ref{eq:devstripans}), $\abs{w\eta^\prime}=2$ indicates a breakdown of injectivity for the mapping (\ref{eq:devstripans}) at the edge of the strip, and the through-width integration leading to (\ref{eq:devenergy}), (\ref{eq:gpar}) in \cite{Wunderlich1962} is no longer valid.  Of course the right side of (\ref{eq:gpar}) is undefined for $\abs{w\eta^\prime}\geq2$  We further observe from Figure \ref{fig:append1} that the larger the width, the larger the value of regularizing coefficient.  In Figures \label{fig:stripplotso8w} and \label{fig:stripplotso8ep}, we plot the computed ``generators of curvature" on the reference strip, viz., the right side of (\ref{eq:devstripans}) at regularly spaced values of $s$ with $v$ ranging over the width.  Observe that as $\epsilon$ is decreased or $w$ increased, the generators almost intersect at the midpoint ($s=\pi$) where the hinge boundary condition is enforced.

In Figure \ref{fig:o8tableresults} we illustrate some key solution fields for various fixed widths (as indicated).  In each field we also demonstrate the robustness of our results for three consecutively decreasing values of the regularizing parameter.  For each of the widths $w=0.2$, $0.4$, and $1.0$, the plots for $\kappa_1$ and $\eta$ demonstrate singular-perturbation behavior in a small neighborhood of the hinge location $s=\pi$. For the larger widths $w=1.6$ and $w=2.0$, this effect is less concentrated, which is undoubtedly due to the large values of $\epsilon$ required.   

\begin{figure}
\centering
\subfigure{\includegraphics[width=.47\textwidth] {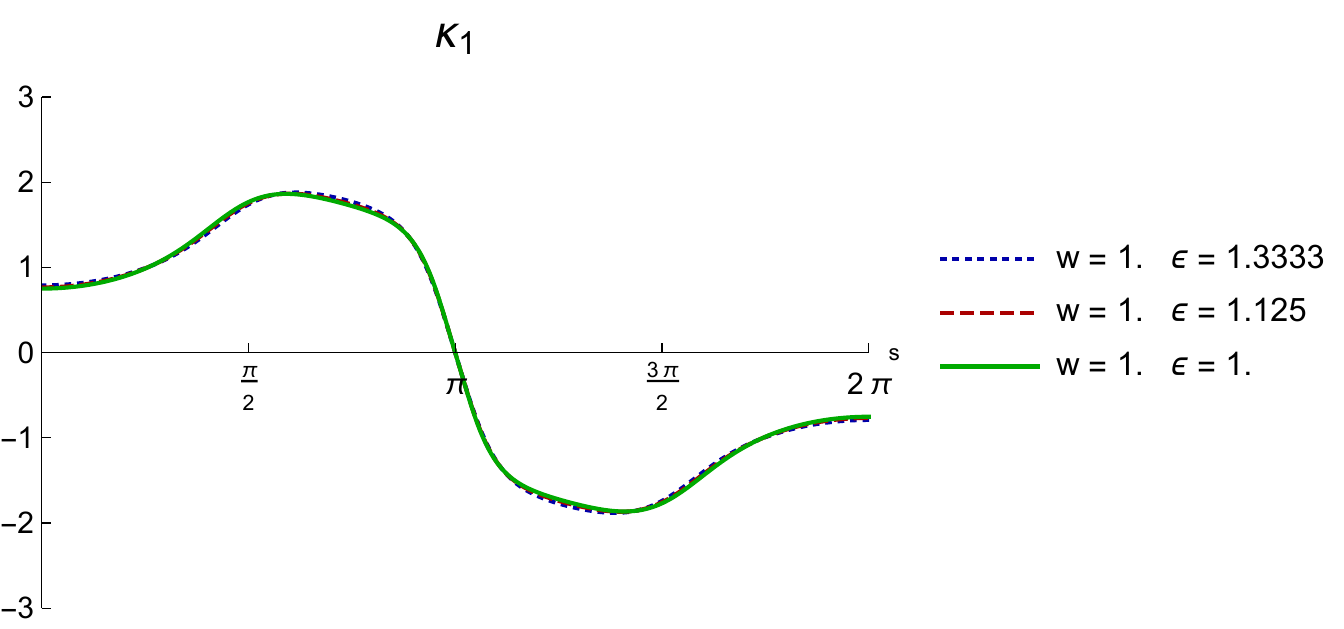}}\quad
\subfigure{\includegraphics[width=.47\textwidth] {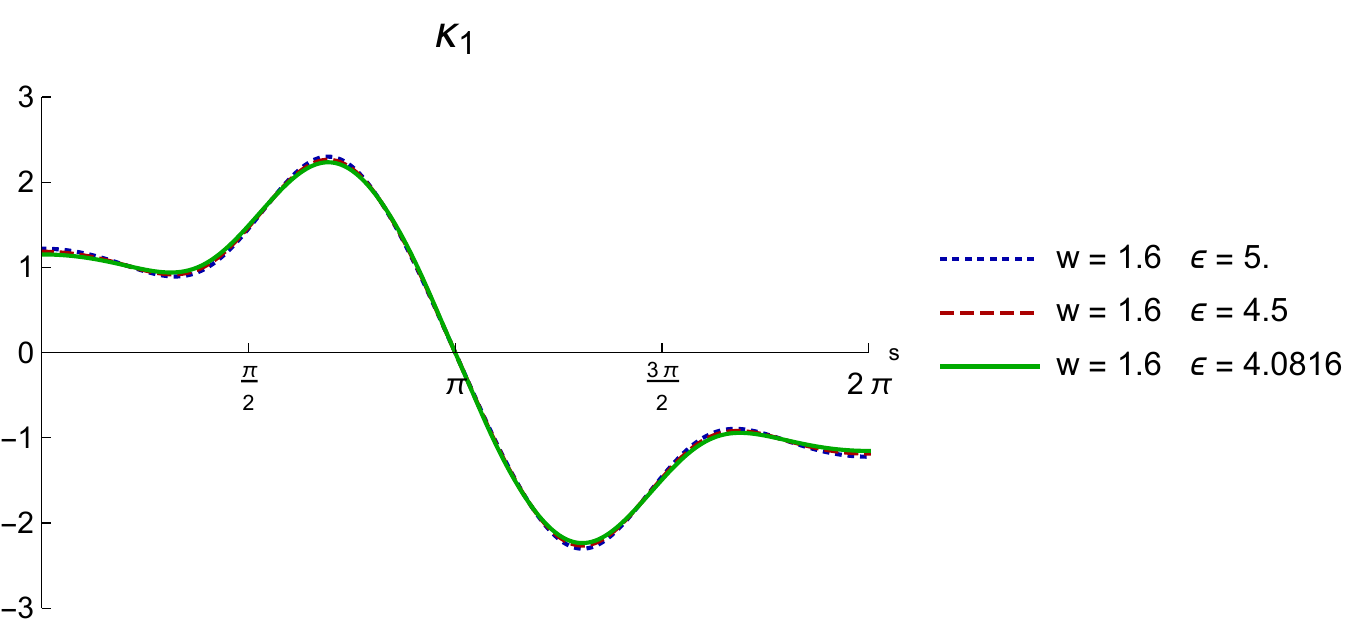}}
\subfigure{\includegraphics[width=.47\textwidth] {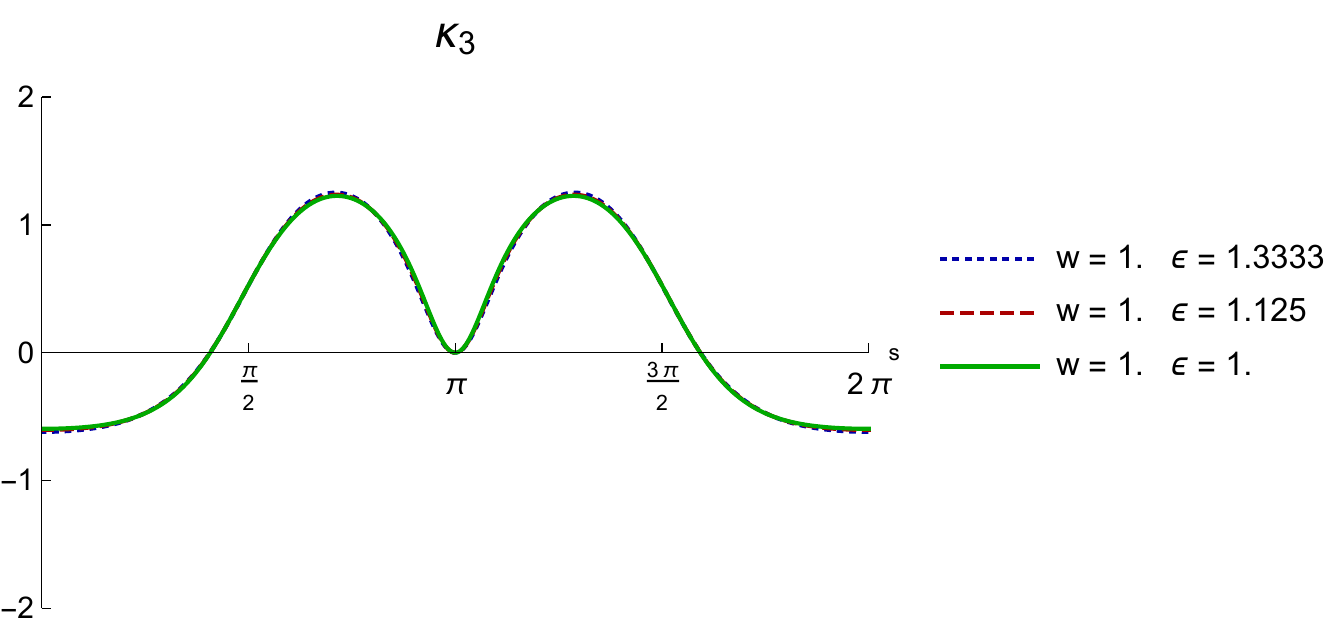}}\quad
\subfigure{\includegraphics[width=.47\textwidth] {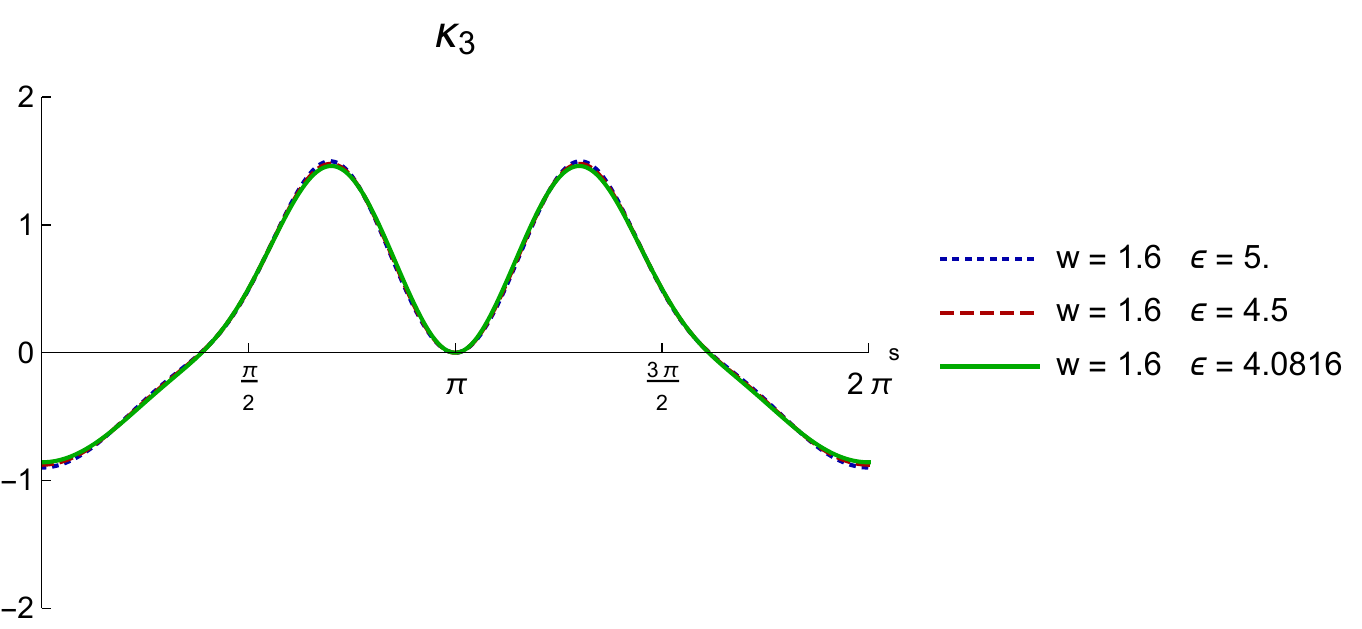}}
\subfigure{\includegraphics[width=.47\textwidth] {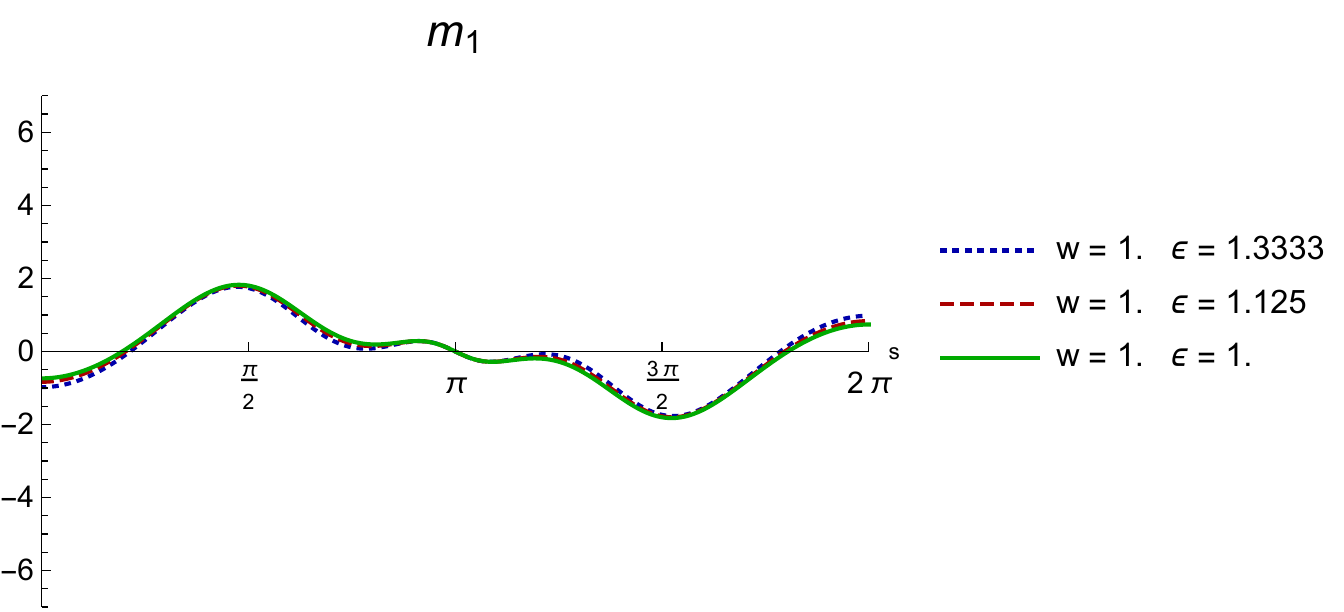}}\quad
\subfigure{\includegraphics[width=.47\textwidth] {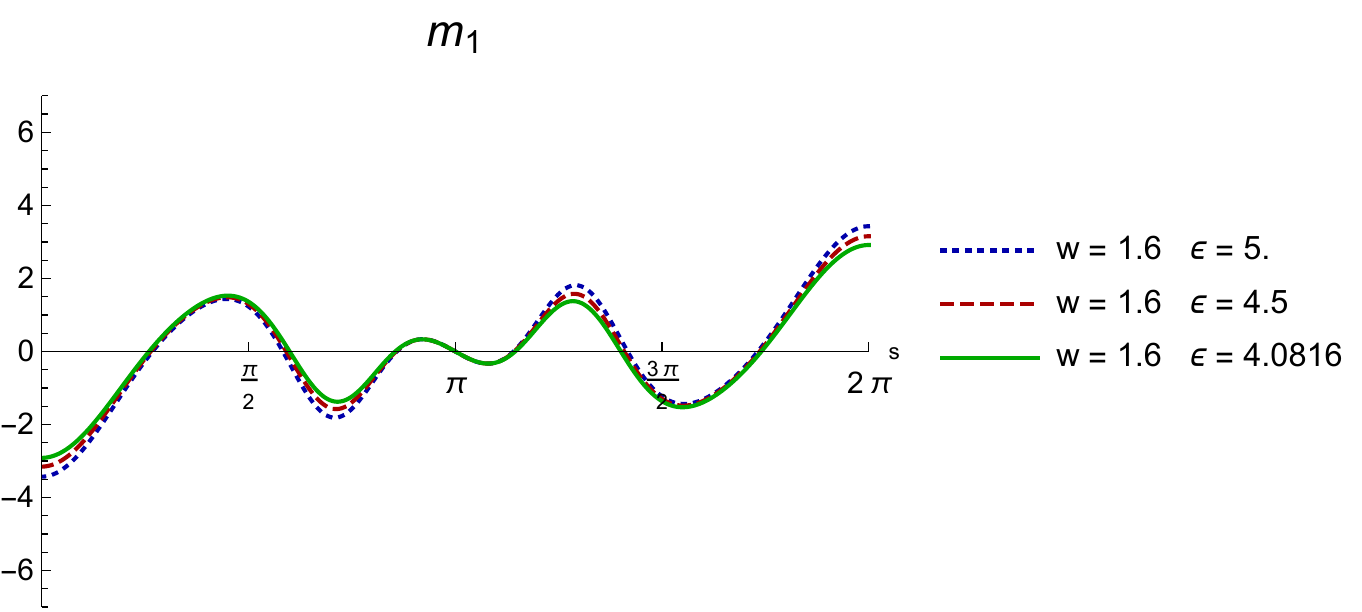}}
\subfigure{\includegraphics[width=.47\textwidth] {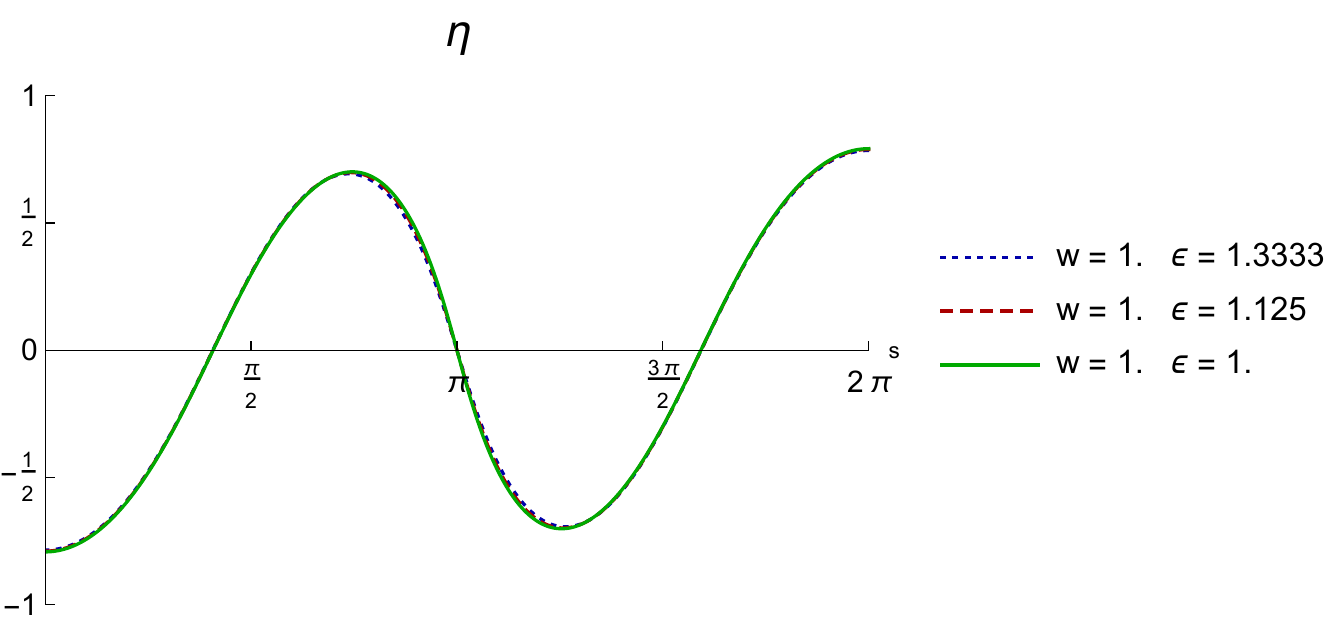}}\quad
\subfigure{\includegraphics[width=.47\textwidth] {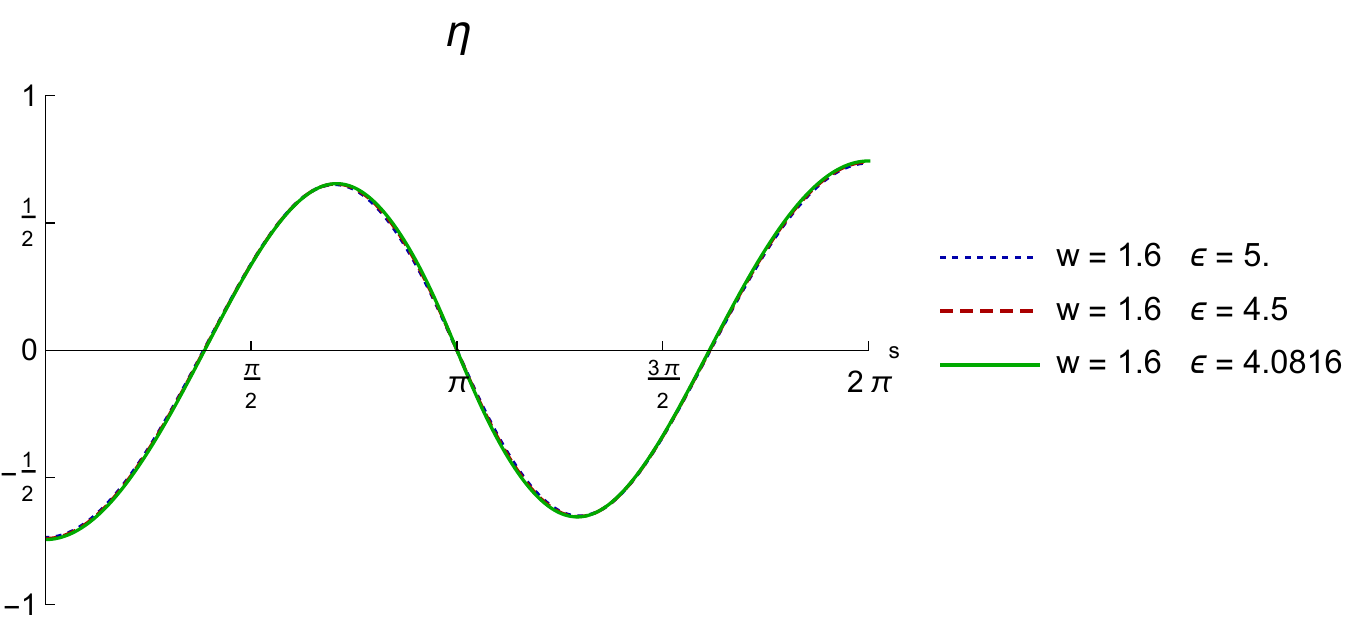}}
\subfigure{\includegraphics[width=.47\textwidth] {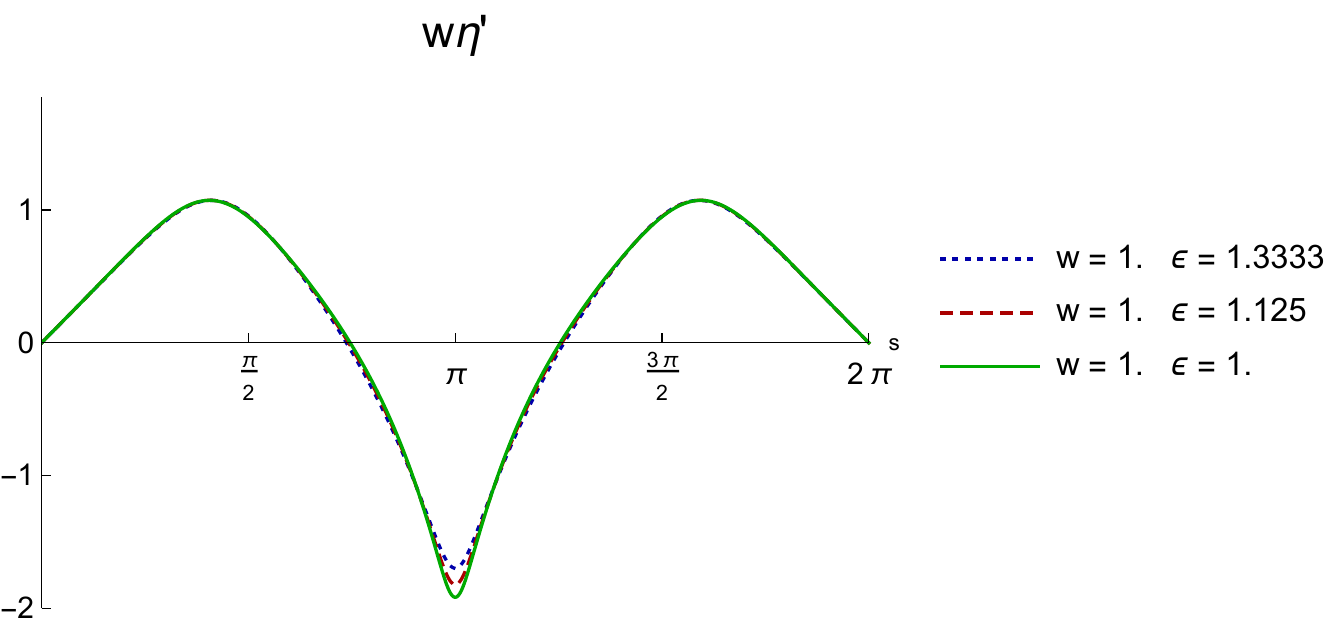}}\quad
\subfigure{\includegraphics[width=.47\textwidth] {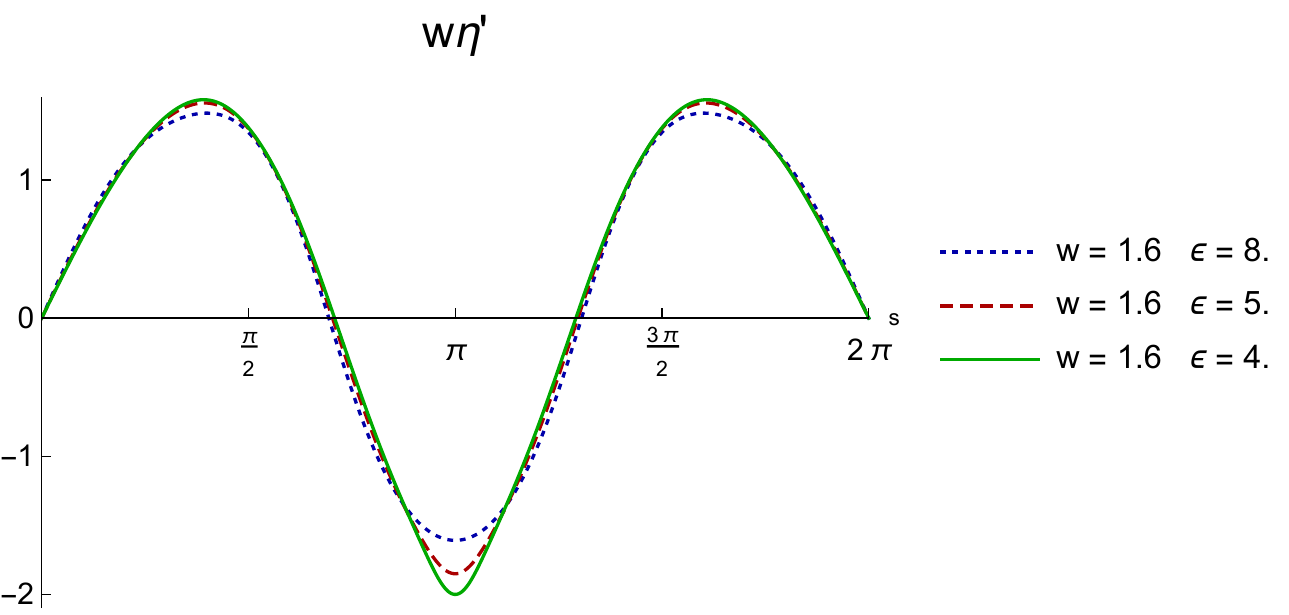}}
\caption{\footnotesize  The curvature, twist, moment, $\eta$, and $\eta\prime w$ values for different continuations configurations for $w=1.0$ and $w=1.6$. Note that $\eta$ shows a boundary layer around the point $s=\pi$. Also note that the energy density in (\ref{eq:gpar}) becomes complex for $w=\pm2$. }
\end{figure}
\begin{figure}
\centering
\subfigure[$w=0.4$, $\epsilon=0.25$]{\includegraphics[width=.9\textwidth] {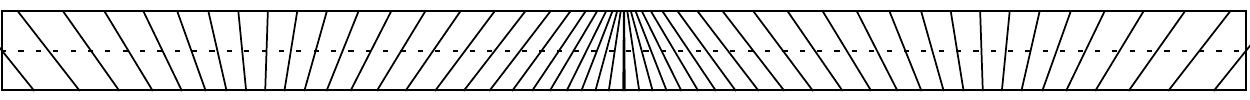}}\\
\subfigure[$w=0.4$, $\epsilon=0.167$]{\includegraphics[width=.9\textwidth] {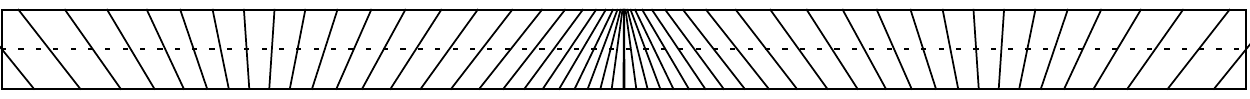}}\\
\subfigure[$w=0.4$, $\epsilon=0.111$]{\includegraphics[width=.9\textwidth]{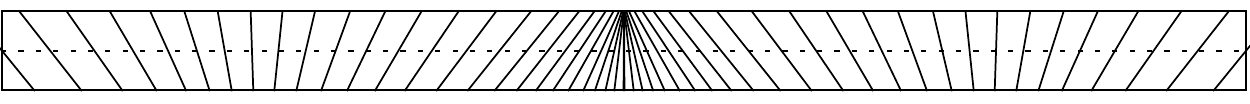}}
\caption{\footnotesize Plots of the generator of curvature, $\uvec{b}$, on the M\"{o}bius strip reference configuration for $w=0.4$ and different values of $\epsilon$, the regularizing term. The left end corresponds to $s=0$ and the right end corresponds to $s=2\pi$.}
\label{fig:stripplotso8w}
\end{figure}
\begin{figure}
\centering
\subfigure[$w=0.2$, $\epsilon=0.033$]{\includegraphics[width=.9\textwidth] {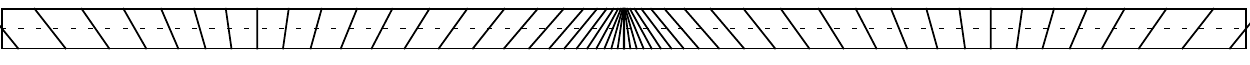}}\
\subfigure[$w=0.4$, $\epsilon=0.11$]{\includegraphics[width=.9\textwidth] {w4cont3.pdf}}\\
\subfigure[$w=1.0$, $\epsilon=1$]{\includegraphics[width=.9\textwidth] {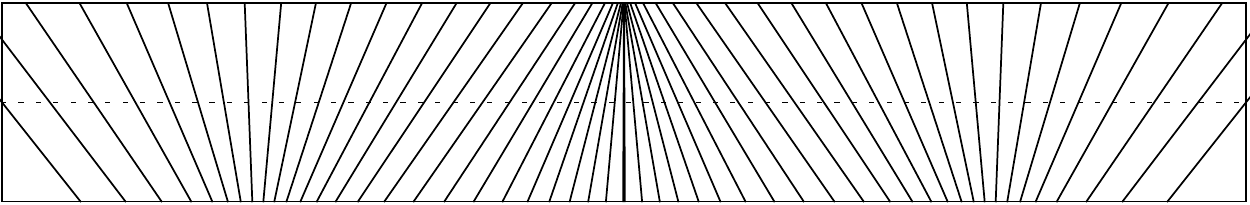}}\\
\subfigure[$w=1.6$, $\epsilon=4.08$]{\includegraphics[width=.9\textwidth] {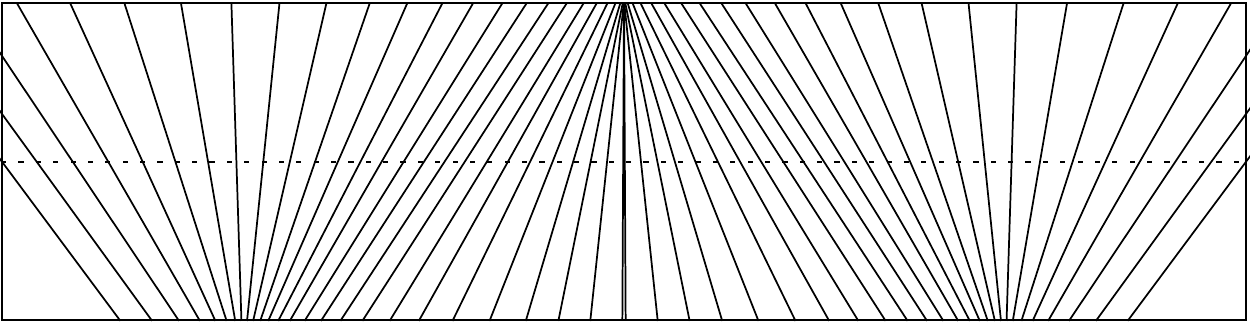}}\\
\subfigure[$w=2.0$, $\epsilon=10$]{\includegraphics[width=.9\textwidth]{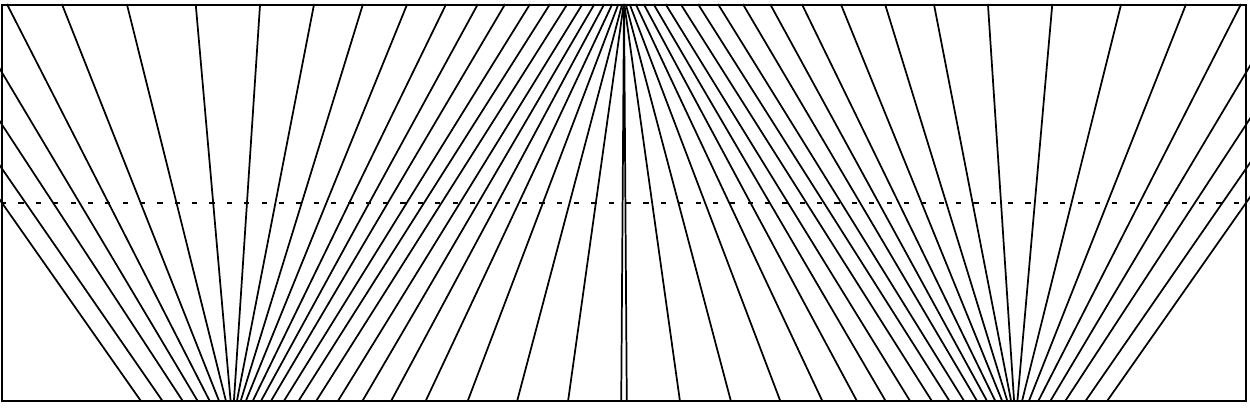}}
\caption{\footnotesize Plots of the generator of curvature, $\uvec{b}$, on the M\"{o}bius strip reference configuration for different values of $w$. The left end corresponds to $s=0$ and the right end corresponds to $s=2\pi$.}
\label{fig:stripplotso8ep}
\end{figure}
\section{Developable Rod Stability}\label{sec:dstab}
As in the case of the Kirchhoff rod in section \ref{sec:kstab}, the local stability of the configurations from section \ref{sec:devrodresults} is determined through linearized dynamic analysis about the equilibrium configurations via an adaptation of the procedure in \cite{Kumar2010}.
\subsection{Derivation of Stability Equations}\label{sec:devsecderiv}
As in section \ref{sec:kstabform}, the spatial weak form of the dynamical equations governing the rod-strip (\ref{eq:sysstartdev})-(\ref{eq:sysend2devend}) is given by
\begin{align}
G&=\int\limits_0^{2\pi}\,\Bigg\{\,\left(p\,A\,\frac{\mathrm{d}^2\uvec{r}}{\mathrm{d}t^2}-\uvec{n}^\prime\right)\cdot\boldsymbol{\rho}+\left(\dot{\mathbf{H}}-\uvec{m}^\prime -\uvec{r}^\prime\times\uvec{n}\right)\cdot\boldsymbol{\psi}+...\nonumber \\
&\hspace{.05\textwidth}...+\boldsymbol{\xi}_l\uvec{r}^\prime\cdot\uvec{d}_l+\boldsymbol{\xi}_3\,\left(\uvec{r}^\prime\cdot\uvec{d}_3-1\right)+ \boldsymbol{\omega}_1 \boldsymbol{\kappa}\cdot \left(\uvec{d}_3-\,\eta\,\uvec{d}_1\right)\,+\boldsymbol{\omega}_2 \boldsymbol{\kappa}\cdot\uvec{d}_2+...  \nonumber \\
&\hspace{.1\textwidth}...+\Bigg[-\epsilon\,\eta^{\prime\prime}-\frac{2\pi^2\,w}{L^2}\,\frac{\mathrm{d}}{\mathrm{d}s}\left[ \kappa_1^2\left(1+\eta^2\right)^2\,\dot{g}\left(w\,\eta^\prime\right) \right]+...\label{eq:bigolweakform}\\
&\hspace{.25\textwidth}...+\frac{8\,\pi^2}{L^2}\,\kappa_1^2\eta\left(1+\eta^2\right)\,g\left(w\,\eta^\prime\right)-\dvec{m}_3\,\kappa_1 \Bigg]\,\chi\,\Bigg\}\,\mathrm{d}s\,,\nonumber
\end{align}
where $p$ is the density of the rod, $A$ is the cross-sectional area, $\mathbf{H}$ is the cross-sectional angular momentum, $\boldsymbol{\rho},\boldsymbol{\psi}$, $\boldsymbol{\xi}$, $\boldsymbol{\omega}_1$, $\boldsymbol{\omega}_2$, and $\chi$ correspond to smooth variations in $\uvec{r},\uvec{R}$, $\uvec{n}$, $\dvec{m}_2$, $\dvec{m}_3$, and $\eta$ respectively. As in the Kirchhoff rod case, $\mathbf{H}$ and the dynamic pieces are not needed to calculate the stability of a conservative system, so they are not explicitly derived. After integrating by parts, $G$ decomposes into dynamic, static, and boundary contributions with the static piece given by
\begin{align}\label{eq:Gdymdev}
G_{static} &=\int\limits_0^{2\pi}\,\Bigg\{\, \uvec{n}\cdot\left(\boldsymbol{\rho}^\prime-\boldsymbol{\psi}\times\uvec{r}^\prime\right)+\uvec{m}\cdot\boldsymbol{\psi}^\prime+\boldsymbol{\xi}_l\uvec{r}^\prime\cdot\uvec{d}_l+\boldsymbol{\xi}_3\,\left(\uvec{r}^\prime\cdot\uvec{d}_3-1\right)\,+\boldsymbol{\omega}_1 \boldsymbol{\kappa}\cdot \left(\uvec{d}_3-\,\eta\,\uvec{d}_1\right)+...\nonumber
\\&\hspace{.1\textwidth}...+\,\boldsymbol{\omega}_2 \boldsymbol{\kappa}\cdot\uvec{d}_2+\,\epsilon\,\eta^\prime\,\chi^\prime  +\frac{2\pi^2\,w}{L^2}\,\left[\kappa_1^2\left(1+\eta^2\right)^2\,\dot{g}\left(w\,\eta^\prime\right)\right]\,\chi^\prime+...\\
&\hspace{.2\textwidth}\nonumber...+\left[ \frac{8\pi^2}{L^2}\,\kappa_1^2\eta\left(1+\eta^2\right)\,g\left(w\,\eta^\prime\right)-\dvec{m}_3\,\kappa_1\right] \,\chi\label{eq:Gstatdev}  \Bigg\}\,\mathrm{d}s\,.
\end{align}
The contact couple, $\dvec{m}_1$, is constitutively determined, and hence, static solutions corresponding to the zeros of (\ref{eq:Gdymdev}) satisfy the dynamical, spatial weak form
\begin{align}
G\left(\uvec{r},\uvec{R},\uvec{n},\dvec{m}_2,\dvec{m}_3,\eta\,\,;\,\,\boldsymbol{\rho},\boldsymbol{\psi},\boldsymbol{\xi},\omega_{2},\omega_3,\chi\right) = 0\,,
\end{align}
at a static equilibrium. The linear perturbations
\begin{align}
\tilde{\uvec{r}} = \uvec{r} + \alpha\,\Delta\uvec{r}\,,\quad
\tilde{\uvec{R}}  = \text{exp}\left(\alpha\,\Delta \Theta\right)\uvec{R}\, , \quad
\tilde{\eta}=\eta+\alpha\,\Delta\,\eta\,,\label{eq:Rotvardev}
\end{align}
are applied where $\Delta\uvec{r}$, $\Delta\Theta$, $\Delta\eta$ are smooth admissible variations, $\text{exp}\left(\cdot\right)$ denotes the matrix exponential, and $\Delta\Theta$ is a smooth admissible skew-symmetric matrix. We define the vector $\Delta\boldsymbol{\theta}=\text{axial}\left(\Delta\Theta\right)$ and note that (\ref{eq:Rotvardev}) induces the variations
\begin{align}
\tilde{\uvec{d}}_i=\uvec{d}_i+\alpha\,\left(\Delta\boldsymbol{\theta}\times\uvec{d}_i\right)\,,\quad
\tilde{\kappa}_i=\kappa_i+\alpha\left[\Delta\boldsymbol{\theta}^\prime\cdot\uvec{d}_1 \right]\,.
\end{align}

In addition, the Lagrange multiplier fields $\left\{\dvec{n}_1,\dvec{n}_2,\dvec{n}_3\right\}$ and $\left\{\dvec{m}_2,\dvec{m}_3\right\}$, are linearly perturbed by the expressions
\begin{align}
\tilde{\uvec{n}} & = \left(\dvec{n}_i + \alpha\,\Delta\dvec{n}_i\right)\,\left(\mathbf{d}_i+\alpha\left[\,\Delta\boldsymbol{\theta}\times\uvec{d}_i\,\right]\right)\,, \quad\quad i=1,2,3   \,\,,\\
\tilde{\dvec{m}}_\ell&= \left(\dvec{m}_\ell + \alpha\,\Delta\dvec{m}_\ell\right)\,\left(\mathbf{d}_\ell+\alpha\left[\,\Delta\boldsymbol{\theta}\times\uvec{d}_\ell\,\right]\right)\,, \quad\quad \ell=2,3\,.  \end{align}
We define the vector of unknowns, $\Delta\boldsymbol{\zeta}_0$ and their time dependent perturbations, $\Delta\boldsymbol{\zeta}$, as
\begin{align}\label{eq:devunknown}
\Delta\boldsymbol{\zeta}_0 = \begin{pmatrix} \left[ \Delta\uvec{r} \right]& \left[ \Delta\boldsymbol{\theta} \right]&\Delta\eta&\left[ \Delta\dvec{n} \right]&\left[\Delta\dvec{m}\right]\end{pmatrix}^T\,,\quad\quad
\Delta\boldsymbol{\zeta}=\Delta\boldsymbol{\zeta}_0\,e^{\,\sigma\,t}\,,
\end{align}
As in section \ref{sec:kstab} Taylor's expansion about an equilibrium point and substitution of (\ref{eq:devunknown}) into the linear part produces the generalized eigenvalue problem
\begin{align}
DG_{static}\,\Delta\boldsymbol{\zeta}_0 = -\sigma^2\,DG_{dynamic}\,\Delta\boldsymbol{\zeta}_0 \label{eq:eigenstabdev}\,.
\end{align}
where $\mu:=-\sigma^2$ is the eigenvalue. As discussed in section \ref{sec:kstab}, a negative eigenvalue, $\mu<0$, indicates instability, and the solution is stable if all eigenvalues are positive.

Since the problem is conservative, the explicit form for $DG_{dynamic}$ is not needed, so it is not derived here. The derivation of the explicit form for $DG_{static}$ is provided in appendix \ref{sec:derivation}. The explicit form of $DG_{static}$ is
\begin{align}
D&G_{static}=\int\limits_0^{2\pi}\mathbf{R}\,\Delta\dvec{n}\cdot\left(\boldsymbol{\rho}^\prime-\boldsymbol{\psi}\times\uvec{r}^\prime\right)+\left(\uvec{n}\times\Delta\uvec{r}^\prime\right)\cdot\boldsymbol{\psi}-\left(\uvec{n}\times\Delta\boldsymbol{\theta}\right)\cdot\left(\boldsymbol{\rho}^\prime-\boldsymbol{\psi}\times\uvec{r}^\prime\right)...+\nonumber\\
&\hspace{.05\textwidth}...+\mathbf{R}\left[\begin{array}{c}\frac{4\pi^2\left(1+\eta^2\right)^2}{L^2}\,g\left(w\eta^\prime\right)\,\mathbf{R}^T\left[ \Delta\boldsymbol{\theta}^\prime \right]\cdot \uvec{e}_1-\eta\,\Delta\dvec{m}_3     \\ \Delta\,\dvec{m}_2\\ \Delta\,\dvec{m}_3\end{array}\right]\cdot\boldsymbol{\psi}^\prime+...\nonumber\\
&\hspace{.05\textwidth}...+\frac{4\pi^2\,w\,\kappa_1 \left(1+\eta^2\right)^2}{L^2}\,g^{\prime}\left(w\eta^\prime\right)\,\Delta\eta^\prime\,\mathbf{R}^T\boldsymbol{\psi}^\prime\cdot\uvec{e}_1+...\label{eq:DGstaticdev}
\\
&\hspace{.05\textwidth}...+\left(\frac{16\pi^2\,\kappa_1\,\eta\,\left(1+\eta^2\right)\,g\left(w\,\eta^\prime\right)}{L^2}-\dvec{m}_3\right)\,\Delta\eta\,\mathbf{R}^T\boldsymbol{\psi}^\prime\cdot\uvec{e}_1
+\Delta\boldsymbol{\theta}\times\mathbf{R}\left[\begin{array}{c}\dvec{m}_1\\ \dvec{m}_2\\ \dvec{m}_3\end{array}\right]\cdot\boldsymbol{\psi}^\prime+...\nonumber
\\
& ...+\mathbf{R}\,\boldsymbol{\xi}\cdot\left(\Delta\uvec{r}^\prime+\left[\uvec{r}^\prime\times\Delta\boldsymbol{\theta}\right]\right)
+\mathbf{R}\left[\begin{array}{c}-\boldsymbol{\omega}_1\,\eta\\\boldsymbol{\omega}_2\\\boldsymbol{\omega}_1 \end{array}\right]\cdot\Delta\boldsymbol{\theta}^\prime+\,\kappa_1\,\mathbf{R}\left[\begin{array}{c}\boldsymbol{\omega}_2\eta\\\boldsymbol{\omega}_1\left(1-\eta^2\right)\\-\boldsymbol{\omega}_2 \end{array}\right]\cdot\Delta\boldsymbol{\theta}+...\nonumber
\\
&...+\chi\left\{\frac{16\pi^2\,\kappa_1\,\eta\,\left(1+\eta^2\right)\,g\left(w\eta^\prime\right)}{L^2}-\dvec{m}_3\right\}\,\mathbf{R}^T\left[ \Delta\boldsymbol{\theta}^\prime \right]\cdot \uvec{e}_1-\chi\kappa_1\,\Delta\dvec{m}_3+\epsilon\,\chi^\prime\,\Delta\eta^\prime+...\nonumber\\
&\hspace{.05\textwidth}...+\frac{8\pi^2\,w\,\chi}{L^2}\,\kappa_1^2\,\eta\,\left(1+\eta^2\right)\dot{g}\left(w\eta^\prime\right)\,\Delta\eta^\prime+\frac{8\,\pi^2\,\chi}{L^2}\kappa_1^2\,\left(1+3\eta^2\right)\,g\left(w\eta^\prime\right)\,\Delta\eta+...\nonumber \\
&\hspace{.15\textwidth}-\boldsymbol{\omega}_1\,\kappa_1\,\Delta\eta +\frac{4\,\pi^2\chi^\prime\,w\,\kappa_1}{L^2}\Big\{ \frac{w}{2}\,\kappa_1\,\left(1+\eta^2\right)^2\ddot{g}\left(w\eta^\prime\right) \Delta\eta^\prime+...\nonumber\\
&\hspace{.02\textwidth}...+2\kappa_1\,\eta\left(1+\eta^2\right)\dot{g}\left(w\eta^\prime\right)\,\Delta\eta+ \left(1+\eta^2\right)^2\,\dot{g}\left(w\eta^\prime\right) \mathbf{R}^T\left[ \Delta\boldsymbol{\theta}^\prime \right]\cdot \uvec{e}_1\Big\} \, \mathrm{d}s\nonumber\,.
\end{align}
In addition, the boundary terms are given by
\begin{align}
G_{bdry}&=\label{eq:Gbdrydev}
\Bigg[ \uvec{n}\cdot\boldsymbol{\rho}+\uvec{m}\cdot\boldsymbol{\psi}+\epsilon\,\eta^\prime\,\chi+\frac{2\,\pi^2\,w}{L^2}\,\left[\kappa_1^2\left(1+\eta^2\right)^2\,\dot{g}\left(w\,\eta^\prime\right)\right]\,\,\chi\Bigg]_{0}^{2\pi} \,.
\end{align}
\subsection{Numerical Implementation}
As with the Kirchhoff rod in section \ref{sec:kstab}, the eigenvalues in (\ref{eq:eigenstabdev}) are calculated via the finite element method as implemented in \cite{Kumar2010,Simo1986}. The smooth test functions $\left(\boldsymbol{\rho},\boldsymbol{\psi},\boldsymbol{\xi},\boldsymbol{\omega},\chi\right)$ and spatial perturbations $(\Delta\uvec{r}, \Delta\boldsymbol{\theta},\Delta\uvec{n},\Delta\dvec{m},\Delta\eta)$ are approximated with piecewise linear functions. Nodal values for the variables $(\uvec{r},\uvec{R},\uvec{n},\uvec{m},\eta)$ are obtained from the continuation results on $[0,\pi]$ and symmetry transformations detailed in section \ref{sec:bcconstruct}. For $N$ elements, this discretization transforms (\ref{eq:eigenstabdev}) into the matrix eigenvalue problem
\begin{align}
\left[\begin{array}{cc}\mathbf{K}_{m\times m}&\mathbf{C}_{m\times p}\\\\\mathbf{C}^T_{p\times m}&\mathbf{0}_{p\times p} \end{array}\right]\left[\begin{array}{c}\left[\begin{array}{c}\Delta\uvec{r}\\\Delta\boldsymbol{\theta}\\\Delta\eta\end{array}\right]\\\left[\begin{array}{c}\Delta\dvec{n}\\\Delta\dvec{m}\end{array}\right]\end{array}\right]=\mu\left[\begin{array}{cc}\mathbf{M}_{m\times m}&\mathbf{0}\\\mathbf{0}&\mathbf{0}\end{array}\right]\left[\begin{array}{c}\left[\begin{array}{c}\Delta\uvec{r}\\\Delta\boldsymbol{\theta}\\\Delta\eta\end{array}\right]\\\left[\begin{array}{c}\Delta\dvec{n}\\\Delta\dvec{m}\end{array}\right]\end{array}\right]\end{align}
where $\mathbf{K}$ is the global stiffness matrix, $\mathbf{C}$ is the global constraint matrix, $\mathbf{M}$ is the global mass matrix, $m=7N$, and $p=5N$. Note that $p$ represents the total number of point-wise constraints acting on the discretized rod.

This problem has the same structure as the stability calculation for the Kirchhoff rod in section \ref{sec:kstab}, and it again results in the eigenvalue problem (\ref{eq:methend}). Employing the same solution procedure from \cite{Kumar2010}, we find the eigenvalues of $\tilde{\mathbf{K}}$ where positive eigenvalues indicate stability and negative eigenvalues indicate unstable perturbations. 
\subsection{Full Strip Construction and Boundary Conditions}\label{sec:bcconstruct}
For the stability calculation, the closed developable strip is generated from AUTO's half solution differently than the Kirchhoff rod case. If a solution on $\left[0,\pi\right]$ was extended via (\ref{eq:rflip})-(\ref{eq:mflip}) to a solution on $\left[0,2\pi\right]$, a node would be placed exactly at the $s=\pi$ singular point. Instead, we extend the solution on $\left[0,\pi\right]$ to a solution on $\left[-\pi,\pi\right]$ via a flip rotation by $180$ degrees about the $\uvec{e}_2$-axis to avoid placing a node exactly at this singular point. As in the Kirchhoff rod case, the following procedure is rigorously detailed in \cite{Domokos2001}.

 Denote the calculated solution to (\ref{eq:sysstartdev})-(\ref{eq:sysenddev}) for $s\in[0,\pi]$ by a superscript $``c"$, e.g. $\uvec{r}^c\left(s\right)$ for the calculated rod centerline position. The position of the centerline for $s\in[0,2\pi]$ is given by
\begin{align}
\uvec{r}\left(s\right)=\begin{cases}
\mathbf{E}\,\uvec{r}^c\left(-s\right)\quad& s\in[-\pi,0]\\
\uvec{r}^c\left(s\right)\quad& s\in[0,\pi]
\end{cases}\,,
\end{align}
where $\mathbf{E}$ is defined in (\ref{eq:operator}).
The extension of the rod's orientation on $\left[-\pi,0\right]$ is defined in terms of the director fields $\uvec{d}_i\left(s\right)$:
\begin{align}
\uvec{d}_1\left(s\right)&=\begin{cases}
-\mathbf{E}\,\uvec{d}_1^c\left(-s\right)\quad& s\in[-\pi,0]\\
\uvec{d}_1^c\left(s\right)\quad& s\in[0,\pi]\\
\end{cases}\,,\\
\uvec{d}_2\left(s\right)&=\begin{cases}
\mathbf{E}\,\uvec{d}_2^c\left(-s\right)\quad& s\in[-\pi,0]\\
\uvec{d}_2^c\left(s\right)\quad& s\in[0,\pi]\\
\end{cases}\,,\\
\uvec{d}_3\left(s\right)&=\begin{cases}
-\mathbf{E}\,\uvec{d}_3^c\left(-s\right)\quad& s\in[-\pi,0]\\
\uvec{d}_3^c\left(s\right)\quad& s\in[0,\pi]\\
\end{cases}\,.
\end{align}
Observe that $\uvec{d}_i\left(\cdot\right)$, $i=1,2,3$ is continuous on $\left[-
\pi,\pi\right]$.
Following the results in \cite{Domokos2001}, the required extensions are given by:
\begin{align}
\uvec{n}\left(s\right)&=\begin{cases}
-\mathbf{E}\,\uvec{n}^c\left(-s\right)\quad& s\in[-\pi,0]\\
\uvec{n}^c\left(s\right)\quad& s\in[0,\pi]
\end{cases}\,,\\
\uvec{m}\left(s\right)&=\begin{cases}
-\mathbf{E}\,\uvec{m}^c\left(-s\right)\quad& s\in[-\pi,0]\\
\uvec{m}^c\left(s\right)\quad& s\in[0,\pi]
\end{cases}\,.
\end{align}
In addition, the transformation for the scalar parameter $\eta$ and its derivative are
\begin{align}
\eta\left(s\right)&=\begin{cases}
\eta^c\left(-s\right)\quad& s\in[-\pi,0]\\
\eta^c\left(s\right)\quad& s\in[0,\pi]
\end{cases}\,,\\
\eta^\prime\left(s\right)&=\begin{cases}
-\left(\eta'\right)^c\left(-s\right)\quad& s\in[-\pi,0]\\
\left(\eta'\right)^c\left(s\right)\quad& s\in[0,\pi]
\end{cases}\,.
\end{align}
In our closed strip, both the position and the orientation of the rod at $s=-\pi$ and $s=\pi$ are clamped. Assuming the rod is divided into $N$ elements with $N+1$ nodes, the boundary conditions are
\begin{align}\label{eq:stabbc1dev}
\Delta\uvec{r}^{\left(0\right)} = 0\,,\hspace{.075\textwidth}
\Delta\uvec{\boldsymbol{\theta}}^{\left(0\right)} &= 0\,,\hspace{.075\textwidth}
\Delta\eta^{\left(0\right)}=0\, ,\\
\Delta\uvec{r}^{\left(N+1\right)} = 0\,,\hspace{.05\textwidth}
\Delta\uvec{\boldsymbol{\theta}}^{\left(N+1\right)} &= 0\,,\hspace{.05\textwidth}
\Delta\eta^{\left(N+1\right)}=0\,.\label{eq:stabbc2dev}
\end{align}
These $14$ boundary conditions ensure that the $s=-\pi$ and $s=\pi$ ends of the rod will remain smoothly connected under any perturbation and satisfy the variation of (\ref{eq:Gbdrydev}). As before in Section \ref{sec:bcz}, (\ref{eq:stabbc1dev})-(\ref{eq:stabbc2dev}) eliminate the six neutrally stable rigid-body modes corresponding to uniform translation and rotation of the closed rod and the additional degeneracy associated with axial motion of the strip acting through its own fixed configuration.
\subsection{Results and Summary}\label{sec:initialresults}
The numerical equilibrium solutions from AUTO calculated in Section \ref{sec:devrodresults} are extended to the full M\"{o}bius strip on $[0,2\pi]$ and used for the finite element calculation. This results in a mesh resolution of 1000 elements for the full strip. As shown in Table \ref{table:devstabeigs}, all the eigenvalues of $DG_{static}$ are positive. Observe that the smallest (positive) eigenvalue consistently increases as the regularizing parameter is made as small as possible.   We conclude that the developable-rod configurations are stable with respect to all sufficiently small perturbations \textendash symmetric and non-symmetric.
\begin{table}
\caption{\footnotesize The four smallest eigenvalues of $DG_{static}$ for a developable rod with $N=1000$ elements.}
\centering\label{table:devstabeigs}\footnotesize
\vspace{.1in}
\begin{tabular}{cc|c|c|c|c}
  &&Smallest&2nd smallest &3rd smallest&4th smallest\\
\hline\hline
$w = 0.2$& $\epsilon = 0.25$&0.0073&0.0335&0.0409&0.0589\\
$w = 0.2$& $\epsilon = 0.1$&0.0079&0.0359&0.0376&0.0508\\
$w = 0.2$& $\epsilon = 0.033$&0.0087&0.0329&0.0339&0.0483\\\hline
$w = 0.4$& $\epsilon = 0.25$&0.0077&0.035&0.0415&0.0608\\
$w = 0.4$& $\epsilon = 0.167$&0.0082&0.0372&0.0403&0.0577\\
$w = 0.4$& $\epsilon = 0.111$&0.0089&0.0391&0.0396&0.0574\\\hline
$w = 1$& $\epsilon = 1.333$&0.0099&0.0365&0.0625&0.109\\
$w = 1$& $\epsilon = 1.125$&0.0102&0.0386&0.0603&0.1049\\
$w = 1$& $\epsilon = 1$&0.0104&0.0406&0.0587&0.1019\\\hline
$w = 1.6$& $\epsilon = 5$&0.0164&0.0412&0.0928&0.1681\\
$w = 1.6$& $\epsilon = 4.5$&0.0167&0.0436&0.0907&0.1643\\
$w = 1.6$& $\epsilon = 4.082$&0.0173&0.0478&0.0875&0.1573\\\hline
$w = 2$& $\epsilon = 12$&0.0245&0.0484&0.1254&0.2206\\
$w = 2$& $\epsilon = 11$&0.0245&0.0508&0.1216&0.2147\\
$w = 2$& $\epsilon = 10$&0.0255&0.0575&0.1155&0.2008\\
\end{tabular}
\end{table}
\section{Concluding Remarks}
     Here we present the first evidence for the local stability of flip-symmetric configurations of complete elastic M\"{o}bius bands. We employ two distinct models   \textendash \,the Kirchhoff rod model and the developable-surface model of Wunderlich. For the latter we present a novel strategy for the computation of \emph{complete unconstrained} M\"{o}bius bands.  To the best of our knowledge, it is the only known method for accomplishing such.   Our introduction of a small elliptic regularization, cf. (\ref{eq:regenergy}), similar to what is often done in phase-transition problems, avoids the inevitable singularity associated with the rod-like formulation \cite{Wunderlich1962}, \cite{Dias2014} for M\"{o}bius bands, cf. \cite{Starostin2014}, \cite{Freddi}. 
     
     Unlike the numerical approach discussed in \cite{Starostin2014}, ours presented here delivers complete-loop configurations in the absence of extraneous external fields.  Moreover the solutions presented in Figures \ref{fig:append1}-\ref{fig:append2} and the eigenvalue results in Table \ref{table:devstabeigs} all demonstrate the robustness of our results in the small regularizing parameter $\epsilon$. Finally we mention that our implementation of the Wunderlich model, for both computing equilibria and assessing their stability, is applicable to many other thin-strip problems. 
\section*{Acknowledgements}
This work was supported in part by the National Science Foundation through grant DMS-1312377, which is gratefully acknowledged.  We also thank Roberto Paroni for useful discussions related to this work.
\appendix
\section{Derivation of $DG_{static}$ for a Developable rod}\label{sec:derivation}
This appendix contains the work for the derivation of $DG_{static}$ in (\ref{eq:DGstaticdev}) for a strip of length $L$. The quantity $G_{static}$ is given by
\begin{align}
G_{static} &=\int\limits_0^{L}\,\Bigg\{\, \uvec{n}\cdot\left(\boldsymbol{\rho}^\prime-\boldsymbol{\psi}\times\uvec{r}^\prime\right)+\uvec{m}\cdot\boldsymbol{\psi}^\prime+\boldsymbol{\xi}_l\uvec{r}^\prime\cdot\uvec{d}_l+\boldsymbol{\xi}_3\,\left(\uvec{r}^\prime\cdot\uvec{d}_3-1\right)\,+...\nonumber
\\&\hspace{.1\textwidth}\nonumber...+\boldsymbol{\omega}_1 \boldsymbol{\kappa}\cdot \left(\uvec{d}_3-\,\eta\,\uvec{d}_1\right)+\,\boldsymbol{\omega}_2 \boldsymbol{\kappa}\cdot\uvec{d}_2 +...\\
& \hspace{.15\textwidth}... 
\,+\epsilon\,\eta^\prime\,\chi^\prime +\frac{w}{2\,L^2}\,\left[\kappa_1^2\left(1+\eta^2\right)^2\,\dot{g}\left(w\,\eta^\prime\right)\right]\,\chi^\prime+...\label{eq:appenDGstatic}\\
&\hspace{.2\textwidth}\nonumber...+\left[ \frac{2}{L^2}\,\kappa_1^2\eta\left(1+\eta^2\right)\,g\left(w\,\eta^\prime\right)-\dvec{m}_3\,\kappa_1\right] \,\chi  \Bigg\}\,\mathrm{d}s\,.
\end{align}
The quantity $DG_{static}$ is the linearization of $G_{static}$ about an equilibrium point, viz.
\begin{align}
DG_{static}=\fvar{G_{static}\left(\uvec{r}+\alpha\Delta\uvec{r},\text{exp}\left(\alpha\Delta\Theta\right)\uvec{R},\uvec{n}+\alpha\Delta\uvec{n},\uvec{m}+\alpha\Delta\uvec{m},\eta+\alpha\Delta\eta\right)}
\end{align}
Using the notation $\delta\left\{\cdot\right\}=\fvar{\cdot}$,  we derive $DG_{static}$ term by term starting with the first component in (\ref{eq:appenDGstatic}):
\begin{align}
\delta\left\{\uvec{n}\cdot\left(\boldsymbol{\rho}^\prime-\boldsymbol{\psi}\times\uvec{r}^\prime\right)\right\}&=
\fvar{\left(\dvec{n}_i+\alpha\Delta\dvec{n}_i\right)\,\left(\mathbf{d}_i+\alpha\left[\,\Delta\boldsymbol{\theta}\times\uvec{d}_i\,\right]\right)\,\cdot\left(\boldsymbol{\rho}^\prime-\boldsymbol{\psi}\times\left(\uvec{r}^\prime+\alpha\Delta\uvec{r}\right)\right)}\\
&=\Delta\dvec{n}_i\,\uvec{d}_i\cdot\left(\boldsymbol{\rho}^\prime-\boldsymbol{\psi}\times\uvec{r}^\prime\right)-\dvec{n}_i\uvec{d}_i\cdot\left(\boldsymbol{\psi}\times\Delta\uvec{r}^\prime\right)+\dvec{n}_i\left(\Delta\boldsymbol{\theta}\times\uvec{d}_i\right)\cdot\left(\boldsymbol{\rho}^\prime-\boldsymbol{\psi}\times\uvec{r}^\prime\right)\\
&=\mathbf{R}\,\Delta\dvec{n}\cdot\left(\boldsymbol{\rho}^\prime-\boldsymbol{\psi}\times\uvec{r}^\prime\right)+\left(\uvec{n}\times\Delta\uvec{r}^\prime\right)\cdot\boldsymbol{\psi}-\left(\uvec{n}\times\Delta\boldsymbol{\theta}\right)\cdot\left(\boldsymbol{\rho}^\prime-\boldsymbol{\psi}\times\uvec{r}^\prime\right)\label{eq:sixthtolastpart}
\end{align}

The second term in (\ref{eq:appenDGstatic}) is significantly more complicated because it involves quantities subject to variations in the rod orientation. The variation of the second term starts as
\begin{align}
\delta\left\{\uvec{m}\cdot\boldsymbol{\psi}^\prime\right\}&=\fvar{\tilde{\dvec{m}}_1\tilde{\uvec{d}_1}+\tilde{\dvec{m}}_2\,\tilde{\uvec{d}}_2+\tilde{\dvec{m}}_3\,\tilde{\uvec{d}}_3}\cdot\boldsymbol{\psi}^\prime
\end{align}\,.
Since $\dvec{m}_2$ and $\dvec{m}_3$ are both Lagrange multipliers, their variations are straightforward yielding
\begin{align}
\fvar{\tilde{\dvec{m}}_2\,\tilde{\uvec{d}}_2+\tilde{\dvec{m}}_3\,\tilde{\uvec{d}}_3}&=\fvar{\left(\dvec{m}_2+\alpha\Delta\dvec{m}_2\right)\,\left(\mathbf{d}_2+\alpha\left[\,\Delta\boldsymbol{\theta}\times\uvec{d}_2\,\right]\right)}  +...\nonumber \\
&...+\fvar{\left(\dvec{m}_3+\alpha\Delta\dvec{m}_3\right)\,\left(\mathbf{d}_3+\alpha\left[\,\Delta\boldsymbol{\theta}\times\uvec{d}_3\,\right]\right)  }\\
&=\mathbf{R}\left[\begin{array}{c}0\\ \Delta\,\dvec{m}_2\\ \Delta\,\dvec{m}_3\end{array}\right]+\Delta\boldsymbol{\theta}\times\mathbf{R}\left[\begin{array}{c}0\\ \dvec{m}_2\\ \dvec{m}_3\end{array}\right]\label{eq:mvarp1}
\end{align}
For the variation in $\dvec{m}_1$, the moment is constitutively determined, so it is significantly more complicated. From (\ref{eq:sysend2}) $\tilde{\dvec{m}}_1$ is
\begin{align}
&\tilde{\dvec{m}}_1\tilde{\uvec{d}}_1 = \frac{1}{L^2}\,\left(1+\tilde{\eta}^2\right)^2\,g\left(w\tilde{\eta}^\prime\right)\,\tilde{\kappa}_1\,\tilde{\uvec{d}}_1\,-\,\tilde{\eta}\,\tilde{\dvec{m}}_3\,\tilde{\uvec{d}_1}\\
&= \frac{\,\left(\kappa_1+\alpha\left[\Delta\boldsymbol{\theta}^\prime\cdot\mathbf{d}_1\right]\right)}{L^2}\,\left(\mathbf{d}_1+\alpha\,\left[\Delta\boldsymbol{\theta}\times\uvec{d}_1\right]\right)\,\left(1+\left[\eta+\alpha\Delta\eta\right]^2\right)^2\,g\left(w\left[\eta^\prime+\alpha\Delta\eta\prime\right]\right)\,+...\nonumber\\
&\hspace{.3\textwidth}...-\,\left[\eta+\alpha\Delta\eta\right]\,\left[\dvec{m}_3+\alpha\Delta\dvec{m}_3\right]\,\left[\mathbf{d}_1+\alpha\left(\Delta\boldsymbol{\theta}\times\uvec{d}_1\right)\right]
\end{align}
Note the vector quantities of the variations. Taking the derivative with respect to $\alpha$ yields
\begin{align}
\fvar{\tilde{m}_1\tilde{\uvec{d}}_1}&=\left\{ \frac{\left(1+\eta^2\right)^2}{L^2}\,g\left(w\eta^\prime\right)\,\left[\Delta\boldsymbol{\theta}^\prime\cdot\uvec{d}_1\right]+\frac{w\,\kappa_1 \left(1+\eta^2\right)^2}{L^2}\,g^{\prime}\left(w\eta^\prime\right)\,\Delta\eta^\prime +...\right.\nonumber\\
&\hspace{.1\textwidth}\left. ...+\left[\frac{4\,\kappa_1\,\eta\,\left(1+\eta^2\right)\,}{L^2}g\left(w\eta^\prime\right)-\dvec{m}_3\right]\,\Delta\eta-\eta\,\Delta\dvec{m}_3\,\right\}\uvec{d}_1+
...\nonumber\\
& \hspace{.2\textwidth}...+
\left\{\frac{\kappa_1\left(1+\eta^2\right)^2}{L^2}\,g\left(w\eta^\prime\right)-\dvec{m}_3\,\eta\right\}\left[\Delta\boldsymbol{\theta}\times\uvec{d}_1\right]\,. \label{eq:m1stabeq}
\end{align}
Since $\left[ \Delta\boldsymbol{\theta}^\prime \right]$ is written with respect to the fixed basis we use
\begin{align}
\left[\Delta\boldsymbol{\theta}^\prime\cdot\uvec{d}_1\right]=\mathbf{R}^T\left[ \Delta\boldsymbol{\theta}^\prime \right]\cdot \uvec{e}_1
\end{align}
Adding this piece to (\ref{eq:m1stabeq}) and combining with (\ref{eq:mvarp1}), the total variation is
\begin{align}
&\delta\left\{\uvec{m}\cdot\boldsymbol{\psi}^\prime\right\}=\nonumber\\
&\quad...+\mathbf{R}\left[\begin{array}{c}\frac{\left(1+\eta^2\right)^2}{L^2}\,g\left(w\eta^\prime\right)\,\mathbf{R}^T\left[ \Delta\boldsymbol{\theta}^\prime \right]\cdot \uvec{e}_1 -\eta\,\Delta\dvec{m}_3 \\ \Delta\,\dvec{m}_2\\ \Delta\,\dvec{m}_3\end{array}\right]\cdot\boldsymbol{\psi}^\prime+...\nonumber \\
&\quad...+\mathbf{R}\left[\begin{array}{c}\frac{w\,\kappa_1 \left(1+\eta^2\right)^2}{L^2}\,g^{\prime}\left(w\eta^\prime\right)\,\Delta\eta^\prime  +\left[\frac{4\,\kappa_1\,\eta\,\left(1+\eta^2\right)\,}{L^2}g\left(w\eta^\prime\right)-\dvec{m}_3\right]\,\Delta\eta     \\ 0\\ 0\end{array}\right]\cdot\boldsymbol{\psi}^\prime+...\nonumber \\
&\hspace{.3\textwidth}...+\Delta\boldsymbol{\theta}\times\mathbf{R}\left[\begin{array}{c}\frac{\kappa_1\left(1+\eta^2\right)^2}{L^2}\,g\left(w\eta^\prime\right)-\dvec{m}_3\,\eta\\ \dvec{m}_2\\ \dvec{m}_3\end{array}\right]\cdot\boldsymbol{\psi}^\prime\,.
\end{align}
Substituting (\ref{eq:sysend2}) into the cross product term and simplifying yields
\begin{align}
&\delta\left\{\uvec{m}\cdot\boldsymbol{\psi}\right\}=\nonumber\\
&\quad...+\mathbf{R}\left[\begin{array}{c}\frac{\left(1+\eta^2\right)^2}{L^2}\,g\left(w\eta^\prime\right)\,\mathbf{R}^T\left[ \Delta\boldsymbol{\theta}^\prime \right]\cdot \uvec{e}_1 -\eta\,\Delta\dvec{m}_3 \\ \Delta\,\dvec{m}_2\\ \Delta\,\dvec{m}_3\end{array}\right]\cdot\boldsymbol{\psi}^\prime+...\nonumber \\
&\quad...+\mathbf{R}\left[\begin{array}{c}\frac{w\,\kappa_1 \left(1+\eta^2\right)^2}{L^2}\,g^{\prime}\left(w\eta^\prime\right)\,\Delta\eta^\prime  +\left[\frac{4\,\kappa_1\,\eta\,\left(1+\eta^2\right)\,}{L^2}g\left(w\eta^\prime\right)-\dvec{m}_3\right]\,\Delta\eta     \\ 0\\ 0\end{array}\right]\cdot\boldsymbol{\psi}^\prime+...\nonumber \\
&\hspace{.45\textwidth}...+\Delta\boldsymbol{\theta}\times\mathbf{R}\left[\begin{array}{c}\dvec{m}_1\\ \dvec{m}_2\\ \dvec{m}_3\end{array}\right]\cdot\boldsymbol{\psi}^\prime\,.\label{eq:fifthtolastpart}
\end{align}

The third term of (\ref{eq:appenDGstatic}) is relatively straightforward and yields
\begin{align}
\delta\left\{\boldsymbol{\xi}_l\uvec{r}^\prime\cdot\uvec{d}_l+\boldsymbol{\xi}_3\,\left(\uvec{r}^\prime\cdot\uvec{d}_3-1\right)\right\} &=\fvar{\boldsymbol{\xi}_1\left(\uvec{r}^\prime+\alpha\Delta\uvec{r}^\prime\right)\cdot\left(\mathbf{d}_1+\alpha\left[\,\Delta\boldsymbol{\theta}\times\uvec{d}_1\,\right]\right)}\nonumber+...\\
...+&\fvar{\boldsymbol{\xi}_2\left(\uvec{r}^\prime+\alpha\Delta\uvec{r}^\prime\right)\cdot\left(\mathbf{d}_2+\alpha\left[\,\Delta\boldsymbol{\theta}\times\uvec{d}_2\,\right]\right)}\nonumber+...\\
...+&\fvar{\boldsymbol{\xi}_3\left\{\left(\uvec{r}^\prime+\alpha\Delta\uvec{r}^\prime\right)\cdot\left(\mathbf{d}_3+\alpha\left[\,\Delta\boldsymbol{\theta}\times\uvec{d}_3\,\right]\right)-1\right\}\,}\,,\\
&=\boldsymbol{\xi}_i\Delta\,\mathbf{r}^\prime\cdot\uvec{d}_i+\boldsymbol{\xi}_i\,\mathbf{r}^\prime\cdot\left[\,\Delta\boldsymbol{\theta}\times\uvec{d}_i\,\right]\,,\\
&=\mathbf{R}\,\boldsymbol{\xi}\cdot\left(\Delta\uvec{r}^\prime+\left[\uvec{r}^\prime\times\Delta\boldsymbol{\theta}\right]\right)\,.\label{eq:fourthtolast}
\end{align}

The second line, or fourth term, of (\ref{eq:appenDGstatic}) is
\begin{align}
&\delta\left[ \boldsymbol{\omega}_1 \boldsymbol{\kappa}\cdot \left(\uvec{d}_3-\,\eta\,\uvec{d}_1\right)+\boldsymbol{\omega}_2 \boldsymbol{\kappa}\cdot\uvec{d}_2\right]=\nonumber\\
&\hspace{.1\textwidth}-\boldsymbol{\omega}_1\left(\boldsymbol{\kappa}+\alpha\,\Delta\boldsymbol{\theta}^\prime\right)\cdot\left(\eta+\alpha\Delta\eta\right)\left(\uvec{d}_1+\alpha\,\left[\Delta\boldsymbol{\theta}\times\uvec{d}_1\right]\right)+...\nonumber
\\
&\hspace{.2\textwidth}...+\boldsymbol{\omega}_1\left(\boldsymbol{\kappa}+\alpha\,\Delta\boldsymbol{\theta}^\prime\right)\cdot\left( \uvec{d}_3+\alpha\,\left[\Delta\boldsymbol{\theta}\times\uvec{d}_3\right]\right)+...\nonumber \\
&\hspace{.3\textwidth}...+\boldsymbol{\omega}_2\left(\boldsymbol{\kappa}+\alpha\,\Delta\boldsymbol{\theta}^\prime\right)\cdot\left( \uvec{d}_2+\alpha\,\left[\Delta\boldsymbol{\theta}\times\uvec{d}_2\right]\right)\,.
\end{align}
\begin{align}
&\nonumber\delta\left[ \boldsymbol{\omega}_1 \boldsymbol{\kappa}\cdot \left(\uvec{d}_3-\,\eta\,\uvec{d}_1\right)+\boldsymbol{\omega}_2 \boldsymbol{\kappa}\cdot\uvec{d}_2\right]=\\&\hspace{.1\textwidth}-\boldsymbol{\omega}_1\boldsymbol{\kappa}\cdot\Delta\eta\,\uvec{d}_1+\Delta\boldsymbol{\theta}^\prime\cdot\left(-\boldsymbol{\omega}_1\eta\uvec{d}_1+\boldsymbol{\omega}_2\uvec{d}_2+\boldsymbol{\omega}_1\uvec{d}_3\right)+...
\nonumber\\
&\hspace{.2\textwidth}...+\boldsymbol{\omega}_1\boldsymbol{\kappa}\cdot\left(\left[\Delta\boldsymbol{\theta}\times\uvec{d}_3\right]-\eta\,\left[\Delta\boldsymbol{\theta}\times\uvec{d}_1\right]\right)+...\nonumber \\&\hspace{.3\textwidth}...+\boldsymbol{\omega}_2\boldsymbol{\kappa}\cdot\left[\Delta\boldsymbol{\theta}\times\uvec{d}_2\right]\\\nonumber
\\
&\delta\left[ \boldsymbol{\omega}_1 \boldsymbol{\kappa}\cdot \left(\uvec{d}_3-\,\eta\,\uvec{d}_1\right)+\boldsymbol{\omega}_2 \boldsymbol{\kappa}\cdot\uvec{d}_2\right]=\nonumber\\&
\hspace{.05\textwidth}\Delta\boldsymbol{\theta}\,\cdot\,\left(-\boldsymbol{\omega}_2\kappa_1\,\uvec{d}_3+\boldsymbol{\omega}_1\kappa_1\,\uvec{d}_2\right)+\Delta\boldsymbol{\theta}\,\cdot\,\left(-\boldsymbol{\omega}_1\kappa_1\eta^2\,\uvec{d}_2+\boldsymbol{\omega}_2\,\eta\,\kappa_1\,\uvec{d}_1\right)
\end{align}
Putting into matrix form
\begin{align}
&\delta\left[ \boldsymbol{\omega}_1 \boldsymbol{\kappa}\cdot \left(\uvec{d}_3-\,\eta\,\uvec{d}_1\right)+\boldsymbol{\omega}_2 \boldsymbol{\kappa}\cdot\uvec{d}_2\right]=\nonumber\\&\hspace{.1\textwidth}-\boldsymbol{\omega}_1\,\kappa_1\,\Delta\eta+\mathbf{R}\left[\begin{array}{c}-\boldsymbol{\omega}_1\,\eta\\\boldsymbol{\omega}_2\\\boldsymbol{\omega}_1 \end{array}\right]\cdot\Delta\boldsymbol{\theta}^\prime+\,\kappa_1\,\mathbf{R}\left[\begin{array}{c}\boldsymbol{\omega}_2\eta\\\boldsymbol{\omega}_1\left(1-\eta^2\right)\\-\boldsymbol{\omega}_2 \end{array}\right]\cdot\Delta\boldsymbol{\theta}\label{eq:thirdtolastpart}
\end{align}

The variation of the Euler-Lagrange equation for the terms $\chi^\prime$ is
\begin{align}
&\delta\left\{\epsilon\,\eta^\prime\,\chi^\prime +\frac{w}{2\,L^2}\,\left[\kappa_1^2\left(1+\eta^2\right)^2\,\dot{g}\left(w\,\eta^\prime\right)\right]\,\chi^\prime\right\}=\epsilon\,\chi^\prime\Delta\eta^\prime+...\\
&...+\frac{w\,\chi^\prime}{2L^2}\fvar{\left(\kappa_1+\alpha\left[\Delta\boldsymbol{\theta}^\prime\cdot\uvec{d}_1\right]\right)^2\left(1+\alpha\left[\eta+\Delta\eta\right]^2\right)^2\dot{g}\left(w\left[\eta^\prime+\alpha\Delta\eta^\prime\right]\right)}\nonumber\\\nonumber\\
&=\epsilon\,\chi^\prime\,\Delta\eta^\prime +\frac{\chi^\prime\,w\,\kappa_1}{L^2}\left\{  \left(1+\eta^2\right)^2\,\dot{g}\left(w\eta^\prime\right) \mathbf{R}^T\left[ \Delta\boldsymbol{\theta}^\prime \right]\cdot \uvec{e}_1\right\}\nonumber+...\\\label{eq:secondtolastpart}
&...+\frac{\chi^\prime\,w\,\kappa_1}{L^2}\left\{ \frac{w}{2}\,\kappa_1\,\left(1+\eta^2\right)^2\ddot{g}\left(w\eta^\prime\right) \Delta\eta^\prime+2\kappa_1\,\eta\left(1+\eta^2\right)\dot{g}\left(w\eta^\prime\right)\,\Delta\eta\right\}
\end{align}

Finally, the variation of the Euler-Lagrange equation proportional to $\chi$ is
\begin{align}
&\delta\left\{\left[ \frac{2}{L^2}\,\kappa_1^2\eta\left(1+\eta^2\right)\,g\left(w\,\eta^\prime\right)-\dvec{m}_3\,\kappa_1\right] \,\chi\right\}=\nonumber\\
&\hspace{.05\textwidth}-\chi\,\fvar{\left(\dvec{m}_3+\alpha\Delta\dvec{m}_3\right)\left(\kappa_1+\alpha\left[\Delta\boldsymbol{\theta}^\prime\cdot\uvec{d}_1\right]\right)}+...\\
&...+\frac{2\,\chi}{L^2}\fvar{\left(\kappa_1+\alpha\left[\Delta\boldsymbol{\theta}^\prime\cdot\uvec{d}_1\right]\right)^2\left(\eta+\alpha\Delta\eta\right)\left(1+\left[\eta+\alpha\Delta\eta\right]^2\right)g\left(w\left[\eta^\prime+\alpha\Delta\eta^\prime\right]\right)}\nonumber
\end{align}
\begin{align}
&=\chi\left\{\frac{4\,\kappa_1\,\eta\,\left(1+\eta^2\right)\,g\left(w\eta^\prime\right)}{L^2}-\dvec{m}_3\right\}\,\mathbf{R}^T\left[ \Delta\boldsymbol{\theta}^\prime \right]\cdot \uvec{e}_1-\chi\kappa_1\,\Delta\dvec{m}_3+...\nonumber
\\ &
\hspace{.05\textwidth}...+\frac{2\,w\,\chi}{L^2}\,\kappa_1^2\,\eta\,\left(1+\eta^2\right)\dot{g}\left(w\eta^\prime\right)\,\Delta\eta^\prime+\frac{2\,\chi}{L^2}\kappa_1^2\,\left(1+3\eta^2\right)\,g\left(w\eta^\prime\right)\,\Delta\eta\label{eq:lastpart}\end{align}
Combining the six terms in (\ref{eq:sixthtolastpart}), (\ref{eq:fifthtolastpart}), (\ref{eq:fourthtolast}),(\ref{eq:thirdtolastpart}), (\ref{eq:secondtolastpart}), and (\ref{eq:lastpart}) yields the formula for $DG_{static}$ given in (\ref{eq:DGstaticdev}) in section \ref{sec:devsecderiv}.
\section*{References}
\bibliographystyle{elsarticle-num} 

\end{document}